\documentstyle{livrev}
\RequirePackage{epsf,psfig}
\RequirePackage{longtable}

\newcommand{\mb}[1]{\mbox{\boldmath $#1$}}
\newcommand{\himpc}{{\hbox {$h^{-1}$}{\rm Mpc}} }

\newcommand{\0}{ {\scriptscriptstyle {0}} }
\newcommand{\E}{{\scriptscriptstyle E}}

\newcommand{\simgt}{\lower.5ex\hbox{$\; \buildrel > \over \sim \;$}}
\newcommand{\simlt}{\lower.5ex\hbox{$\; \buildrel < \over \sim \;$}}
\newcommand{\Omegam}{\Omega_{\rm\scriptscriptstyle m}}
\newcommand{\Omegal}{\Omega_{\scriptscriptstyle \Lambda}}

\def\hompc{\ifmmode {\,h\,\rm Mpc^{-1}} \else {$h^{-1}$~Mpc}\fi}
\def\s8m{{\sigma_{8{\rm m}}}}
\def\s8g{{\sigma_{8{\rm g}}}}

\def\Mpc{\, h^{-1} \, {\rm Mpc}}

\def\Mpc{\ifmmode {\, h^{-1} \, {\rm Mpc}}
\else {$h^{-1}\,$ Mpc}\fi}

\def\s8{{\sigma_8}}

\newcommand{\omegam}{\Omega_{\rm\scriptscriptstyle m}}
\newcommand{\omegal}{\Omega_{\scriptscriptstyle \Lambda}}
 
\def\omegab{{\Omega_{\rm\scriptscriptstyle b}}}

\def\omegabh2{{\omegab h^2}} 

\begin{document}

\title{Measuring our universe \\ from galaxy redshift surveys}

\author{Ofer Lahav$^{(1),(2)}$ and Yasushi Suto$^{(3),(4)}$\\
        $^{(1)}$ Department of Physics and Astronomy, University of London, \\ 
        Gower Street, London WC1E 6BT, UK \\
        $^{(2)}$ Institute of Astronomy, University of Cambridge, \\ 
        Madingley Road, Cambridge CB3 0HA, UK \\
        $^{(3)}$ Department of Physics, The University of Tokyo, \\
        Tokyo  113-0033, Japan \\
        $^{(4)}$ Research Center for the Early Universe, School of
        Science, \\  The University of Tokyo, 
        Tokyo  113-0033, Japan \\ ~ \\
        e-mail:~lahav@star.ucl.ac.uk, suto@phys.s.u-tokyo.ac.jp \\
\\
\small{(last modified: \today)}
}

\date{}
\maketitle

\begin{abstract}
Galaxy redshift surveys have achieved significant progress over the last
couple of decades. Those surveys tell us 
in the most straightforward way 
what our local  universe looks like. While the galaxy
distribution traces the {\it bright} side of the universe, detailed
quantitative analyses of the data have even revealed the {\it dark} side
of the universe dominated by non-baryonic dark matter as well as more
mysterious dark energy (or Einstein's cosmological constant). We
describe several methodologies of using galaxy redshift surveys as
cosmological probes, and then summarize the recent results from
the existing surveys. Finally we present our views on the future of
redshift surveys in the era of Precision Cosmology.
\end{abstract}

\keywords{cosmology: large-scale structure of universe --- 
cosmology: dark matter --- methods: statistical}

\clearpage

\tableofcontents

\clearpage

\section{Introduction}
\label{section:introduction}

Nowadays the exploration of the universe can be performed by a variety
of observational probes and methods over a wide range of the
wavelengths; temperature anisotropy map of the cosmic microwave
background (CMB), the Hubble diagrams (i.e., redshift--magnitude
relations) of nearby galaxies and distant Type Ia supernovae, wide-field
photometric and spectroscopic surveys of galaxies, power spectrum and
abundances of galaxy clusters in optical and X-ray bands combined with
the radio observation through the Sunyaev--Zel'dovich effect, deep
surveys of galaxies in submm, infrared, and optical bands, quasar
surveys in radio and optical, strong and weak lensing of distant
galaxies and quasars, high-energy cosmic rays and so on. Undoubtedly
gamma-rays, neutrinos, gravitational radiations will join the above
already crowded list.

Among those, optical galaxy redshift surveys are the most
classical. Indeed one may phrase that the modern observational cosmology
started with a sort of a galaxy redshift survey by Edwin Hubble. Still
galaxy redshift surveys are of vital importance in cosmology in the
21st century for various reasons:
\begin{description}
\item[Redshift surveys have unprecedented quantity and quality:]
\hfil\par
The numbers of galaxies and quasars in the spectroscopic sample of Two
Degree Field (2dF: see \S \ref{subsubsec:2dF}) are
$\sim 250,000$ and $\sim 30,000$, and will reach $\sim 800,000$ and
$100,000$ upon completion of the on-going Sloan Digital Sky Survey
(SDSS: see \S \ref{subsubsec:SDSS}). 
These unprecedented numbers of the objects as well as the
homogeneous selection criteria enable the precise statistical analysis
of their distribution.
\item[The universe at $z\approx1000$ is well specified:] \hfil\par The
first-year WMAP ({\it Wilkinson Microwave Anisotropy Probe}) data
~\cite{wmap-bennett} among others have established a set of
cosmological parameters. This may be taken as the {\it initial
condition} of the universe from the point-of-view of the structure
evolution toward $z=0$.  In a sense, the origin of the universe at
$z\approx1000$ and the evolution of the universe after the epoch are now
equally important, but well separable questions that particle and
observational cosmologists focus on, respectively.
\item[Gravitational growth of dark matter component is well
understood:] \hfil\par 
In addition, extensive numerical simulations of structure formation in
the universe has significantly advanced our understanding of the
gravitational evolution of dark matter component in the standard
gravitational instability picture. In fact, we even have very accurate
and useful analytic formulae to describe the evolution deep in its
nonlinear regime. Thus we can now directly address the evolution of {\it
visible objects} from the analysis of their redshift surveys separately
from the nonlinear growth of the underlying dark matter gravitational
potentials.
\item[Formation and evolution of galaxies:] \hfil\par In the era of
precision cosmology among others, scientific goals of researches out of
galaxy redshift surveys are gradually shifting from inferring a set of
values of cosmological parameters using galaxy as their probes to
understanding the origin and evolution of galaxy distribution given a
set of parameters accurately determined by the other probes like CMB and
supernovae.

\end{description}

With the above in mind, we will attempt to summarize what we have
learned so far from galaxy redshift surveys and then to describe what
will be done with future data. The review is organized as follows. We
first present a brief overview of the Friedmann model and gravitational
instability theory in \S \ref{section:clustering}. Then we describe the
non-Gaussian nature of density fluctuations generated by the nonlinear
gravitational evolution of the primordial Gaussian field (\S
\ref{section:statistics}).  Next we discuss the spatial biasing of
galaxies relative to the underlying dark matter distribution in \S
\ref{section:bias}. Our understanding of biasing is still far from
complete, and its description is necessarily empirical and very
approximate. Nevertheless this is one of the most important ingredients
for proper interpretation of galaxy redshift surveys.  Section
\ref{section:lightcone} introduces general relativistic effects which
become important especially for galaxies at high redshifts.  We present
the latest results from the two currently largest galaxy redshift
surveys, 2dF (Two Degree Field) and SDSS (Sloan Digital Sky Survey), in
\S \ref{section:2dFSDSS}. Finally \S \ref{section:discussion} is devoted
to summary of the present knowledge of our universe and our personal
view of the future direction of cosmological researches in the new
millennium.

\clearpage
\section{Clustering in the expanding universe}
\label{section:clustering}

\subsection{The cosmological principle}

Our current universe exhibits a wealth of nonlinear structures, but the
zero-th order description of our universe is based on the assumption
that the universe is homogeneous and isotropic smoothed over
sufficiently large scales. This statement is usually referred to as {\it
the cosmological principle}.  In fact, the cosmological principle was
first adopted when observational cosmology was in its infancy; it was
then little more than a conjecture, embodying 'Occam's razor' for the
simplest possible model.

Rudnicki (1995)~\cite{Rudnicki} summarized various forms of cosmological
principles in modern-day language, which were stated over different
periods in human history based on philosophical and aesthetic
considerations rather than on fundamental physical laws.
\begin{description}
\item[(i) The ancient Indian cosmological principle:] \hfil\par
{\it The Universe is infinite in space and time and is 
infinitely heterogeneous}.
\item[(ii) The ancient Greek cosmological principle:] \hfil\par
{\it Our Earth is the natural center of the Universe}.
\item[(iii) The Copernican cosmological principle:] \hfil\par
{\it The Universe as 
observed from any planet looks much the same}. 
\item[(iv) The (generalized) cosmological principle:] \hfil\par
{\it The Universe is (roughly) homogeneous and isotropic}.
\item[(v) The perfect cosmological principle:] \hfil\par
{\it The Universe is (roughly) homogeneous in space and time,
and is isotropic in space}.
\item[(vi) The anthropic principle:] \hfil\par
{\it A human being, as he/she is, can exist only in the Universe
as it is.}
\end{description}
We note that the ancient Indian principle may be viewed as a sort of
`fractal model'.  The perfect cosmological principle led to the steady
state model, which although more symmetric than the (generalized)
cosmological principle, was rejected on observational grounds.  The
anthropic principle is becoming popular again, e.g. in `explaining' the
non-zero value of the cosmological constant. 

Like with any other idea about the physical world, we cannot prove a
model, but only falsify it.  Proving the homogeneity of the Universe is
in particular difficult as we observe the Universe from one point in
space, and we can only deduce isotropy indirectly.  The practical
methodology we adopt is to assume homogeneity and to assess the level of
fluctuations relative to the mean, and hence to test for consistency
with the underlying hypothesis.  If the assumption of homogeneity turns
out to be wrong, then there are numerous possibilities for inhomogeneous
models, and each of them must be tested against the observations.

For that purpose, one needs observational data with good quality and
quantity extending up to high redshifts. Let us mention some of those.
\begin{description}
 \item[The CMB Fluctuations] 
Ehlers, Garen and Sachs (1968)~\cite{EGS-68} 
showed that by combining the CMB isotropy
with the Copernican principle one can deduce homogeneity. More formally
their theorem (based on the Liouville theorem) states that ``If the
fundamental observers in a dust space-time see an isotropic radiation
field, then the space-time is locally given by the
	    Friedman-Robertson-Walker (FRW) metric''.  The COBE (COsmic
	    Background Explorer) measurements of temperature
	    fluctuations
$\Delta T/T = 10^{-5} $ on scales of $10^\circ$ give via the Sachs Wolfe
effect ($\Delta T/T = \frac {1}{3} \Delta \phi/c^2$) and Poisson
equation rms density fluctuations of ${{\delta \rho} \over {\rho}} \sim
10^{-4} $ on $1000 \Mpc$ (e.g. \cite{WLR-99}),
which implies that the deviations from a smooth Universe are tiny.
 \item[Galaxy Redshift Surveys] 
The distribution of galaxies in local redshift surveys is highly clumpy,
with the Supergalactic Plane seen in full glory.  However, deeper
surveys like 2dF and SDSS (see \S \ref{section:2dFSDSS}) show that the
fluctuations decline as the length-scales increase. Peebles
(1993)~\cite{peebles-93} has shown that the angular correlation
functions for the Lick and APM (Automatic Plate Measuring) surveys scale
with magnitude as expected in a universe which approaches homogeneity on
large scales.  While redshift surveys can provide interesting estimates
of the fluctuations on intermediate scales (e.g. ~\cite{percival-02}),
the problems of biasing, evolution and $K$-correction, would limit the
ability of those redshift surveys to `prove' the Cosmological Principle.
Despite these worries the measurement of the power spectrum of galaxies
derived on the assumption of an underlying FRW metric shows good
agreement with the $\Lambda$-CDM (cold dark matter) model.
 \item[Radio Sources] 
Radio sources in surveys have typical median redshift ${\bar z} \sim 1$,
and hence are useful probes of clustering at high redshift.
Unfortunately, it is difficult to obtain distance information from these
surveys: the radio luminosity function is very broad, and it is
difficult to measure optical redshifts of distant radio sources.
Earlier studies claimed that the distribution of radio sources supports
the `Cosmological Principle'.  However, the wide range in intrinsic
luminosities of radio sources would dilute any clustering when projected
on the sky.  Recent analyses of new deep radio surveys
suggest that radio sources are actually clustered at least as strongly
as local optical galaxies.  Nevertheless, on
the very large scales the distribution of radio sources seems nearly
isotropic.
 \item[The XRB] The X-ray Background (XRB) is likely to be due to
sources at high redshift.  The XRB sources
are probably located at redshift $z < 5$, making them convenient tracers
of the mass distribution on scales intermediate between those in the CMB
as probed by COBE, and those probed by optical and IRAS redshift
surveys. The interpretation of the results depends somewhat on the
nature of the X-ray sources and their evolution.  By comparing the
predicted multipoles to those observed by HEAO1,
Scharf et al. (2000)~\cite{scharf-00} estimate the amplitude of
	    fluctuations for an assumed
shape of the density fluctuations.  The observed
fluctuations in the XRB are roughly as expected from interpolating
between the local galaxy surveys and the COBE and other CMB experiments.
The rms fluctuations ${ {\delta \rho} \over {\rho} }$ on a scale of
$\sim 600 h^{-1}$Mpc are less than 0.2 \%.
\end{description}

Since the (generalized) cosmological principle is now well
supported by the above observations, we shall assume below
that it holds over scales  $ l > 100 h^{-1}$ Mpc.

The rest of the current section is devoted to a brief review of the
homogeneous and isotropic cosmological model. Further details may be
easily found in standard cosmology textbooks~\cite{weinberg-72, 
pad-93, peebles-93, peacock-99, coles-lucchin-02, padmanabhan-02}.

The cosmological principle is mathematically paraphrased as that the
metric of the universe (in its zero-th order approximation) is given by
\begin{eqnarray}
\label{eq:rw}
 ds^2 = - dt^2 + a(t)^2 \left( {dx^2 \over 1-Kx^2} 
        + x^2d\theta^2 + x^2{\rm sin}^2\theta d\phi^2 \right) ,
\end{eqnarray}
where $x$ is the comoving coordinate, and we use the units in which the
light velocity $c=1$.  The above Robertson -- Walker metric is specified
by a constant $K$, the spatial curvature, and a function of time $a(t)$,
the scale factor.

The homogeneous and isotropic assumption also implies that the energy
momentum tensor of the matter field, $T_{\mu\nu}$, should take the form
of the ideal fluid:
\begin{eqnarray}
\label{eq:emtensor}
 T_{\mu\nu}  = ( \rho +  p)u_\mu u_\nu +  p  \, g_{\mu\nu} ,
\end{eqnarray}
where $u_\mu$ is the 4-velocity of the matter, $\rho$ is the mean energy
density, and $p$ is the mean pressure.

\subsection{From the Einstein equation to the Friedmann equation}

The next task is to write down the  Einstein equation:
\begin{equation}
\label{eq:einstein}
R_{\mu\nu} - {1 \over 2}R \, g_{\mu\nu} + \Lambda \, g_{\mu\nu} 
  =  8\pi G \, T_{\mu\nu}
\end{equation}
using equations (\ref{eq:rw}) and (\ref{eq:emtensor}). In this case one
is left with the following two independent equations:
\begin{eqnarray}
\label{eq:da}
\left( {1 \over a} {d a \over dt} \right)^2  
   & = & {8\pi G \over 3} \rho  - {K \over a^2} + {\Lambda \over 3} , \\ 
\label{eq:d2a}
 {1 \over a} {d^2 a \over dt^2} 
   & = & - {4\pi G \over 3} ( \rho + 3p) + {\Lambda \over 3} ,
\end{eqnarray}
for the three independent functions, $a(t)$, $\rho(t)$, and $p(t)$.

Differentiation of equation (\ref{eq:da}) with respect to $t$ yields
\begin{eqnarray}
\frac{\ddot a}{a} = \frac{4\pi G}{3}
 \left(\dot\rho \frac{a}{\dot a}+2\rho\right) + \frac{\Lambda}{3} .
\end{eqnarray}
Then eliminating $\ddot a$ with equation (\ref{eq:d2a}), one obtains
\begin{eqnarray}
\label{eq:drhodt}
\dot \rho = -3 \frac{\dot a}{a}(\rho+p) .
\end{eqnarray}
This can be easily interpreted as the 1st law of the thermodynamics
$dQ=dU-pdV= d(\rho a^3) -pd(a^3)=0$ in the present context.  Equations
(\ref{eq:da}) and (\ref{eq:drhodt}) are often used as the two
independent basic equations for $a(t)$, instead of equations
(\ref{eq:da}) and (\ref{eq:d2a}).  

In either case, however, one needs another independent equation to solve
for $a(t)$. This is usually given by the equation of state of the form
$p=p(\rho)$.  In cosmology, the following simple relation:
\begin{equation}
p=w\rho .
\end{equation}
While the value of $w$ may change as redshift in principle, it is often
assumed that $w$ is independent of time just for simplicity.  Then
substituting this equation of state into equation (\ref{eq:drhodt})
immediately yields
\begin{eqnarray}
\label{eq:rho-a}
\rho \propto a^{-3(1+w)} .
\end{eqnarray}
The non-relativistic matter (or dust), ultra-relativistic matter (or
radiation), and the cosmological constant correspond to $w=0$, $1/3$,
and $-1$, respectively. 

If the universe consists of different fluid species with $w_i$
($i=1,\ldots,N$), equation (\ref{eq:rho-a}) still holds independently as
long as they do not interact with each other. If one denotes the present
energy density of the $i$-th component by $\rho_{i,\0}$, then the total
energy density of the universe at the epoch corresponding to the scale
factor of $a(t)$ is given by
\begin{eqnarray}
\rho = \sum_{i=1}^N \frac{\rho_{i,\0}}{a^{3(1+w_i)}} ,
\end{eqnarray}
where the present value of the scale factor, $a_\0$, is set to be unity
without loss of generality.  Thus, equation (\ref{eq:da}) becomes
\begin{eqnarray}
\label{eq:da_sumrhoi}
\left( {1 \over a} {d a \over dt} \right)^2  
 =  {8\pi G \over 3} \sum_{i=1}^N \frac{\rho_{i,\0}}{a^{3(1+w_i)}}
 - {K \over a^2} + {\Lambda \over 3} .
\end{eqnarray}
Note that those components with $w_i=-1$ may be equivalent to the
conventional cosmological constant $\Lambda$ at this level, although
they may exhibit spatial variation unlike $\Lambda$.

Evaluating equation (\ref{eq:da_sumrhoi}) at the present epoch, one finds
\begin{eqnarray}
H_\0^2 =  {8\pi G \over 3} \sum_{i=1}^N \rho_{i,\0}
 - K + {\Lambda \over 3} ,
\end{eqnarray}
where $H_\0$ is the Hubble constant at the present epoch. The above
equation is usually rewritten as follows:
\begin{eqnarray}
\label{eq:komega}
K \equiv H_\0^2 \Omega_K  =  H_\0^2 
\left( \sum_{i=1}^N \Omega_{i,\0} + \Omegal -1 \right),
\end{eqnarray}
where the density parameter for the $i$-th component is defined as
\begin{eqnarray}
 \Omega_{i,\0} \equiv \frac{8\pi G}{3H_\0^2} \,  \rho_{i,\0},
\end{eqnarray}
and similarly the dimensionless cosmological constant is
\begin{eqnarray}
 \Omegal \equiv \frac{\Lambda}{3H_\0^2} .
\end{eqnarray}
Incidentally equation (\ref{eq:komega}) clearly illustrates the Mach
principle in the sense that the space curvature is simply determined by
the amount of matter components in the universe. In particular, the flat
universe $(K=0)$ implies that the sum of the density parameters is
unity:
\begin{eqnarray}
 \sum_{i=1}^N \Omega_{i,\0} + \Omegal = 1 .
\end{eqnarray}

Finally  the cosmic expansion is described by
\begin{equation}
\label{eq:dadt_all}
\left(\frac{da}{dt} \right)^2 =  H_\0^2
\left(
 \sum_{i=1}^N \frac{\Omega_{i,\0}}{a^{1+3w_i}}
+1 - \sum_{i=1}^N \Omega_{i,\0} - \Omegal
+ \Omegal a^2
\right) .
\end{equation}
As will be shown below, the present universe is supposed to be dominated
by non-relativistic matter (baryons and collisionless dark matter) and
the cosmological constant. So in the present review, we approximate
equation (\ref{eq:dadt_all}) as
\begin{equation}
\label{eq:dadt}
\left(\frac{da}{dt} \right)^2 =  H_\0^2
\left(
\frac{\Omegam}{a} +1 - \Omegam - \Omegal + \Omegal a^2
\right) 
\end{equation}
unless otherwise stated.

\subsection{Expansion law and age of the universe \label{subsec:expand}}

\begin{figure}[h]
\begin{center}
    \leavevmode\epsfxsize=10cm \epsfbox{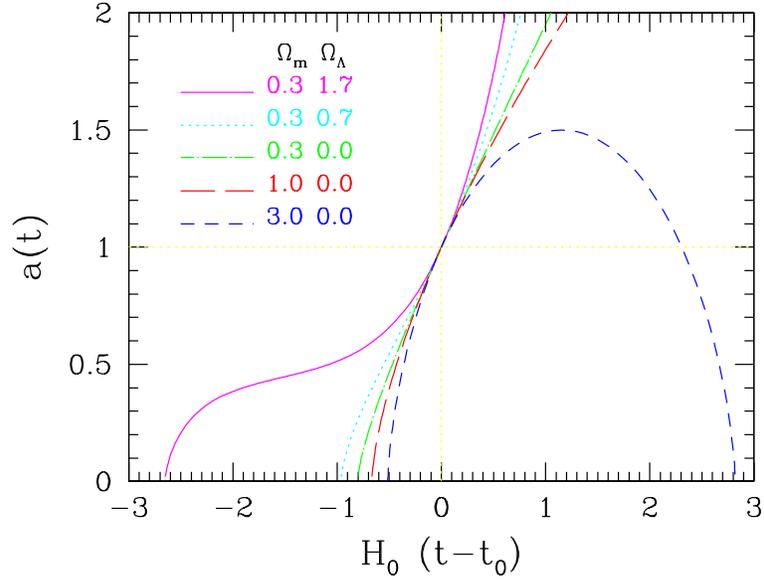}
\caption{Evolution of the cosmic scale factor as a function of
$H_\0(t-t_\0)$.  The present value of the scale factor $a_\0$ is set to
unity; solid line: $(\Omegam, \Omegal) = (0.3,1.7)$, dotted line:
$(0.3,0.7)$, dot-dashed line: $(0.3,0.0)$, long dashed line:
$(1.0,0.0)$, short dashed line: $(3.0,0.0)$.}  \label{fig:rvst}
\end{center}
\end{figure}

Equation (\ref{eq:dadt}) has the following analytic solutions in several
simple but practically important cases.
\begin{itemize}
\item[(a)] Einstein -- de Sitter model $(\Omegam=1, \Omegal=0)$
  \begin{equation}
 a(t) = \left({t \over t_\0}\right)^{2/3}, \qquad t_\0 = {2 \over 3 H_\0}
  \end{equation}
\item[(b)] Open model with vanishing cosmological constant  
$(\Omegam<1, \Omegal=0)$
\begin{eqnarray}
a &=& {{\displaystyle {\Omegam \over 2(1-\Omegam)}}}
    \left( \cosh \theta -1 \right), \quad
H_\0 t = {{\displaystyle {\Omegam \over 2(1-\Omegam)^{3/2}}}} 
   \left( \sinh \theta - \theta \right) \qquad \\
H_\0 t_\0 &=& {\displaystyle {1 \over 1-\Omegam} -
 {\Omegam \over 2(1-\Omegam)^{3/2}} 
   \ln {2-\Omegam+2\sqrt{1-\Omegam} \over \Omegam} }
  \end{eqnarray}
\item[(c)] Spatially-flat model with  cosmological constant
  $(\Omegam<1, \Omegal=1-\Omegam)$
  \begin{eqnarray}
a(t) &=& \left({\Omegam \over 1-\Omegam}\right)^{1/3} 
\left[ \sinh {3\sqrt{1-\Omegam} \over 2} H_\0 t \right]^{2/3} , \\
H_\0 t_\0 &=& {\displaystyle {1 \over 3 \sqrt{1-\Omegam} } 
   \ln {2-\Omegam+2\sqrt{1-\Omegam} \over \Omegam} }
  \end{eqnarray}
\end{itemize}
In the above, $t_\0$ denotes the present age of the universe:
\begin{equation}
\label{eq:t0}
t_\0 = {1 \over H_\0} 
\int_\0^1 { x\, dx \over 
\sqrt {\Omegam x + (1-\Omegam-\Omegal) x^2 + \Omegal x^4} } ,
\end{equation}
and we adopt the initial condition that $a=0$ at $t=0$.  The expression
clearly indicates that $t_\0$ increases as $\Omegam$ decreases and/or
$\Omegal$ increases.  Figure~\ref{fig:rvst} plots the scale factor as a
function of $H_\0(t-t_\0)$, and Table \ref{tab:t0} summarizes the age of
the universe.
\begin{table}[h]
\caption{The present age of the universe [$(h/0.7)^{-1}$ Gyr].}
\label{tab:t0}
\begin{center}
\begin{tabular}{l|c|c} 
$\Omegam$& Open model ($\Omegal=0$) 
& Spatially-flat model ($\Omegal=1-\Omegam$)\\ \hline
1.0 & ~9.3 & ~9.3 \\ 
0.5 & 10.5 & 11.6 \\ 
0.3 & 11.3 & 13.5 \\ 
0.1 & 12.5 & 17.8 \\ 
0.01 & 13.9 & 28.0\\
\end{tabular} 
\end{center}
\end{table}

\subsection{Einstein's static model and Lema\^{\i}tre's model}

So far we have shown that solutions of the Einstein equation are {\it
dynamical} in general, i.e., the scale factor $a$ is time-dependent. As
a digression, let us examine why Einstein once introduced the
$\Lambda$-term to obtain a static cosmological solution. This is mainly
important for historical reasons, but also is interesting to observe how
the operationally identical parameter (the $\Lambda$-term, the
cosmological constant, the vaccum energy, the dark energy) shows up in
completely different contexts in the course of the development of
cosmological physics.

Consider first the case of $\Lambda=0$ in equations (\ref{eq:da}) and
(\ref{eq:d2a}). Clearly the necessary and sufficient condition that the
equations admit the solution of $a=const.$ is given by
\begin{eqnarray}
\rho = -3p = \frac{3K}{8\pi G a^2} .
\end{eqnarray}
Namely, any static model requires that the universe is dominated by
matter with either negative pressure or negative density. This is
physically unacceptable as long as one considers {\it normal} matter in
the standard model of particle physics. If $\Lambda \not= 0$ on the
other hand, the condition for the static solution is
\begin{eqnarray}
K = 4\pi G (\rho + p)a^2 , \qquad \Lambda = 4\pi G (\rho + 3p) ,
\end{eqnarray}
yielding
\begin{eqnarray}
\rho = \frac{1}{8\pi G}\left(\frac{3K}{a^2} - \Lambda \right) ,
\qquad p = \frac{1}{8\pi G}\left(\Lambda - \frac{K}{a^2}\right) .
\end{eqnarray}
Thus both $\rho$ and $p$ can be positive if
\begin{eqnarray}
\frac{K}{a^2} \le \Lambda \le \frac{3K}{a^2} .
\end{eqnarray}
In particular, if $p=0$,
\begin{eqnarray}
\rho = \frac{\Lambda}{4\pi G}, \qquad K = \Lambda a^2 .
\end{eqnarray}
This represents the closed universe (with positive spatial curvature),
and corresponds to Einstein's static model.

The above static model is a special case of Lema\^{\i}tre's universe
model with $\Lambda>0$ and $K>0$. For simplicity, let us assume that the
universe is dominated by non-relativistic matter with negligible
pressure, and consider the behavior of Lema\^{\i}tre's model. First
define the values of the density and the scale factor corresponding to
Einstein's static model:
\begin{eqnarray}
\rho_\E = \frac{\Lambda}{4\pi G}, \qquad K = \Lambda a_\E^2 .
\end{eqnarray}
In order to study the stability of the model around the static model,
consider a model in which the density at $a=a_\E$ is a factor of
$\alpha(>1)$ larger than $\rho_\E$. Then
\begin{eqnarray}
\rho = \alpha \rho_\E \left(\frac{a_\E}{a}\right)^3
= \frac{\alpha \Lambda}{4\pi G} \left(\frac{a_\E}{a}\right)^3 ,
\end{eqnarray}
and equations (\ref{eq:da}) and (\ref{eq:d2a}) reduce to
\begin{eqnarray}
\label{eq:dalemaitre}
\dot{a}^2 &=& \frac{\Lambda}{3} \left(\frac{2\alpha a_\E^3}{a}
-3 a_\E^2 + a^2 \right) , \\
\label{eq:d2alemaitre}
\frac{\ddot{a}}{a} &=& \frac{\Lambda}{3} 
\left[1-\alpha \left(\frac{a_\E}{a}\right)^3 \right] .
\end{eqnarray}
For the period of $a \ll a_\E$, equation (\ref{eq:dalemaitre}) indicates
that $a \propto t^{2/3}$ and the universe is decelerating
($\ddot{a}<0$). When $a$ reaches $\alpha^{1/3}a_\E$, $\dot{a}^2$ takes
the minimum value $\Lambda a_\E^2(\alpha^{2/3}-1)$ and the universe
becomes accelerating ($\ddot{a}>0$). Finally the universe approaches the
exponential expansion or de Sitter model: $a \propto
\exp(t\sqrt{\Lambda/3})$. If $\alpha$ becomes closer to unity, the
minimum value reaches zero and the expansion of the universe is
effectively frozen. This phase is called the coasting period, and the
case with $\alpha=1$ corresponds to Einstein's static model in which the
coasting period continues forever. A similar consideration for
$\alpha<1$ indicates that the universe starts collapsing ($\dot{a}^2=0$)
before $\ddot{a}=0$. Thus the behavior of Lema\^{\i}tre's model is
crucially different if $\alpha$ is larger or smaller than unity. This
suggests that Einstein's static model ($\alpha=1$) is unstable.

\subsection{Vacuum energy as an effective cosmological constant}

So far we discussed the cosmological constant introduced in the
l.h.s. of the Einstein equation. Formally one can move the
$\Lambda$-term to the r.h.s. by assigning 
\begin{eqnarray}
\rho_\Lambda = \frac{\Lambda}{8\pi G}, \qquad
p_\Lambda = - \frac{\Lambda}{8\pi G} .
\end{eqnarray}
This {\it effective} matter field, however, should satisfies an equation
of state of $p= -\rho$. Actually the following example presents a
specific example for an {\it effective} cosmological constant.
Consider a real scalar field whose Lagrangian density is given by
\begin{equation}
{\cal L} = \frac{1}{2} g_{\mu\nu}
\partial^\mu \phi\partial^\nu \phi - V(\phi) .
\end{equation}
Its energy-momentum tensor is
\begin{equation}
T_{\mu\nu} = 
\partial_\mu \phi\partial_\nu \phi - {\cal L}g_{\mu\nu} ,
\end{equation}
and if the field is spatially homogeneous, its energy density and
pressure are
\begin{equation}
\rho_\phi = \frac{1}{2} \dot\phi^2  + V(\phi) ,
\qquad p_\phi = \frac{1}{2} \dot\phi^2  - V(\phi) .
\end{equation}
Clearly if the evolution of the field is negligible, i.e., $\dot\phi^2
\ll V(\phi)$, $p_\phi \approx - \rho_\phi$ and the field acts as a
cosmological constant. Of course this model is one of the simplest
examples, and one may play with much more complicated models if needed.

If the $\Lambda$-term is introduced in the l.h.s., its should be
constant to satisfy the energy-momentum conservation
$T_{\mu\nu}^{~~~;\nu}=0$. Once it is regarded as a sort of matter field
in the r.h.s, however, it does not have to be constant. In fact, the
above example shows that the equation of state for the field has $w=-1$
only in special cases. This is why recent literature refers to the
field as {\it dark energy} instead of the cosmological constant.

\subsection{Gravitational instability}

We have presented the zero-th order description of the universe
neglecting the inhomogeneity or spatial variation of matter inside.
Now we are in a position to consider the evolution of matter in the
universe. For simplicity we focus on the non-relativistic regime where
Newtonian approximation is valid. 
Then the basic equations for the self-gravitating fluid are given by
the continuity equation, Euler's equation and the Poisson equation:
\begin{eqnarray}
\label{eq:cont0}
{\partial \rho \over \partial t} +  {\mb \nabla}\cdot
   \left(\rho {\mb u}\right) &=& 0 , \\ 
 \label{eq:euler0}
{\partial \mb u \over \partial t} + 
({\mb u}\cdot{\mb \nabla}){\mb u} &=& 
- {1 \over \rho} {\mb \nabla} p  -  {\mb \nabla} \Phi, \\
\label{eq:field0}
\nabla^2 \Phi &=& 4\pi G \rho . 
\end{eqnarray}
We would like to rewrite those equations in the comoving frame. For this
purpose, we introduce the position $\mb x$ in the comoving coordinate,
the peculiar velocity, $\mb v$, density fluctuations $\delta(t,\mb x)$,
and the gravitational potential $\phi(t, \mb x)$ which are defined
respectively as
\begin{eqnarray}
\mb x &=& \frac{\mb r}{a(t)} , \\
\mb v &=& a(t) \dot{\mb x} , \\
\delta (t,\mb x) &=& \frac{\rho (t,\mb x)}{\bar \rho (t)} -1 ,\\
\phi(t,\mb x) &=& \Phi + {1\over 2}a \ddot a |\mb x|^2 .
\end{eqnarray}
Then equations (\ref{eq:cont0}) to (\ref{eq:field0}) reduce to
\begin{eqnarray}
\label{eq:cont}
\dot \delta &+& {1  \over a} {\mb \nabla}\cdot
   \left[(1+\delta) {\mb v}\right] = 0 , \\ 
 \label{eq:euler}
\dot {\mb v} &+& {1  \over a}({\mb v}\cdot{\mb \nabla}){\mb v}
    + {\dot a \over a}{\mb v} = - {1 \over \rho a} {\mb \nabla} p
    - {1 \over a} {\mb \nabla} \phi, \\
\label{eq:field}
\nabla^2 \phi &=& 4\pi G \bar \rho a^2 \delta  ,
\end{eqnarray}
where the dot and {\mb \nabla} in the above equations are the time
derivative for a given $\mb x$ and the spatial derivative with respect
to $\mb x$, i.e., defined in the comoving coordinate (while those in
eqs.[\ref{eq:cont0}] to [\ref{eq:field0}] are defined in the proper
coordinate).

A standard picture of the cosmic structure formation assumes that the
initially tiny amplitude of density fluctuation grow according to
equations (\ref{eq:cont}) to (\ref{eq:field}). Also the universe
smoothed over large scales approaches a homogeneous model.  Thus at
early epochs and/or on large scales, the nonlinear effect is small and
one can linearize those equations with respect to $\delta$ and $\mb v$:
\begin{eqnarray}
\label{eq:contl}
\dot \delta &+& {1  \over a} {\mb \nabla}\cdot{\mb v} = 0 , \\ 
 \label{eq:eulerl}
\dot {\mb v} &+& {\dot a \over a}{\mb v} 
= - {c_s^2 \over a} {\mb \nabla} \delta
    - {1 \over a} {\mb \nabla} \phi, \\
\label{eq:fieldl}
\nabla^2 \phi &=& 4\pi G \bar \rho a^2 \delta  ,
\end{eqnarray}
where $c_s^2 \equiv (\partial p/\partial \rho)$ is the sound velocity.

As usual, we transform the above equations in $\mb k$ space using
\begin{eqnarray}
\delta_{\mb k}(t) \equiv { 1\over V} \int  
\delta(t, \mb x)\, {\rm exp}(i\mb k\cdot \mb x)\,d\mb x .
\end{eqnarray}
Then the equation for $\delta_{\mb k}$ reduces to
\begin{equation}
\label{eq:delta2k}
\ddot \delta_{\mb k} + 2 {\dot a \over a} \dot \delta_{\mb k}
+ \left( {c_s^2k^2 \over a^2} 
- 4\pi G\bar\rho \right)\delta_{\mb k} = 0 .
\end{equation}
If the signature of the third term is positive, $\delta_{\mb k}$ has an
unstable, or, monotonically increasing solution. This condition is
equivalent to {\it the Jeans criterion}:
\begin{equation}
\label{eq:jeansl}
\lambda \equiv {2\pi \over k} > \lambda_{\rm J} 
\equiv c_s \sqrt{\pi  \over G\bar\rho} ,
\end{equation}
namely, the wavelength of the fluctuation is larger than the Jeans
length $\lambda_{\rm J}$ which characterizes the scale that the sound
wave can propagate within the dynamical time of the fluctuation
$\sqrt{\pi/G\bar\rho}$. Below the scale the pressure wave can suppress
the gravitational instability, and the fluctuation amplitude oscillates.

\subsection{Linear growth rate of the density fluctuation}

Most likely our universe is dominated by collisionless dark matter, and
thus $\lambda_{\rm J}$ is negligibly small. Thus at most scales of
cosmological interest, equation (\ref{eq:jeansl}) is well approximated
as
\begin{equation}
\label{eq:delta2}
\ddot \delta_{\mb k} + 2 {\dot a \over a} \dot \delta_{\mb k}
- 4\pi G\bar\rho \delta_{\mb k} = 0 .
\end{equation}
For a given set of cosmological parameters, one can solve the above
equation by substituting the expansion law for $a(t)$ as described in \S
\ref{subsec:expand}. Since equation (\ref{eq:delta2}) is the
second-order differential equation with respect to $t$, there are two
independent solutions; a decaying mode and a growing mode which
monotonically decreases and increases as $t$, respectively.  The former
mode becomes negligibly small as the universe expands, and thus one is
usually interested in the growing mode alone.

More specifically those solutions are explicitly obtained as follows.
First note that the l.h.s. of equation (\ref{eq:dadt}) is the Hubble
parameter at $t$, $H(t)=\dot a/a$:
\begin{eqnarray}
\label{eq:Ha}
H^2 &=&  H_\0^2
\left(
\frac{\Omegam}{a^3} + \frac{1 - \Omegam - \Omegal}{a^2} + \Omegal
\right) \\ 
\label{eq:Hz}
&=& H_\0^2 \left[\Omegam(1+z)^3 +(1-\Omegam-\Omegal)(1+z)^2
  +\Omegal\right] .
\end{eqnarray}
The first and second differentiation of equation (\ref{eq:Ha}) with
respect to $t$ yields
\begin{eqnarray}
\label{eq:HdotH}
2H \dot H =  H_\0^2
\left(
-3 \frac{\Omegam}{a^3} -2 \frac{1 - \Omegam - \Omegal}{a^2}
\right) H.
\end{eqnarray}
and
\begin{eqnarray}
\label{eq:ddotH}
\ddot H  =  H_\0^2
\left(
9 \frac{\Omegam}{2a^3} +2 \frac{1 - \Omegam - \Omegal}{a^2}
\right) H.
\end{eqnarray}
Thus the differential equation for $H$ reduces to
\begin{equation}
\label{eq:dhdt}
\ddot H + 2H\dot H = H_\0^2H \frac{3\Omegam}{2a^3} 
= 4\pi G \bar \rho H .
\end{equation}
This coincides with the linear perturbation equation for $\delta_{\mb
k}$ (eq.[\ref{eq:delta2}]). Since $H(t)$ is a decreasing function of $t$,
this implies that $H(t)$ is the decaying solution for equation
(\ref{eq:delta2}). Then the corresponding growing solution $D(t)$ can be
obtained according to the standard procedure:
subtracting equation (\ref{eq:delta2}) from equation (\ref{eq:dhdt})
yields
\begin{equation}
\label{eq:wronskian}
a^2{d \over dt}(\dot D H - D\dot H) +
{da^2 \over dt}(\dot D H - D\dot H) 
= \frac{d}{dt} 
\left[a^2H^2 \frac{d}{dt} \left(\frac{D}{H}\right)\right] 
=0 ,
\end{equation}
and therefore the formal expression for the growing solution in linear
theory is
\begin{equation}
\label{eq:growthdt}
D(t) \propto H(t)\int_0^{t} {dt' \over a^2(t') H^2(t')} .
\end{equation}
It is often more useful to rewrite $D(t)$ in terms of the redshift $z$
as follows:
\begin{eqnarray}
\label{eq:growthdz}
D(z) &=& {5\Omegam H^2_0 \over 2} H(z)\int_z^\infty 
\frac{1+z'}{H^3(z')}dz',
\end{eqnarray}
where the proportional factor is chosen so as to reproduce
$D(z) \rightarrow 1/(1+z)$ for $z\rightarrow \infty$. Linear growth
rates for the models described in \S \ref{subsec:expand} are summarized
below.
\begin{itemize}
\item[(a)] Einstein -- de Sitter model $(\Omegam=1, \Omegal=0)$
  \begin{equation}
 D(z) = {1 \over 1+z}   
  \end{equation}
\item[(b)] Open model with vanishing cosmological constant  
$(\Omegam<1, \Omegal=0)$
  \begin{equation}
D(z) \propto 1 + {3\over x} + 3\sqrt{ 1+x \over x^3} 
\ln \left( \sqrt{1+x} - \sqrt{x} \right),  
 \quad x \equiv {1-\Omegam \over \Omegam(1+z)}
  \end{equation}
\item[(c)] Spatially-flat model with  cosmological constant
  $(\Omegam<1, \Omegal=1-\Omegam)$
  \begin{equation}
D(z) \propto  \sqrt{ 1+ {2\over x^3}} 
 \int_0^x \left({u \over 2+u^3}\right)^{3/2} du, 
 \quad x \equiv {2^{1/3}(\Omegam^{-1}-1)^{1/3} \over 1+z} .
  \end{equation}
\end{itemize}
For most purposes, the following fitting formulae~\cite{pd-96} provide
sufficiently accurate approximations.
\begin{eqnarray}
\label{eq:fitDzpd}
 D(z) &=& \frac{g(z)}{1+z} ,  \\
g(z) &=& \frac{5\Omega(z)}{2}
\frac{1}
{\Omega^{4/7}(z) - \lambda(z)+[1+\Omega(z)/2][1+\lambda(z)/70]} , 
\end{eqnarray}
where
\begin{eqnarray}
&&\Omega(z) = \Omegam (1+z)^3 \left[\frac{H_\0}{H(z)}\right]^2
= \frac{\Omegam (1+z)^3}
{\Omegam(1+z)^3 +(1-\Omegam-\Omegal)(1+z)^2+\Omegal}, \qquad \\
&&\lambda(z) = \Omegal \left[\frac{H_\0}{H(z)}\right]^2
= \frac{\Omegal}{\Omegam(1+z)^3 +(1-\Omegam-\Omegal)(1+z)^2+\Omegal} .
\end{eqnarray}
Note that $\Omega_{\rm m}$ and $\Omega_{\Lambda}$ refer to the {\it
present} values of the density parameter and the dimensionless
cosmological constant, respectively, which will be frequently used in
the rest of the review.

\begin{figure}[h]
\begin{center}
    \leavevmode\epsfxsize=8cm \epsfbox{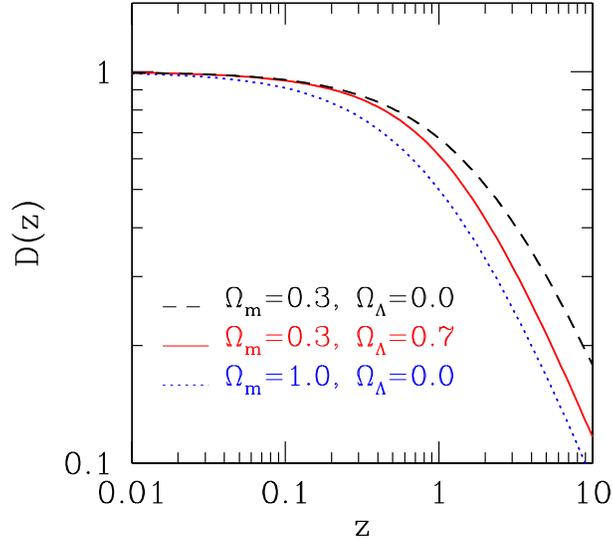}
\caption{Linear growth rate of density fluctuations.} 
\label{fig:growthrate}
\end{center}
\end{figure}

Figure~\ref{fig:growthrate} shows the comparison of the numerically
computed growth rate (thick lines) against the above fitting formulae
(thin lines) which are practically indistinguishable.

\clearpage
\section{Statistics of cosmological density fluctuations}
\label{section:statistics}
\subsection{Gaussian random field}

Consider the density contrast $\delta_i \equiv \delta({\mb x}_i) =
 \rho(\mb x)/\bar \rho -1$ defined at the comoving position ${\mb x}_i$.
The density field is regarded as a stochastic variable, and thus forms
 a random field. The conventional assumption is that the primordial
 density field (in its linear regime) is Gaussian, i.e., its $m$-point
 joint probability distribution obeys the multi-variate Gaussian:
\begin{eqnarray}
\label{eq:mgauss}
&& P(\delta_1, \delta_2, \ldots, \delta_m) \, d\delta_1 d\delta_2 \cdots
d\delta_m \cr
&& \hspace*{1cm}
= {1 \over \sqrt {(2\pi)^m {\rm det}(M)}} 
  \exp \left[- \sum_{i,j=1}^{m} {1\over 2} 
   \delta_i (M^{-1})_{ij}\delta_j \right]
   d\delta_1 d\delta_2 \cdots d\delta_m , \qquad
\end{eqnarray}
for an arbitrary positive integer $m$.  Here $M_{ij} \equiv \langle
\delta_i \delta_j \rangle$ is the covariance matrix, and $M^{-1}$ is its
inverse. Since $M_{ij} =\xi({\mb x}_i,{\mb x}_j)$, equation
(\ref{eq:mgauss}) implies that the statistical nature of the Gaussian
density field is completely specified by the two-point correlation
function $\xi$ and its linear combination (including its derivative and
integral).  For an extensive discussion of the cosmological Gaussian
density field, see ref.~\cite{bbks-86}.

The Gaussian nature of the primordial density field is preserves in its
linear evolution stage, but this is not the case in nonlinear stage.
This is clear even from the definition of the Gaussian distribution:
equation (\ref{eq:mgauss}) formally assumes that the density contrast
distributes symmetrically in the range of $-\infty < \delta_i < \infty$,
but in the real density field $\delta_i$ cannot be less than $-1$.  This
assumption does not make any practical difference as long as the
fluctuations are (infinitesimally) small, but is invalid in nonlinear
regime where the typical amplitude of the fluctuations exceeds unity.

In describing linear theory of cosmological density fluctuations,
the Fourier transform of the spatial density contrast
$\delta(\mb x)$ $\equiv \rho(\mb x)/\langle \bar \rho \rangle -1$
is the most basic variable:
\begin{equation}
  \label{eq:deltakdef}
\delta_{\mb k} = {1 \over V}\int d\mb x
\delta(\mb x) {\rm exp}(i\mb k\cdot \mb x) .
\end{equation}
Since $\delta_{\mb k}$ is a complex variable, it is decomposed by a set
of two real variables, the amplitude $D_{\mb k}$ and the phase
$\phi_{\mb k}$:
\begin{equation}
  \label{eq:deltak}
\delta_{\mb k} \equiv D_{\mb k} {\rm exp}(i\phi_{\mb k}) .
\end{equation}
Then linear perturbation equation reads
\begin{eqnarray}
\label{eq:linamp}
\ddot D_{\mb k} + 2 {\dot a \over a} \dot D_{\mb k}
- (4\pi G\bar\rho + \dot \phi^2) D_{\mb k} = 0 , \\
\label{eq:linphase}
\ddot \phi_{\mb k} + 2\left({\dot a \over a}
+ {\dot D_{\mb k} \over D_{\mb k} }\right)
 \dot \phi_{\mb k} = 0 .
\end{eqnarray}
Equation (\ref{eq:linphase}) yields $\dot\phi(t) \propto
a^{-2}(t)D^{-2}_{\mb k}(t)$, and $\phi(t)$ rapidly converges to a
constant value. Thus $D_{\mb k}$ evolves following the growing solution
in linear theory.

The most popular statistic of clustering in the universe is the power
spectrum of the density fluctuations:
\begin{equation}
P(t, \mb k) \equiv \langle D_{\mb k}(t)^2 \rangle 
\end{equation}
which measures the amplitude of the mode of the wavenumber $\mb k$.
This is the Fourier transform of the two-point correlation function:
\begin{equation}
\label{eq:xipk}
\xi (\mb x, t) = {1 \over 8\pi^3}\int P(t, \mb k) 
{\rm exp}(- i\mb k\cdot \mb x)\,d\mb k .
\end{equation}
If the density field is globally homogeneous and isotropic (i.e., no
preferred position or direction), equation (\ref{eq:xipk}) reduces to
\begin{equation}
\xi (x,t) = {1 \over 2\pi^2}
\int_0^\infty P(t,k) { {\rm sin} kx \over x} kdk .
\end{equation}
Since the above expression obtained after the ensemble average, $x$ does
not denote an amplitude of the position vector, but a comoving
wavelength $2\pi/k$ corresponding to the wavenumber $k = |\mb k|$.  It
should be noted that neither the power spectrum or the two-point
correlation function contains information for the phase $\phi_{\mb k}$.
Thus in principle two clustering patterns may be completely different
even if they have the identical two-point correlation functions.
This implies the practical importance to describe the
statistics of phases $\phi_{\mb k}$ in addition to the amplitude $D_{\mb
k}$ of clustering.

In the Gaussian field, however, one can directly show that equation
(\ref{eq:mgauss}) reduces to the probability distribution function of
$\phi_{\mb k}$ and $D_{\mb k}$ that are explicitly written as 
\begin{equation}
\label{eq:rgauss}
P(|\delta_{\mb k}|, \phi_{\mb k}) 
d|\delta_{\mb k}| d\phi_{\mb k} 
=  {2 |\delta_{\mb k}| \over P(k)} 
{\rm exp}\left(-{|\delta_{\mb k}|^2 \over P(k)}\right) 
d|\delta_{\mb k}| {d\phi_{\mb k} \over 2\pi} .
\end{equation}
mutually independently of $k$.  The phase distribution is uniform, and
thus does not carry information.  The above probability distribution
function is also derived when the real and imaginary parts of the
Fourier components $\delta_{\mb k}$ are uncorrelated and Gaussian
distributed (with the dispersion $P(k)/2$) independently of {\mb k}.  As
is expected, the distribution function (\ref{eq:rgauss}) is completely
fixed if $P(k)$ is specified.  This rephrases the previous statement
that the Gaussian field is completely specified by the two-point
correlation function in real space.

Incidentally the one-point phase distribution turns out to be
essentially uniform even in a strongly non-Gaussian field
\cite{Suginohara1991,Fan1995}. Thus it is unlikely to extract useful
information directly out of it mainly due to the cyclic property of the
phase. Very recently, however, Matsubara\cite{Matsubara2003b} and Hikage
et al.\cite{HMS04} succeeded in detecting a signature of phase
correlations in Fourier modes of mass density fields induced by
nonlinear gravitational clustering using the distribution function of
the phase sum of the Fourier modes for triangle wavevectors.  Several
different statistics which carry the phase information have been also
proposed in cosmology, including the void probability function
\cite{White1979}, the genus statistics \cite{GMD-86}, the Minkowski
functionals \cite{MBW-94,SB1997}.

\subsection{Log-normal distribution}

A probability distribution function (PDF) of the cosmological density
fluctuations is the most fundamental statistic characterizing the
large-scale structure of the universe.  As long as the density
fluctuations are in the linear regime, their PDF remains Gaussian.  Once
they reach the nonlinear stage, however, their PDF significantly
deviates from the initial Gaussian shape due to the strong non-linear
mode-coupling and the non-locality of the gravitational dynamics.  The
functional form for the resulting PDFs in nonlinear regimes are not
known exactly, and a variety of phenomenological models have been
proposed~ \cite{Hubble-34, saslaw-85, CJ-91, GY-93}.

Kayo et al. (2001)~\cite{kayo-01} showed that the one-point log-normal
PDF:
\begin{eqnarray}
 P^{(1)}_{\rm LN}(\delta) =
  \frac{1}{\sqrt{2\pi\sigma_1^2}}\exp\left[ -\frac{\{\ln(1+\delta) +
 \sigma_1^2/2\}^2}{2\sigma_1^2}\right]\frac{1}{1+\delta} .
\label{eq:1pLNPDF}
\end{eqnarray}
describes very accurately the cosmological density distribution even in
the nonlinear regime (the rms variance $\sigma_{\rm nl} \simlt 4$ and
the over-density $\delta \simlt 100$).  The above function is
characterized by a single parameter $\sigma_1$ which is related to the
variance of $\delta$. Since we use $\delta$ to represent the density
fluctuation field smoothed over $R$, its variance is computed from its
power spectrum $P_{\rm nl}$ explicitly as
\begin{equation}
 \sigma_{\rm nl}^2(R)
  \equiv
  \frac{1}{2\pi^2}\int_0^\infty P_{\rm nl}(k)\tilde{W}^2(kR)k^2 dk .
\end{equation}
Here we use subscripts ``lin'' and ``nl'' to
distinguish the variables corresponding to the primordial (linear)
and the evolved (nonlinear) density fields, respectively. Then
$\sigma_1$ depends on the smoothing scale $R$ alone and is given by
\begin{eqnarray}
 \sigma_1^2(R) = \ln\left[1+\sigma_{\rm nl}^2(R)\right]. 
\end{eqnarray}
Given a set of cosmological parameters, one can compute $\sigma_{\rm
nl}(R)$ and thus $\sigma_1(R)$ very accurately using a fitting formula
for $P_{\rm nl}(k)$ (e.g., ref.~\cite{pd-96}).  In this sense, the above
log-normal PDF is completely specified without any free parameter.

\begin{figure}[htb]
\begin{center}
    \leavevmode\epsfxsize=11.5cm \epsfbox{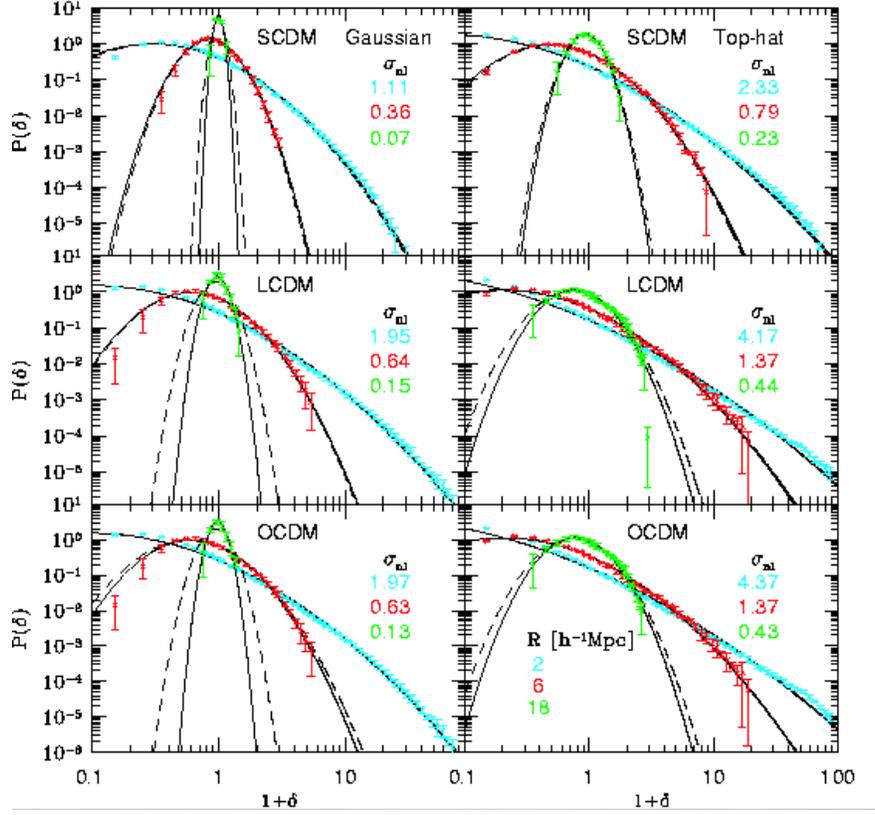} 
\caption{One-point
PDFs in CDM models with Gaussian ({\it left} panels) and top-hat ({\it
right} panels) smoothing windows; $R=2$\himpc ({\it cyan}), 6\himpc
({\it red}), and 18\himpc ({\it green}).  The {\it solid}
and {\it long-dashed} lines represent the log-normal PDF adopting
$\sigma_{\rm nl}$ calculated directly from the simulations and estimated
from the nonlinear fitting formula of ref.~\cite{pd-96},
respectively. (ref.~\cite{kayo-01}) \label{fig:1pLN_CDM}}
\end{center}
\end{figure}

Figure \ref{fig:1pLN_CDM} plots the one-point PDFs computed from
cosmological N-body simulations in SCDM, LCDM and OCDM (for Standard,
Lambda and Open CDM) models, respectively~\cite{JS-98, kayo-01}.  The
simulations employ $N=256^3$ dark matter particles in a periodic
comoving cube $(100h^{-1}{\rm Mpc})^3$.  The density fields are smoothed
over Gaussian ({\it Left panels}) and Top-hat ({\it Right panels})
windows with different smoothing lengths;$R=2h^{-1}$Mpc, $6h^{-1}$Mpc
and $18h^{-1}$Mpc.  Solid lines show the log-normal PDFs adopting the
value of $\sigma_{\rm nl}$ directly evaluated from simulations (shown in
each panel). The agreement between the log-normal model and the
simulation results is quite impressive. A small deviation is noticeable
only for $\delta \simlt -0.5$.

From an empirical point of view, Hubble (1934) ~\cite{Hubble-34} first
noted that the galaxy distribution in angular cells on the celestial
sphere may be approximated by a log-normal distribution, rather than a
Gaussian.  Theoretically the above log-normal function may be obtained
from the one-to-one mapping between the linear random-Gaussian and the
nonlinear density fields~\cite{CJ-91}.  Define a linear density field
$g$ smoothed over $R$ obeying the Gaussian PDF:
\begin{eqnarray}
 P^{(1)}_{\rm G}(g) =
\frac{1}{\sqrt{2\pi\sigma_{\rm lin}^2}}
\exp\left(-\frac{g^2}{2\sigma_{\rm lin}^2}\right),
\end{eqnarray}
where the variance is computed from its linear power spectrum:
\begin{eqnarray}
 \sigma_{\rm lin}^2(R) \equiv
\frac{1}{2\pi^2}\int_0^\infty P_{\rm lin}(k)\tilde{W}^2(kR)k^2dk .
\end{eqnarray}
If one introduces a new field $\delta$  from $g$ as
\begin{equation}
 1+\delta = \frac{1}{\sqrt{1+\sigma_{\rm nl}^2}}
  \exp\left\{\frac{g}{\sigma_{\rm lin}}
\sqrt{\ln(1+\sigma_{\rm nl}^2)}\right\},
  \label{eq:trans1}
\end{equation}
the PDF for $\delta$ is simply given by $(dg/d\delta)P^{(1)}_{\rm
G}(g)$ which reduces to equation (\ref{eq:1pLNPDF}). 

At this point, the transformation (\ref{eq:trans1}) is nothing but a
mathematical procedure to relate the Gaussian and the log-normal
functions. Thus there is no physical reason to believe that the new
field $\delta$ should be regarded as a nonlinear density field evolved
from $g$ even in an approximate sense. In fact it is physically
unacceptable since the relation, if taken at face value, implies that
the nonlinear density field is completely determined by its linear
counterpart locally. We know, on the other hand, that the nonlinear
gravitational evolution of cosmological density fluctuations proceeds in
a quite nonlocal manner, and is sensitive to the surrounding mass
distribution.  Nevertheless the fact that the log-normal PDF provides a
good fit to the simulation data empirically implies that the
transformation (\ref{eq:trans1}) somehow captures an important aspect of
the nonlinear evolution in the real universe.

\subsection{Higher-order correlation functions \label{subsec:higher-order}}

One of the most direct methods to evaluate the deviation from the
Gaussianity is to compute the higher-order correlation functions.
Suppose that ${\mb x}_i$ now labels the position of the $i$-th object
(galaxy), and then the two-point correlation function $\xi_{12} \equiv$
$\xi({\mb x}_1,{\mb x}_2)$ is defined also in terms of the joint
probability of the pair of objects located in the volume elements of
$\delta V_1$ and $\delta V_2$:
\begin{eqnarray}
\label{eq:prob2}
 \delta P_{12} = \bar n^2 \delta V_1 \delta V_2
   ~ \left[ 1 + \xi_{12} \right] ,
\end{eqnarray}
where $\bar n$ is the mean number density of the objects.
This definition is generalized to three- and four-point correlation
functions, $\zeta_{123} \equiv $ $\zeta({\mb x}_1,{\mb x}_2, {\mb x}_3)$
and $\eta_{1234} \equiv$ $\eta({\mb x}_1,{\mb x}_2, {\mb x}_3, {\mb
x}_4)$, in a straightforward manner:
 \begin{eqnarray}
 & \delta P_{123} =& \bar n^3 \delta V_1 \delta V_2 \delta V_3
                ~ \left[ 1 + \xi_{12} + \xi_{23} + \xi_{31} +
                  \zeta_{123} \right] , \label{eq:prob3} \\ 
 & \delta P_{1234} =& \bar n^4 \delta V_1 \delta V_2 \delta V_3 
     \delta V_4 ~ 
      \big[ 1  + \xi_{12} + \xi_{13} + \xi_{14} +\xi_{23} 
           + \xi_{24} + \xi_{34} \cr
     && \qquad + \zeta_{123} + \zeta_{124} + \zeta_{134} + \zeta_{234} \cr
&& \qquad  + \xi_{12} \xi_{34} + \xi_{13} \xi_{24} 
   + \xi_{14} \xi_{23} + \eta_{1234} \big] , \label{eq:prob4} 
 \end{eqnarray}
Apparently $\xi_{12}$, $\zeta_{123}$, and $\eta_{1234}$ are symmetric
with respect to the change of the indices. Define the following
quantities with the same symmetry properties:
\begin{eqnarray}
  Z_{123} &\equiv& \xi_{12}\xi_{23} + \xi_{21}\xi_{13}
                 + \xi_{23}\xi_{31} , \\
A_{1234} &\equiv &  \xi_{12}\xi_{23}\xi_{34}
                 + \xi_{23}\xi_{34}\xi_{41}
                 + \xi_{24}\xi_{41}\xi_{12}
                 + \xi_{13}\xi_{32}\xi_{24}
                 + \xi_{32}\xi_{24}\xi_{41} \cr
                 &+& \xi_{24}\xi_{41}\xi_{13}
                 + \xi_{12}\xi_{24}\xi_{43}
                 + \xi_{24}\xi_{43}\xi_{31}
                 + \xi_{31}\xi_{12}\xi_{24}
                 + \xi_{13}\xi_{34}\xi_{42} \cr
                 &+& \xi_{34}\xi_{42}\xi_{21}
                 + \xi_{42}\xi_{21}\xi_{13} , \\
B_{1234} &\equiv& \xi_{12}\xi_{13}\xi_{14}
                 + \xi_{21}\xi_{23}\xi_{24}
                 + \xi_{31}\xi_{32}\xi_{34}
                 + \xi_{41}\xi_{42}\xi_{43}, \\
C_{1234} &\equiv&
     \zeta_{123} (\xi_{14} + \xi_{24} + \xi_{34})
     +  \zeta_{134} (\xi_{12} + \xi_{32} + \xi_{42}) \cr
     &+& \zeta_{124} (\xi_{31} + \xi_{32} + \xi_{34})
      +  \zeta_{234} (\xi_{12} + \xi_{13} + \xi_{14}).
\end{eqnarray}
Then it is not unreasonable to suspect that the following relations hold
\begin{eqnarray}
  \label{qr1}
   &\zeta_{123} =& Q~ Z_{123} , \\
  \label{qr2}
   &\eta_{1234} =& R_a~ A_{1234} + R_b~ B_{1234} , \\
  \label{qr3}
   &\eta_{1234} =& R_c~ C_{1234} , 
\end{eqnarray}
where $Q$, $R_a$, $R_b$, and $R_c$ are constants. In fact, the analysis
of the two-dimensional galaxy catalogues~\cite{peebles-80} revealed
\begin{eqnarray}
   &&Q = 1.29 \pm 0.21 
     \qquad (0.1 \himpc \simlt r \simlt 10 \himpc) , \\ 
   \label{qrab}
   && R_a= 2.5 \pm 0.6, ~~ R_b = 4.3 \pm 1.2 
     \quad (0.5 \himpc \simlt r \simlt 4 \himpc) . \quad
\end{eqnarray}
The generlization of those relations for $N$-point correlation functions
is suspected to be hold generally:
\begin{equation}
\label{hier}
   \xi_N({\mb r_1},\ldots, {\mb r_N}) 
    = \sum_j Q_{N,j} \sum_{(ab)} \prod^{N-1} \xi(r_{ab}) ,
\end{equation}
and called {\it the hierarchical clustering ansatz}. Cosmological N-body
simulations approximately support the validity of the above ansatz, but
also detect the finite deviation from it ~\cite{suto-93}.

\subsection{Genus statistics}

A complementary approach to characterize the clustering of the universe
beyond the two-point correlation functions is the genus
statistics~\cite{GMD-86}.  This is a mathematical measure of the
topology of the isodensity surface. For definiteness, consider the
density contrast field $\delta(\mb x)$ at the position $\mb x$ in the
survey volume $V_{\rm all}$. This may be evaluated, for instance, by
taking the ratio of the number of galaxies $N(\mb x,V_f)$ in the volume
$V_f$ centered at $\mb x$ to its average value $\overline{N}(V_f)$:
\begin{equation}
  \label{eq:deltaf}
   \delta(\mb x,V_f) = {N(\mb x,V_f) \over \overline{N}(V_f)} -1,
\qquad \sigma(V_f) = \langle |\delta(\mb x,V_f)|^2 \rangle^{1/2} ,
\end{equation}
where $\sigma(V_f)$ is its rms value.  Consider the isodensity surface
parameterized by a value of $\nu \equiv \delta(\mb x,V_f)/\sigma(V_f)$.
Genus is one of the topological numbers characterizing the surface
defined as
\begin{equation}
\label{eq:gaubon}
   g \equiv - {1 \over 4\pi} \int \kappa ~ dA ,
\end{equation}
where $\kappa$ is the Gaussian curvature of the isolated surface.  The
Gauss -- Bonnet theorem implies that the value of $g$ is indeed an
integer and equal to the number of holes minus 1. 
This is qualitatively understood as follows;
expand an arbitrary two-dimensional surface around a point as
\begin{equation}
z = {1 \over 2} a x^2 + b xy + {1 \over 2} c y^2
= {1 \over 2} \kappa_1 x_1^2 +  {1 \over 2} \kappa_2 x_2^2 .
\end{equation}
Then the Gaussian curvature of the surface is defined by
$\kappa=\kappa_1\kappa_2$. A surface topologically equivalent to a
sphere (a torus) has $\kappa=1$ ($\kappa=0$), and thus equation
(\ref{eq:gaubon}) yields $g=-1$ ($g=0$) which coincides with the number
of holes minus 1.

In reality, there are many disconnected isodensity surfaces for a given
$\nu$, and thus it is more convenient to define the genus density in the
survey volume $V_{\rm all}$ using the additivity of the genus:
\begin{equation}
\label{eq:gcurve}
G(\nu) \equiv {1 \over V_{\rm all}} \sum_{i=1}^{I} g_i = - {1 \over
  4\pi V_{\rm all}} \sum_{i=1}^{I} \int_{A_i} \kappa ~ dA ,
\end{equation}
where $A_i$ ($i=1 \sim I$) denote the disconnected isodensity surfaces
with the same value of $\nu =\delta(\mb x,V_f)/\sigma(V_f)$.
Interestingly the Gaussian density field has an analytic expression for
equation (\ref{eq:gcurve}):
\begin{equation}
   \label{eq:genus_rd}
   G(\nu) = \frac{1}{4\pi^2}
            \left(\frac{\langle k^2 \rangle}{3}\right)^{3/2}   
            e^{-\nu^2/2} (1-\nu^2) ,
\end{equation}
where
\begin{equation}
   \label{eq:k2}
   \langle k^2 \rangle \equiv
 { {\displaystyle \int k^2 P(k) \tilde W^2(kR) d^3 k} \over
{\displaystyle \int P(k) \tilde W^2(kR) d^3 k} } 
\end{equation}
is the moment of $k^2$ weighted over the power spectrum of fluctuations
$P(k)$, and the smoothing function $\tilde W^2(kR)$ (e.g.,
ref.~\cite{bbks-86}).  It should be noted that in the Gaussian density field
the information of the power spectrum shows up only in the proportional
constant of equation (\ref{eq:genus_rd}), and its functional form is
deterimined uniquely by the threshold value $\nu$. This $\nu$-dependence
reflects the the phase information which is ignored in the two-point
correlation function and power spectrum. In this sense, genus statistics
is a complementary measure of the clustering pattern of universe.

\begin{figure}[h]
\begin{center}
    \leavevmode\epsfxsize=5.5cm \epsfbox{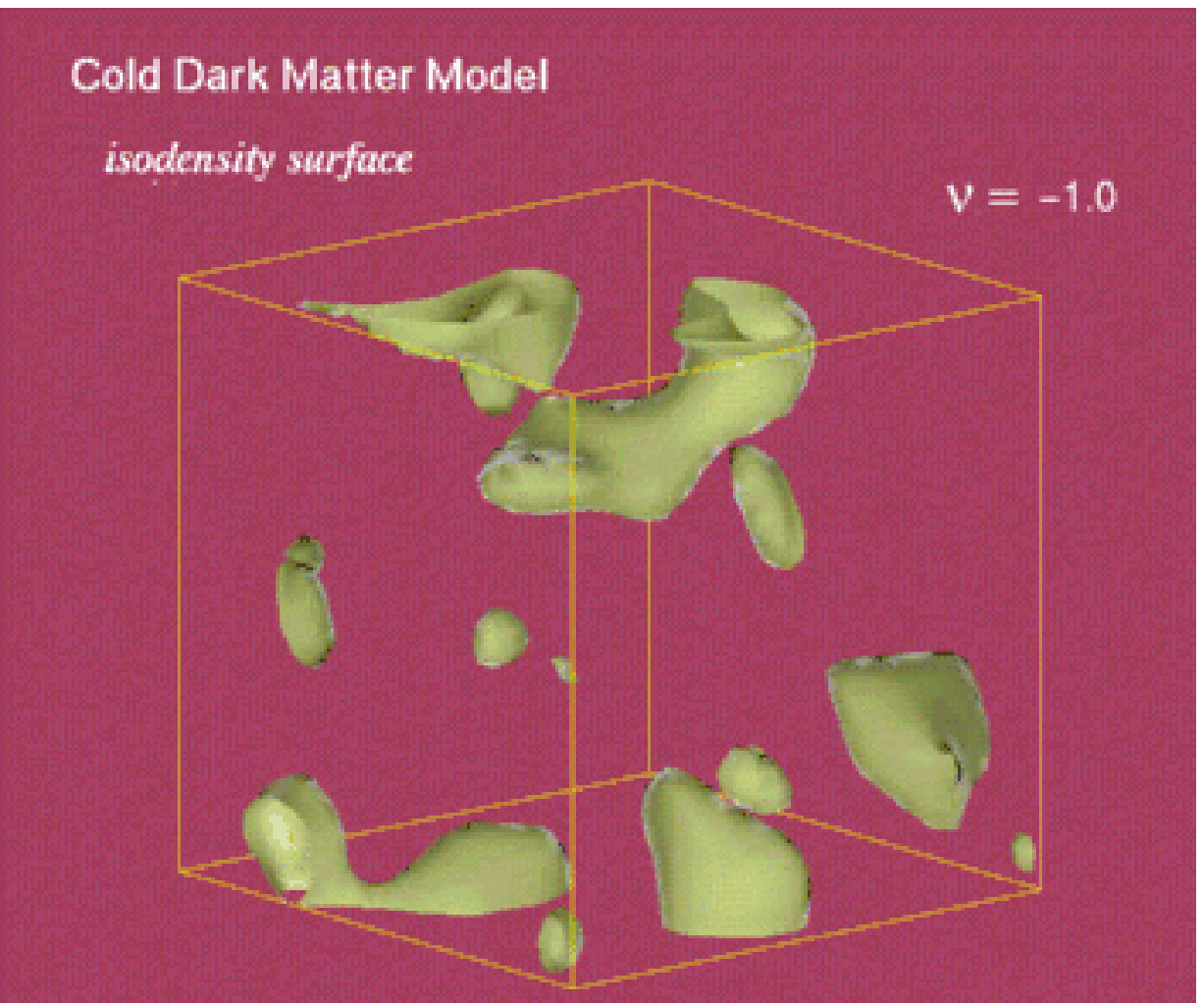}
\hspace*{0.5cm}
    \leavevmode\epsfxsize=5.5cm \epsfbox{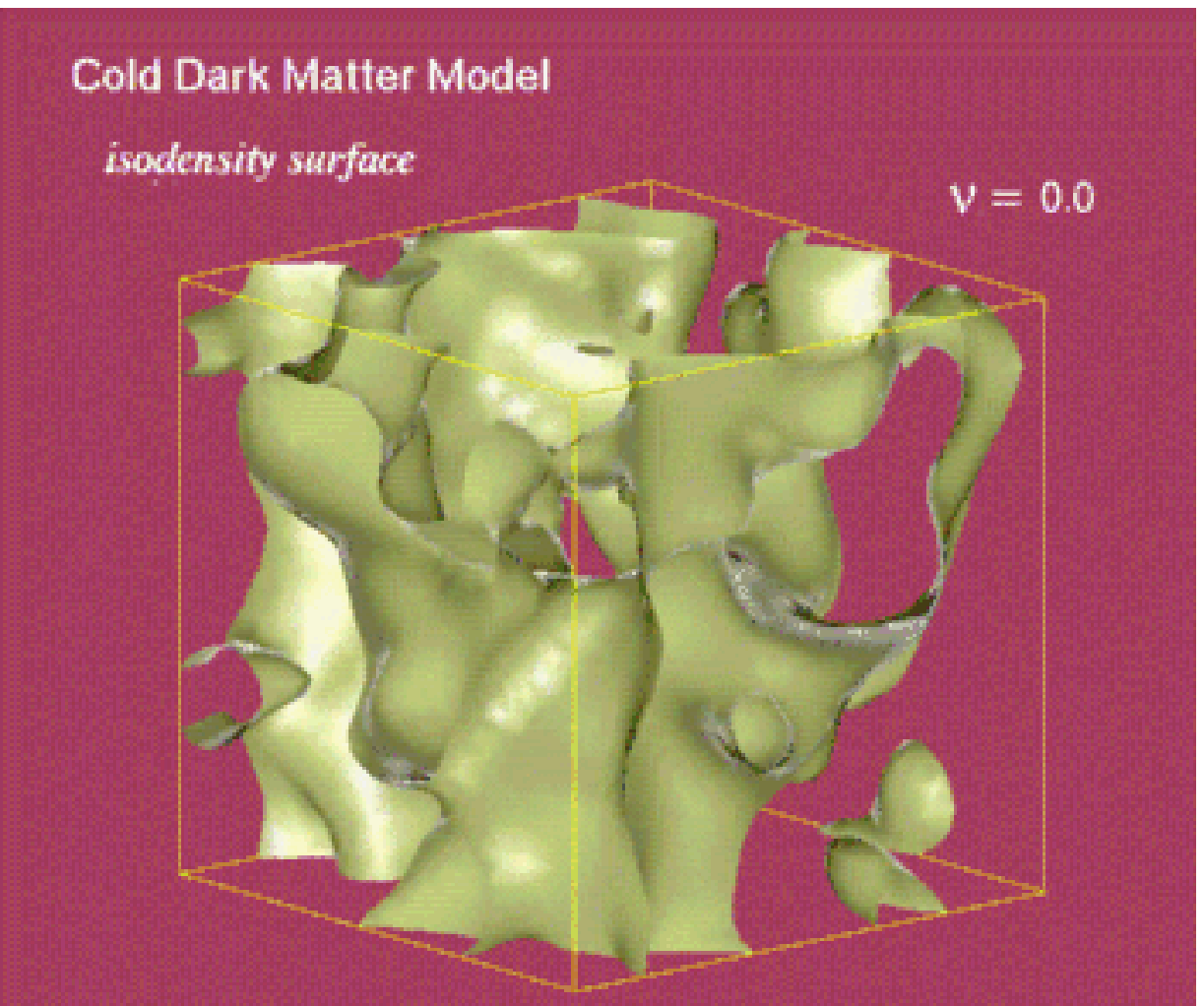}
\vspace*{0.5cm}
    \leavevmode\epsfxsize=5.5cm \epsfbox{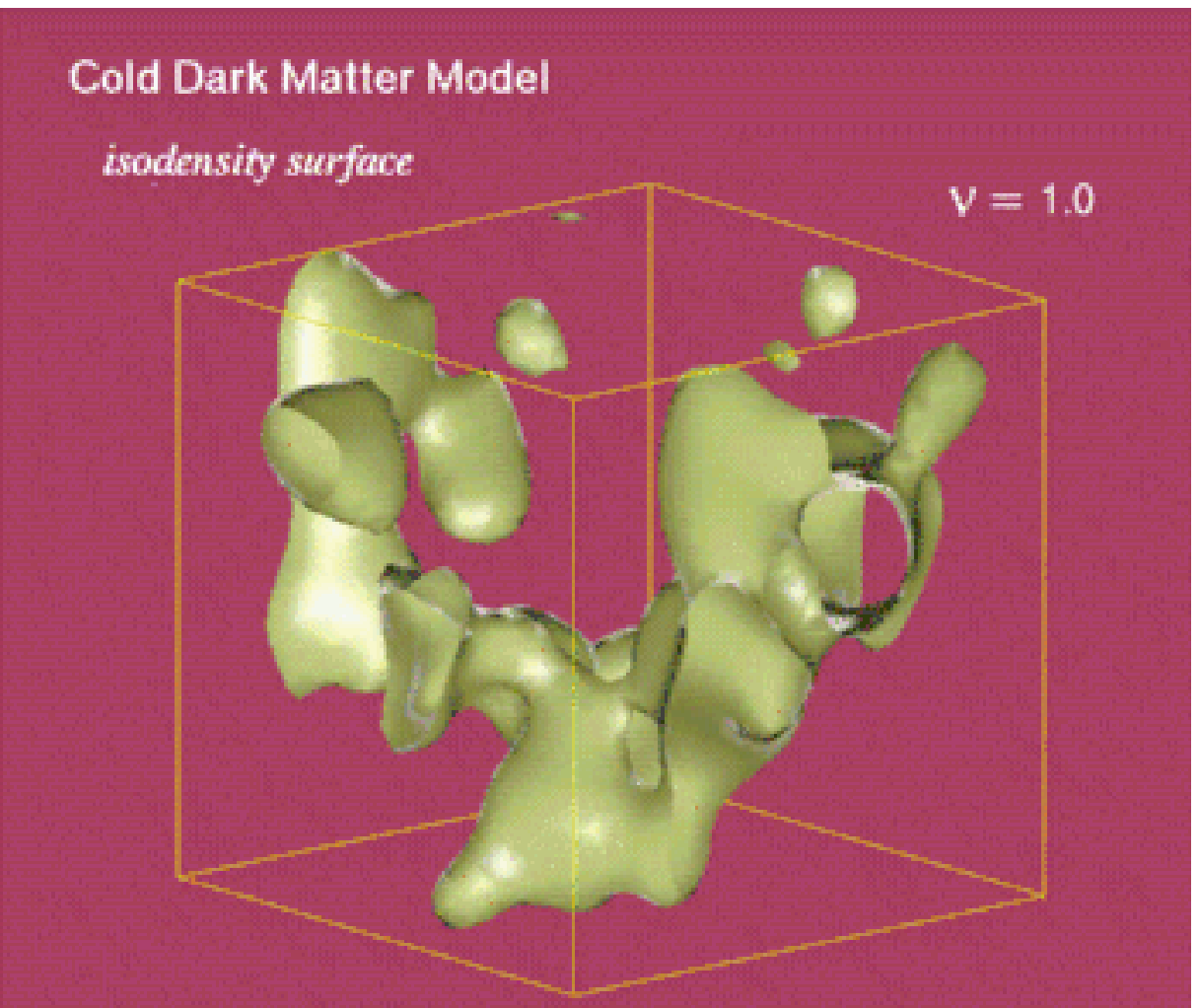}
\hspace*{0.5cm}
    \leavevmode\epsfxsize=5.5cm \epsfbox{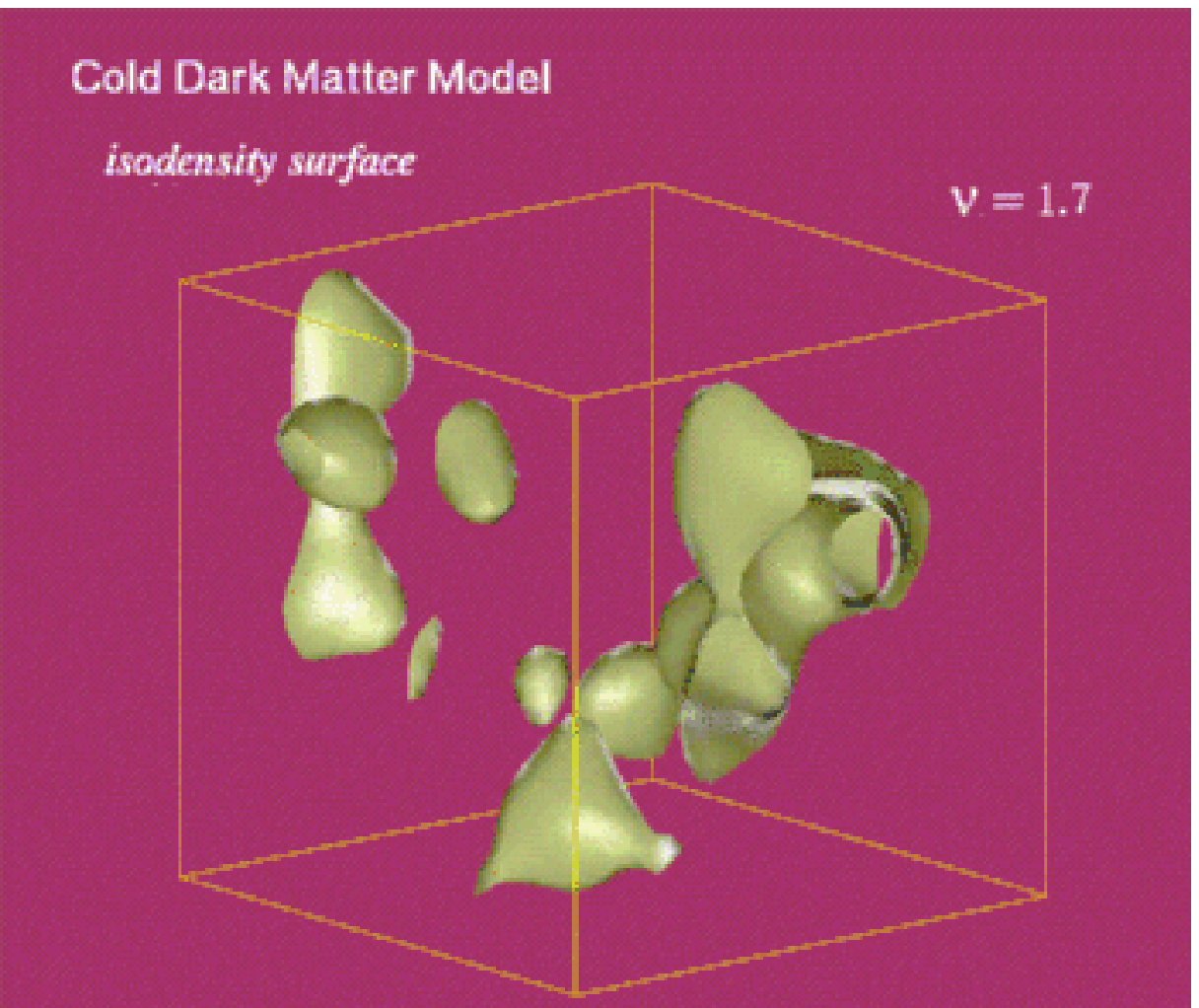}
\caption{Isodensity surfaces of dark matter distribution from N-body
 simulation; LCDM in $(100\himpc)^3$ at $\nu = -1.0$,
  $0.0$, $1.0$, and $1.7$. (ref.~\cite{ms-96})}
\label{fig:isodensity}
\end{center}
\end{figure}

Even if the primordial density field obeys the Gaussian statistics, the
subsequent nonlinear gravitational evolution generates the significant
non-Gaussianity. To distinguish the initial non-Gaussianity from that
aquired by the nonlinear gravity is of fundamental importance in
inferring the initial condition of the universe in a standard
gravitational instability picture of structure formation.  In a weakly
nonlinear regime, Matsubara (1996) derived an analytic expression for
the non-Gaussianity emerging from the primordial Gaussian
field~\cite{matsubara-94}:
\begin{equation}
   \label{eq:genuswnl}
   G(\nu) = - \frac{1}{4\pi^2}
   \left(\frac{\langle k^2 \rangle}{3}\right)^{3/2}
   e^{-\nu^2/2}
   \left[ H_2(\nu)
          + \sigma \left( \frac{S}{6} H_5(\nu)
                          + \frac{3T}{2} H_3(\nu)
                          + 3U H_1(\nu)\right)
   \right] ,
\end{equation}
where
\begin{equation}
   \label{eq:hermite}
H_n(\nu) \equiv (-1)^n e^{\nu^2/2} \left({d \over d\nu}\right)^n
e^{-\nu^2/2}
\end{equation}
are the Hermite polynomials;
$H_1=\nu$, $H_2=\nu^2 -1$, $H_3=\nu^3 -
3\nu$, $H_4=\nu^4 - 6\nu^2 + 3$, $H_5=\nu^5 - 10\nu^3 + 15\nu$,
$\ldots$. The three quantities:
\begin{eqnarray}
   \label{eq:stu}
   S = \frac{1}{\sigma^4} \langle\delta^3\rangle, \quad
   T = - \frac{1}{2\langle k^2 \rangle \sigma^4}
            \langle \delta^2 \nabla^2 \delta \rangle, \quad
   U = - \frac{3}{4\langle k^2 \rangle^2 \sigma^4}
            \langle \nabla\delta\cdot\nabla\delta
                    \nabla^2 \delta \rangle 
\end{eqnarray}
denote the third-order moments of $\delta$. This expression
plays a key role in understanding if the non-Gaussianity in galaxy
distribution is ascribed to the primordial departure from the Gaussian
statistics.

\subsection{Minkowski functionals}

In fact, genus is one of the complete sets of $N+1$ quantities, known as
the Minkowski functionals (MFs), which determine the morphological
properties of a pattern in $N$-dimensional space.  In the analysis of
galaxy redshift survey data, one considers isodensity contours from the
three--dimensional density contrast field $\delta$ by taking its
excursion set $F_{\nu}$, i.e., the set of all points where the density
contrast $\delta$ exceeds the threshold level $\nu$ as was the case in
the case of genus described in the above subsection.  

All MFs can be expressed as integrals over the excursion set.  While
the first MF is simply given by the volume integration of a Heaviside
step function $\Theta$ normalized to the total volume $V_{\rm tot}$,
\begin{equation}
V_0(\nu)=\frac{1}{V_{\rm tot}}\int_V d^3x\Theta(\nu-\nu(x))\;,
\end{equation}
the other MFs, $V_k(k=1,2,3)$, are calculated by the surface integration
of the local MFs, $v^{\rm loc}_k$.  The general expression is
\begin{equation}
\label{eq:local_MFs}
V_k(\nu)=\frac{1}{V_{\rm tot}}\int_{\partial F_{\nu}}
d^2S(\mbox{\boldmath $x$})v^{\rm loc}_k(\nu, \mbox{\boldmath $x$}),
\end{equation}
with the local Minkowski Functionals for $k=1,2,3$ given by
\begin{eqnarray}
v^{\rm loc}_1(\nu,\mbox{\boldmath $x$}) &=& \frac{1}{6}, \\
v^{\rm loc}_2(\nu,\mbox{\boldmath $x$})
&=& \frac{1}{6\pi}\left(\frac{1}{R_1}+\frac{1}{R_2}\right), \\
v^{\rm loc}_3(\nu,\mbox{\boldmath $x$}) &=& \frac{1}{4\pi}\frac{1}{R_1R_2} ,
\end{eqnarray}
where $R_1$ and $R_2$ are the principal radii of curvature of the
isodensity surface.

For a 3--D Gaussian random field, the average MFs per unit volume can be
expressed analytically as follows:
\begin{eqnarray}
\label{eq:minkowski_v0}
V_0(\nu) &=&
\frac{1}{2}-\frac{1}{\sqrt{2\pi}}\int^{\nu}_0 \exp{\left(
-\frac{x^2}{2}\right)}dx, \\
\label{eq:minkowski_v1}
V_1(\nu) &=& 
\frac{2}{3}\frac{\lambda}{\sqrt{2\pi}}\exp\left(-\frac{1}{2}\nu^2\right)
\\ 
\label{eq:minkowski_v2}
V_2(\nu) &=& \frac{2}{3}\frac{\lambda^2}{\sqrt{2\pi}}\nu\exp
\left(-\frac{1}{2}\nu^2\right), \\
\label{eq:minkowski_v3}
V_3(\nu) &=& 
\frac{\lambda^3}{\sqrt{2\pi}}(\nu^2-1)\exp\left(-\frac{1}{2}\nu^2\right) ,
\end{eqnarray}
where $\lambda=\sqrt{\sigma_1^2/6\pi\sigma^2}$,
$\sigma\equiv\langle\delta^2\rangle^{1/2}$,
$\sigma_1\equiv\langle|\nabla\delta|^2\rangle^{1/2}$, and $\delta$ is
the density contrast.

The above MFs can be indeed interpreted as well--known geometric
quantities; the volume fraction $V_0(\nu)$, the total surface area
$V_1(\nu)$, the integral mean curvature $V_2(\nu)$, and the integral
Gaussian curvature, i.e., the Euler characteristic $V_3(\nu)$.  In our
current definitions (eqs.  [\ref{eq:gcurve}] and [\ref{eq:local_MFs}],
or eqs.[\ref{eq:genus_rd}] and [\ref{eq:minkowski_v3}]), one can easily
show that $V_3(\nu)$ reduces simply to $-G(\nu)$. The MFs were first
introduced to cosmological studies by Mecke et al. (1994)~\cite{MBW-94},
and further details may be found in refs.~\cite{MBW-94, hikage-03}.
Analytic expressions of MFs in weakly non-Gaussian fields are derived in
ref.~\cite{matsubara-03}.

\clearpage
\section{Galaxy biasing}
\label{section:bias}

\subsection {Concepts and definitions of biasing}

As discussed above, luminous objects, such as galaxies and quasars, are
not direct tracers of the mass in the universe. In fact, the difference
of the spatial distribution between luminous objects and dark matter, or
the {\it bias}, has been indicated from a variety of observations.
Galaxy ``biasing" clearly exists.  The fact that galaxies of different
types cluster differently (e.g., ref.~\cite{Dressler-80}) implies that
not all of them are exact tracers of the underlying mass
distribution. See also \S \ref{section:2dFSDSS}.

In order to confront theoretical model predictions for the {\it mass}
distribution against observational data, one needs a relation of density
fields of mass and luminous objects.  The biasing of density peaks in a
Gaussian random field is well formulated~\cite{kaiser-84, bbks-86} and
it provides the first theoretical framework for the origin of galaxy
density biasing.  In this scheme, the galaxy--galaxy and mass--mass
correlation functions are related in the linear regime via
\begin{equation}
\xi_{gg}(r) =b^2 \xi_{mm}(r),
\label{eq:xibias}
\end{equation}
where the biasing parameter $b$ is a constant independent of scale $r$.
However, a much more specific linear biasing model is often assumed in
common applications, in which the local density fluctuation fields of
galaxies and mass are assumed to be deterministically related via the
relation
\begin{equation}
\delta_g({\bf x}) = b\, \delta_m({\bf x}).
\label{eq:lineardeltabias}
\end{equation}
Note that equation (\ref{eq:xibias}) follows from equation
(\ref{eq:lineardeltabias}), but the reverse is not true.

The above deterministic linear biasing is not based on a reasonable
physical motivation.  If $b>1$, it must break down in deep voids because
values of $\delta_g$ below $-1$ are forbidden by definition.  Even in
the simple case of no evolution in comoving galaxy number density, the
linear biasing relation is not preserved during the course of
fluctuation growth.  Non-linear biasing, where $b$ varies with
$\delta_m$, is inevitable.

Indeed, an analytical model for biasing of halos on the basis of the
extended Press-Schechter approximation~\cite{mw-96} predicts that the
biasing in nonlinear and provides a useful approximation for its
behavior as a function of scale, time and mass threshold.  $N$-body
simulations provide a more accurate description of the nonlinearity of
the halo biasing confirming the validity of the
Mo \& White model~\cite{jing-98,YTJS-01}.

\subsection{Modeling biasing }

Biasing is likely to be {\it stochastic}, not
deterministic~\cite{DL-99}.  An obvious part of this stochasticity can
be attributed to the discrete sampling of the density field by galaxies,
i.e. the shot noise.  In addition, a statistical, physical scatter in
the efficiency of galaxy formation as a function of $\delta_m$ is
inevitable in any realistic scenario.  For example, the random
variations in the density on smaller scales is likely to be reflected in
the efficiency of galaxy formation. As another example, the local
geometry of the background structure, via the deformation tensor, must
play a role too.  Such `hidden variables' would show up as physical
scatter in the density-density relation~\cite{TS-00}.

Consider the density contrasts of visible objects and mass, $\delta_{\rm
\scriptscriptstyle obj}({\bf x},z|R)$ and $\delta_{\rm
\scriptscriptstyle m}({\bf x},z|R)$, at a position ${\bf x}$ and a
redshift $z$ smoothed over a scale $R$~\cite{TMJS-01}. In general, the
former should depend on various other auxiliary variables $\vec{\cal A}$
defined at different locations ${\bf x'}$ and redshifts $z'$ smoothed
over different scales $R'$ in addition to the mass density contrast at
the same position, $\delta_{\rm \scriptscriptstyle m}({\bf x},z|R)$.
While this relation can be schematically expressed as
\begin{equation}
  \label{eq:generalbias}
  \delta_{\rm \scriptscriptstyle obj}({\bf x},z|R)=
{\cal F}[{\bf x},z, R, 
  \delta_{\rm \scriptscriptstyle m}({\bf x},z|R),
\vec{\cal A}({\bf x'},z'|R'), \ldots ] ,
\end{equation}
it is impossible even to specify the list of the astrophysical
variables $\vec{\cal A}$, and thus hopeless to predict the functional
form in a rigorous manner. Therefore if one simply focuses on the
relation between $\delta_{\rm \scriptscriptstyle obj}({\bf x},z|R)$
and $\delta_{\rm \scriptscriptstyle m}({\bf x},z|R)$, the relation
becomes inevitably {\it stochastic} and {\it nonlinear} due to the 
dependence on unspecified auxiliary variables $\vec{\cal A}$. 

For illustrative purposes, define the {\it biasing} factor as the
ratio of the density contrasts of luminous objects and mass:
\begin{equation}
  \label{eq:bgeneral}
  B_{\rm \scriptscriptstyle obj}({\bf x},z|R)
\equiv \frac{\delta_{\rm \scriptscriptstyle obj}({\bf x},z|R)}
{\delta_{\rm \scriptscriptstyle m}({\bf x},z|R)}
= \frac
{{\cal F}[{\bf x},z, R, 
  \delta_{\rm \scriptscriptstyle m}({\bf x},z|R),
\vec{\cal A}({\bf x'},z'|R'), \ldots ] }
{\delta_{\rm \scriptscriptstyle m}({\bf x},z|R)} .
\end{equation}
Only in very idealized situations, the above {\it nonlocal stochastic
  nonlinear} factor in terms of $\delta_{\rm \scriptscriptstyle m}$
may be approximated by
\begin{enumerate}
\item a {\it local stochastic nonlinear} bias:
\begin{equation}
  \label{eq:localbias}
  B_{\rm \scriptscriptstyle obj}({\bf x},z|R)=
b_{\rm \scriptscriptstyle obj}^{\rm \scriptscriptstyle (sn)}[{\bf x},z, R, 
  \delta_{\rm \scriptscriptstyle m}({\bf x},z|R),
\vec{\cal A}({\bf x},z|R), \ldots ] ,
\end{equation}
\item a {\it local deterministic nonlinear} bias:
\begin{equation}
  \label{eq:ldbias}
  B_{\rm \scriptscriptstyle obj}({\bf x},z|R)=
b_{\rm \scriptscriptstyle obj}^{\rm \scriptscriptstyle (dn)}[z, R, 
  \delta_{\rm \scriptscriptstyle m}({\bf x},z|R)] ,
\end{equation}
and
\item a {\it local deterministic linear} bias:
\begin{equation}
  \label{eq:linearbias}
  B_{\rm \scriptscriptstyle obj}({\bf x},z|R) 
= b_{\rm \scriptscriptstyle obj}(z, R) 
\end{equation}
\end{enumerate}

From the above point of view, the local deterministic linear bias is
obviously unrealistic, but is still a widely used conventional model for
biasing. In fact, the time- and scale-dependence of the linear bias
factor $b{\rm \scriptscriptstyle obj}(z, R)$ was neglected in many
previous studies of biased galaxy formation until very
recently. Currently, however, various models beyond the deterministic
linear biasing have been seriously considered with particular emphasis
on the nonlinear and stochastic aspects of the biasing~\cite{pen-98,
DL-99, TS-00, TMJS-01}.

\subsection{Density peaks and dark matter halos as toy models for
galaxy biasing}

Let us illustrate the biasing from numerical simulations by considering
two specific and popular models; primordial density peaks and dark
matter halos~\cite{TMJS-01}.  We use the N-body simulation data of
$L=100h^{-1}$Mpc again for this purpose~\cite{JS-98}.  We select
density peaks with the threshold of the peak height $\nu_{\rm th}=1.0$,
2.0, and 3.0.  As for the dark matter halos, these are identified using
the standard friend-of-friend algorithm with a linking length of 0.2 in
units of the mean particle separation. We select halos of mass larger
than the threshold $M_{\rm th}=2.0\times 10^{12}$, $5.0\times 10^{12}$
and $1.0\times 10^{13}\,h^{-1}\,M_{\odot}$.

\begin{figure}[htb]
larger amplitudes than those of mass. In nonlinear regimes ($\xi > 1$)
 the finger-of-god effect suppresses the amplitude of $\xi^{\rm (S)}$
 relative to $\xi^{\rm (R)}$, while $\xi^{\rm (S)}$ is larger than
 $\xi^{\rm (R)}$ in linear regimes ($\xi < 1$) due to the coherent
 velocity field.
\begin{center}
   \leavevmode\epsfxsize=13cm \epsfbox{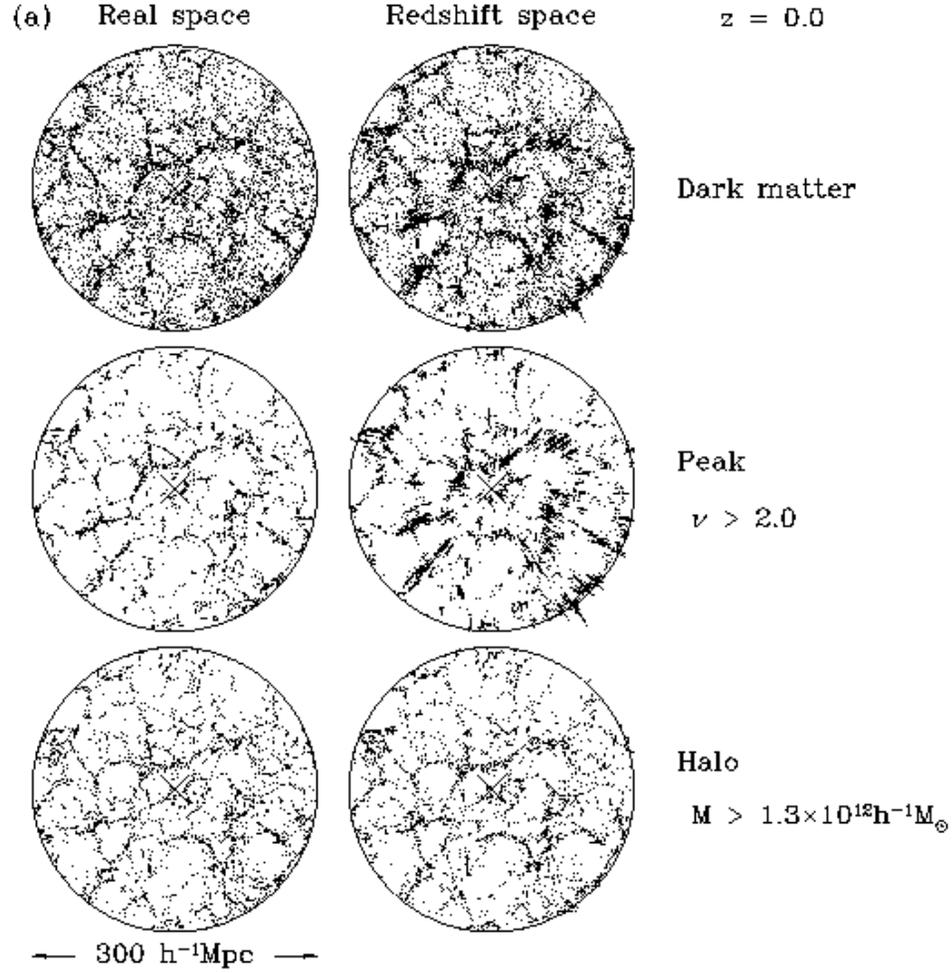}
\caption{Top-view of distribution of objects at $z=0$ in real (left
panels) and redshift (right panels) spaces around the fiducial observer
at the center; dark matter particles (top panels), peaks with $\nu>2$
(middle panels) and halos with $M>1.3\times10^{12}{\rm M}_\odot$ (bottom
panels) in LCDM model. The thickness of those slices is 15$h^{-1}$Mpc.
(ref.~\cite{TMJS-01})
\label{fig:peakhaloz0}}
\end{center}
\end{figure}

\begin{figure}[tbh]
\begin{center}
   \leavevmode\epsfxsize=13cm \epsfbox{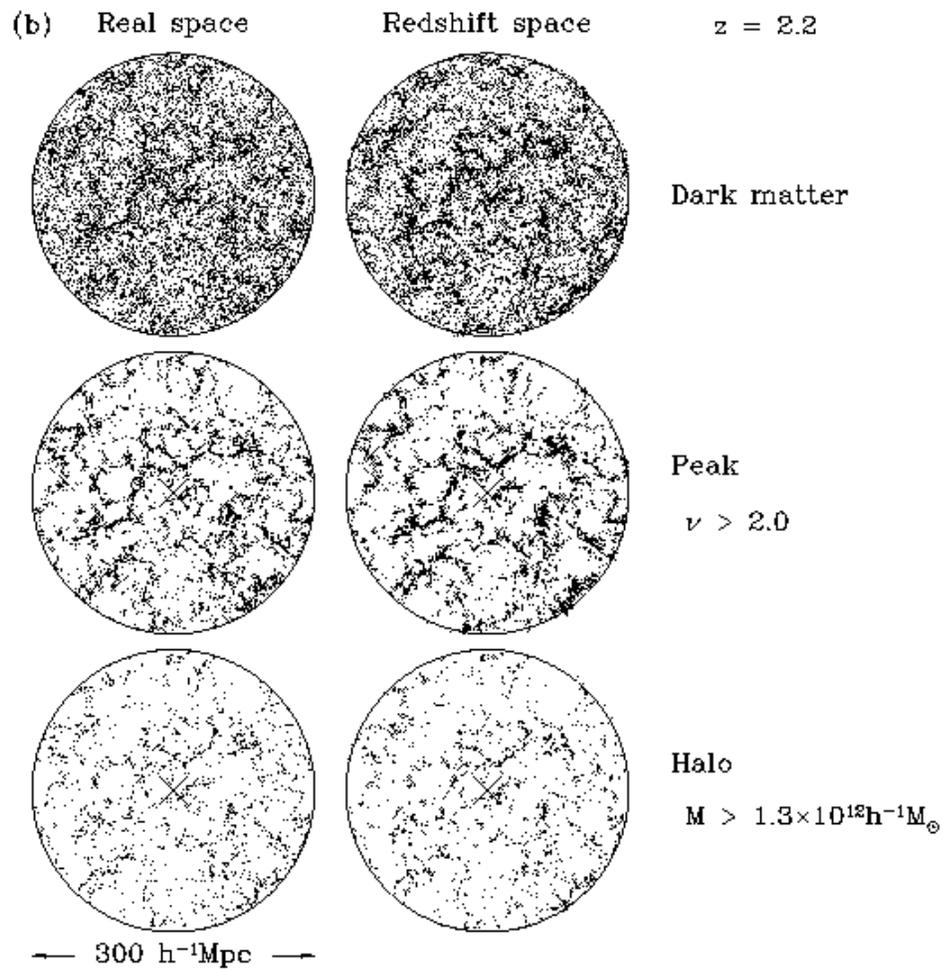}
\caption{Same as Fig.\ref{fig:peakhaloz0} but at
 $z=2.2$. (ref.~\cite{TMJS-01}) \label{fig:peakhaloz2}}
\end{center}
\end{figure}

Figures \ref{fig:peakhaloz0} and \ref{fig:peakhaloz2} depicts the
 distribution of dark matter particles ({\it upper-panel}), peaks ({\it
 middle-panel}) and halos ({\it lower-panel}) in LCDM model at $z=0$ and
 $z=2.2$ within a circular slice ({\it comoving} radius of
 $150h^{-1}$Mpc and thickness of $15h^{-1}$Mpc).  We locate a fiducial
 observer in the center of the circle. Then the {\it comoving} position
 vector {\bf r} for a particle with a {\it comoving} peculiar velocity
 {\bf v} at a redshift $z$ is observed at the position {\it s} in
 redshift space:
\begin{equation}
  \label{eq:s-r}
  {\bf s}={\bf r}+\frac{1}{H(z)}\, \frac{{\bf r}\cdot{\bf v}}{|\bf r|}
\, \frac{{\bf r}}{|\bf r|} ,
\end{equation}
where $H(z)$ is the Hubble parameter at $z$.  The right panels in
Figures \ref{fig:peakhaloz0} and \ref{fig:peakhaloz2} plot the observed
distribution in redshift space, where the redshift-space distortion is
quite visible; the coherent velocity field enhances the structure
perpendicular to the line-of-sight of the observer ({\it squashing})
while the virialized clump becomes elongated along the line-of-sight
({\it finger-of-God}). 

We use two-point correlation functions to quantify stochasticity and
nonlinearity in biasing of peaks and halos, and explore the signature of
the redshift-space distortion. Since we are interested in the relation
of the biased objects and the dark matter, we introduce three different
correlation functions; the auto-correlation functions of dark matter and
the objects, $\xi_{\rm \scriptscriptstyle mm}$ and $\xi_{\rm
\scriptscriptstyle oo}$, and their cross-correlation function $\xi_{\rm
\scriptscriptstyle om}$. In the present case, the subscript o refers to
either h (halos) or $\nu$ (peaks). We also use the superscripts R and S
to distinguish quantities defined in real and redshift spaces,
respectively.  We estimate those correlation functions using the
standard pair-count method.  The correlation function $\xi^{(S)}$ is
evaluated under the distant-observer approximation.

\begin{figure}[tph]
\begin{center}
   \leavevmode\epsfxsize=8cm \epsfbox{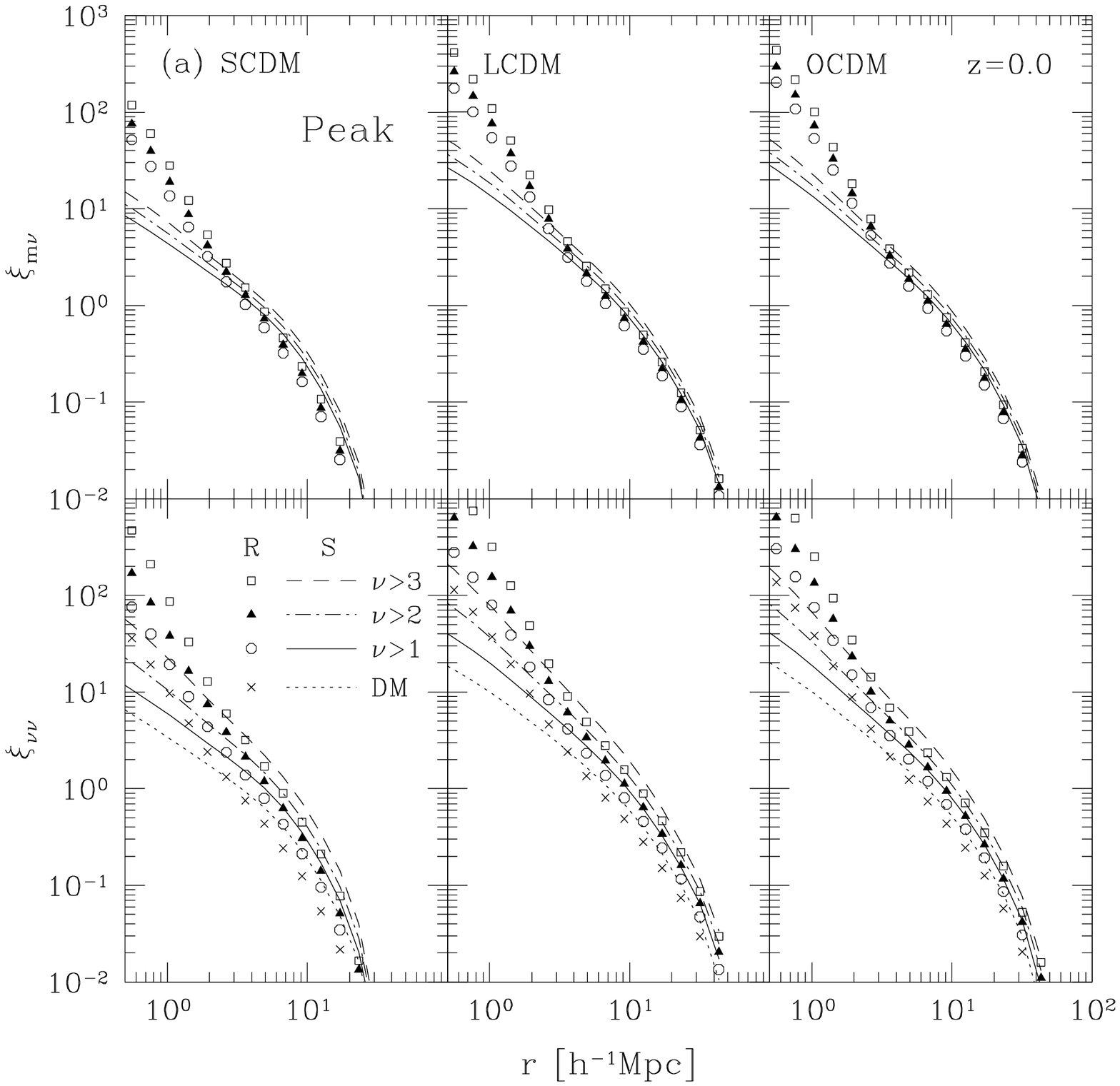}
\vspace*{0.2cm}
   \leavevmode\epsfxsize=8cm \epsfbox{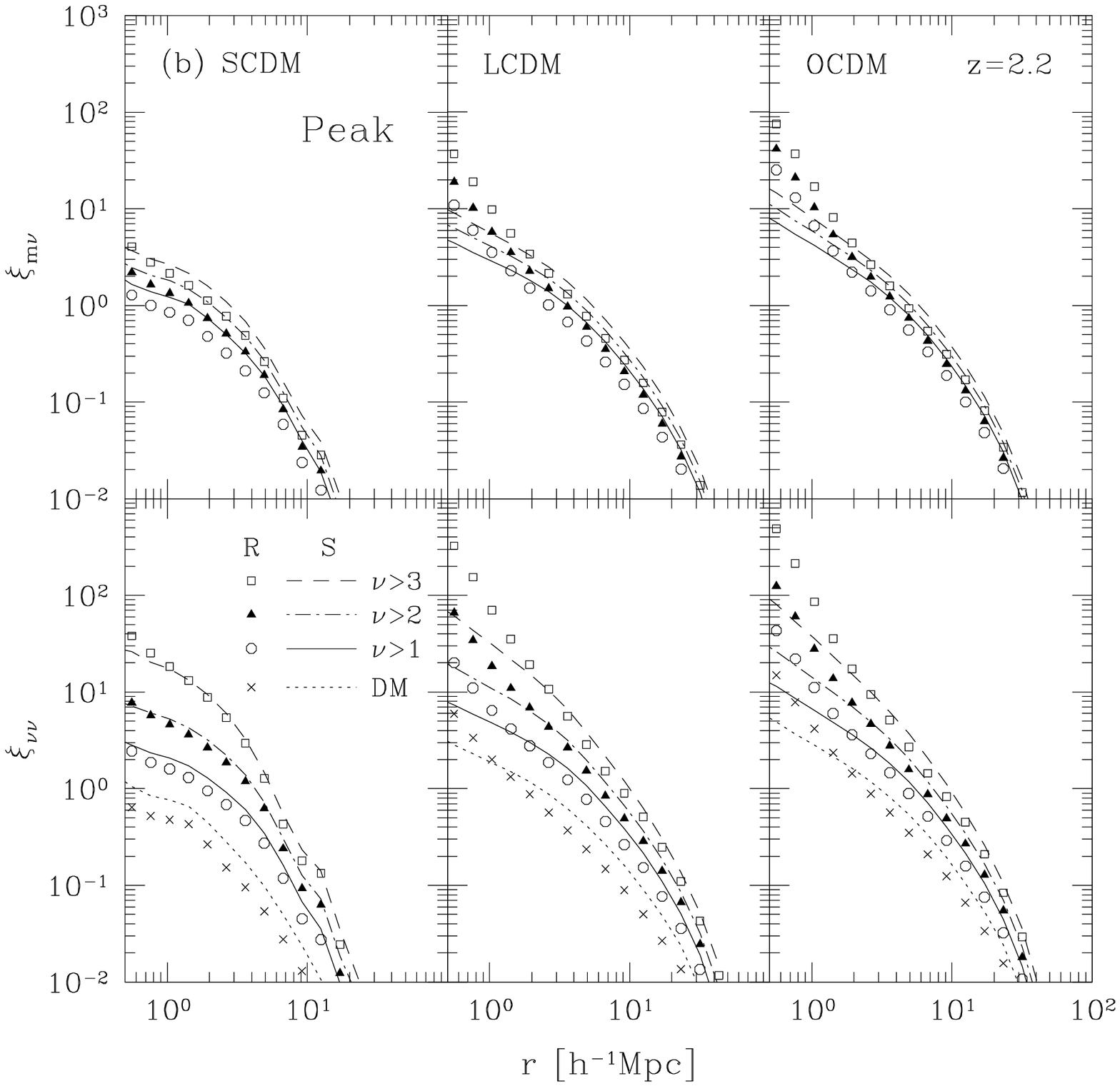}
\caption{Auto- and cross-correlation functions of dark matter and peaks
  in SCDM (left panels), LCDM (middle panels) and OCDM (right panels).
  Different symbols indicate the results in real space (open squares
  for $\nu>3$, filled triangles for $\nu>2$, open circles for $\nu>1$,
  and crosses for dark matter), while different curves indicate those in
  redshift space (dashed for $\nu>3$, dot--dashed for $\nu>2$,
solid for $\nu>1$, and dotted for dark matter).
(a) $z=0$, (b) $z=2.2$. (ref.~\cite{TMJS-01}) \label{fig:peakbias}
}
\end{center}
\end{figure}

\begin{figure}[tph]
\begin{center}
   \leavevmode\epsfxsize=8cm \epsfbox{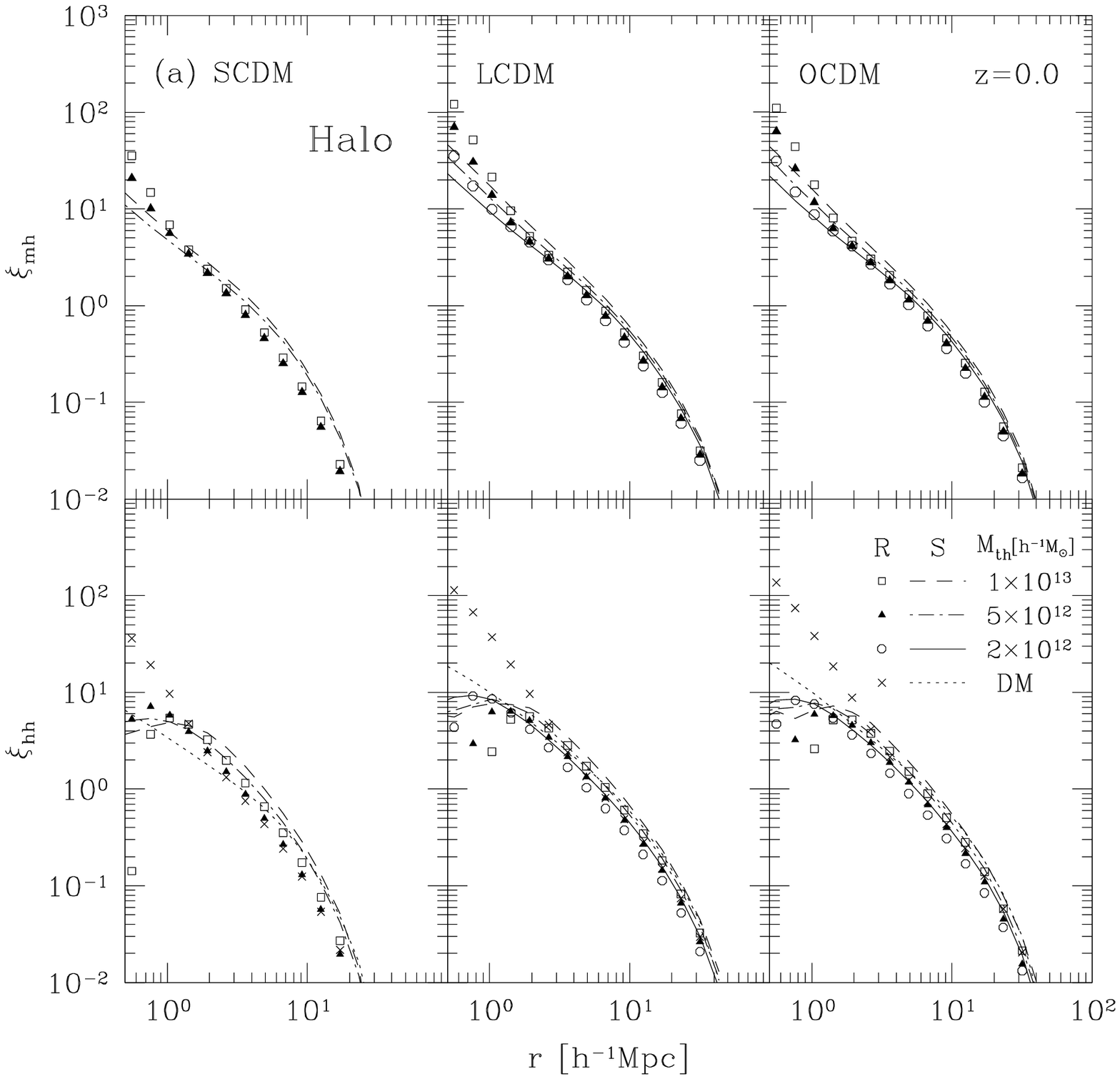}
\vspace*{0.2cm}
   \leavevmode\epsfxsize=8cm \epsfbox{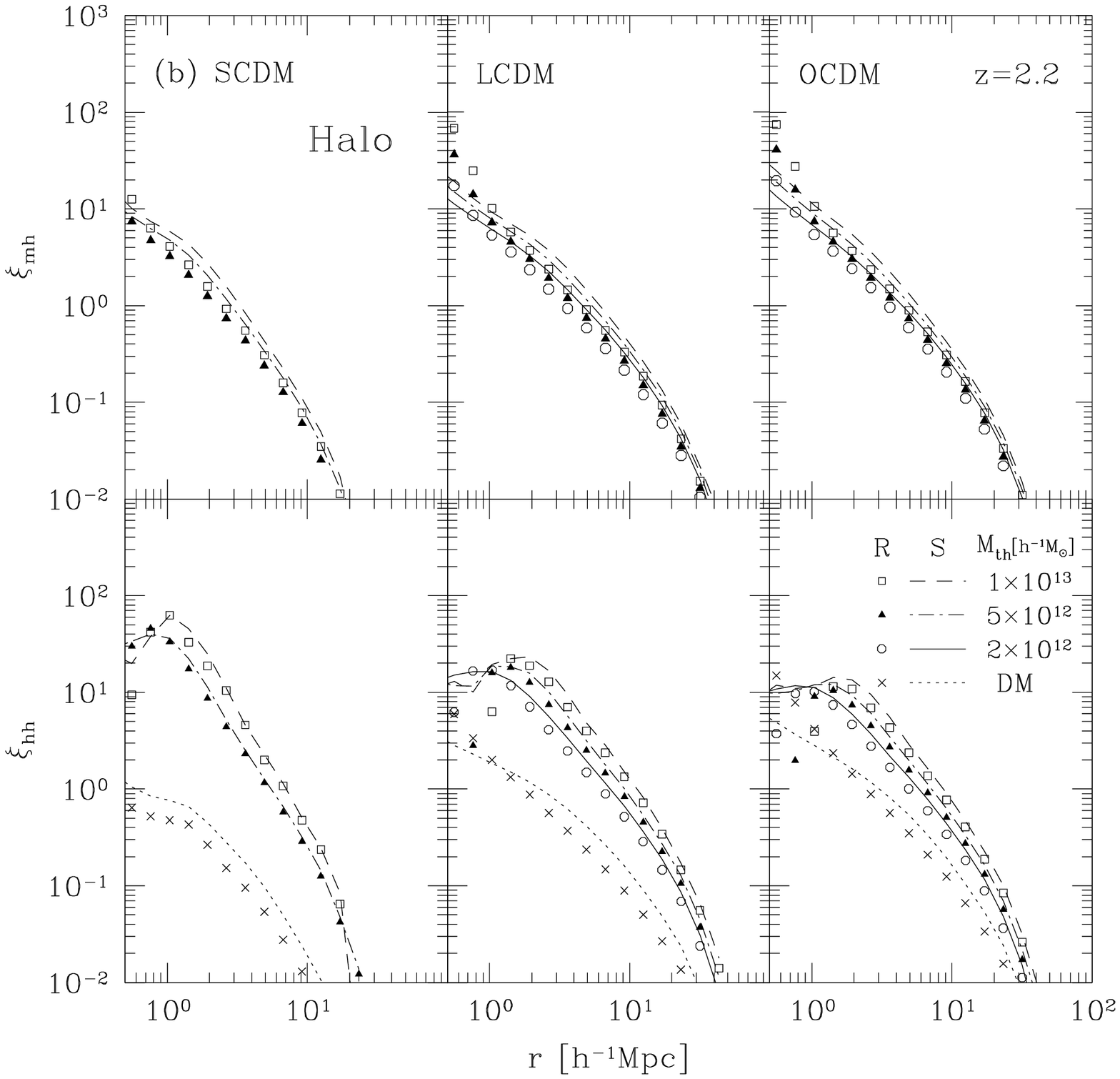}
\caption{Same as Fig. \ref{fig:peakbias} but for halo model;
  open squares and dashed lines for $M > 10^{13}h^{-1}M_\odot$, filled
  triangles and dot--dashed lines for $M > 5\times
  10^{12}h^{-1}M_\odot$, open circles and solid lines for $M > 2\times
  10^{12}h^{-1}M_\odot$, and crosses and dotted lines for dark matter. 
  For SCDM model, we only plot the correlation functions with 
  $M_{\rm th}=5\times 10^{12},\,\,10^{13}h^{-1}M_\odot$. 
(a) $z=0$, (b) $z=2.2$. (ref.~\cite{TMJS-01}) \label{fig:halobias}}
\end{center}
\end{figure}

Those correlation functions are plotted in Figures \ref{fig:peakbias}
 and \ref{fig:halobias} for peaks and halos, respectively.  The
 correlation functions of biased objects generally have 

\clearpage

\clearpage

\subsection{Biasing of galaxies in cosmological hydrodynamic simulations} 

Popular models of the biasing based on the peak or the dark halos are
successful in capturing some essential features of biasing.  None of the
existing models of bias, however, seems to be sophisticated enough for
the coming precision cosmology era.  Development of a more detailed
theoretical model of bias is needed.  A straightforward next step is to
resort to numerical simulations which take account of galaxy formation
even if phenomenological at this point.  We show an example of such
approaches from Yoshikawa et al. (2001)~\cite{YTJS-01} who apply
cosmological smoothed particle hydrodynamic (SPH) )simulations in LCDM
model with particular attention to the comparison of the biasing of dark
halos and simulated galaxies (see also ref.~\cite{somerville-01}).
\begin{figure}[htb]
\begin{center}
\leavevmode\epsfxsize=9cm \epsfbox{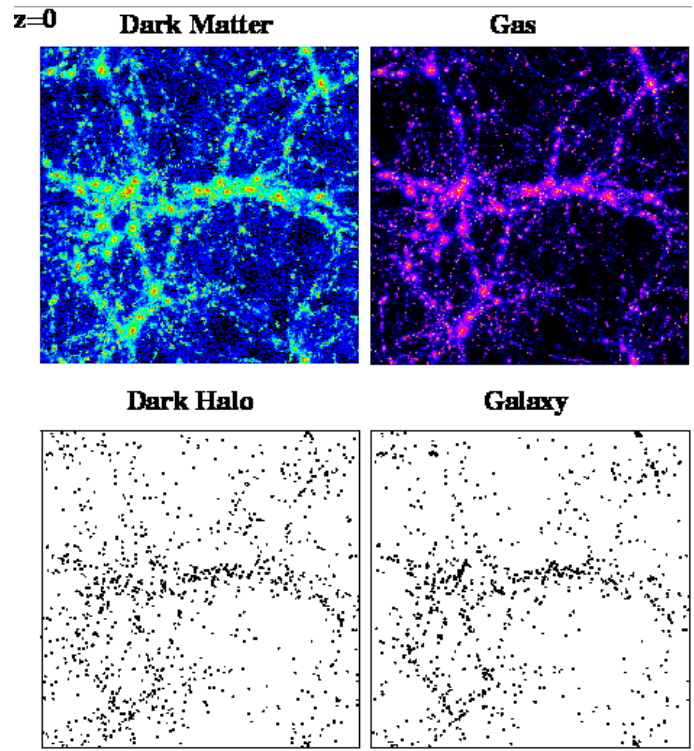} 
\caption{Distribution of gas particles, dark matter particles, galaxies
and dark halos in the volume of $75h^{-1}\times 75h^{-1}\times
30h^{-1}$Mpc$^3$ model at $z=0$. {\it Upper-right:}gas particles; {\it
Upper-left:} dark matter particles; {\it Lower-right:} galaxies; {\it
Lower-left:} DM cores. (ref.~\cite{YTJS-01}) \label{fig:LCDM_Z00}}
\end{center}
\end{figure}
\clearpage

Galaxies in their simulations are identified as clumps of cold and dense
gas particles which satisfy the Jeans condition and have the SPH density
more than 100 times the mean baryon density at each redshift.  Dark
halos are identified with a standard friend-of-friend algorithm; the
linking length is 0.164 times the mean separation of dark matter
particles, for instance, at $z=0$.  In addition, they identify the
surviving high-density substructures in dark halos, DM cores.
See ref.~\cite{YTJS-01} for further details.

\begin{figure}[htb]
\begin{center}
\leavevmode\epsfxsize=9cm \epsfbox{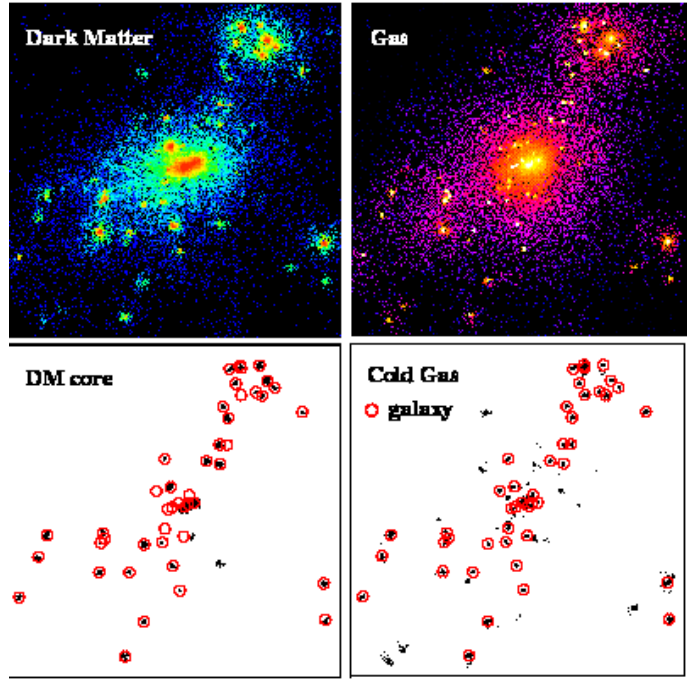}
\caption{Snapshots of the most massive cluster ($M \simeq 8\times
  10^{14}M_{\odot}$) in the simulation at $z=0$. {\it Upper-left:} dark
  matter; {\it Upper-right:} gas; {\it Lower-left:} DM cores; {\it
  Lower-right:} cold gas. Circles in lower panels indicate the positions
  of galaxies identified according to our criteria. The comoving size of
  the box is $6.25h^{-1}$Mpc per side.  (ref.~\cite{YTJS-01})
  \label{fig:region_1}}
\end{center}
\end{figure}

Figure~\ref{fig:LCDM_Z00} illustrates the distribution of dark matter
particles, gas particles, dark halos and galaxies at $z=0$ where
galaxies are more strongly clustered than dark halos.
Figure~\ref{fig:region_1} depicts a close-up snapshot of the most
massive cluster at $z=0$ with mass $M\simeq 8\times 10^{14}M_{\odot}$.
Circles in lower panels indicate the positions of galaxies identified in
our simulation.

\begin{figure}[htb]
\begin{center}
\leavevmode\epsfxsize=10cm \epsfbox{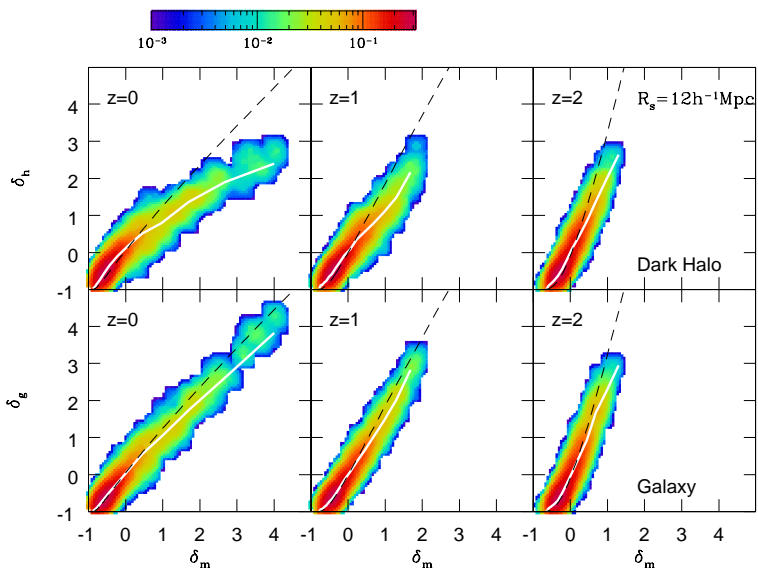}
\caption{Joint probability distributions of overdensity fields for
  dark halos and galaxies with dark matter overdensity smoothed over
  $R_s=12h^{-1}$Mpc  at redshift $z=0$, 1 and 2. Solid lines indicate the
  conditional mean $\bar{\delta}_i(\delta_{\rm m})$ for each
  object. Dashed lines in each panel depict the theoretical prediction
  of conditional mean by Taruya \& Suto (2000).  (ref.~\cite{YTJS-01}) 
\label{fig:sm412}}
\end{center}
\end{figure}

Figure~\ref{fig:sm412} shows the joint distribution of $\delta_{\rm h}$
and $\delta_{\rm g}$ with mass density field $\delta_{\rm m}$ at
redshift $z=0$, 1 and 2 smoothed over $R_s=12h^{-1}$Mpc.  The
conditional mean relation $\bar{\delta}_i(\delta_{\rm m})$ computed
directly from the simulation is plotted in solid lines while dashed
lines indicate theoretical predictions of halo biasing by Taruya \& Suto
(2000)~\cite{TS-00}.  For a given smoothing scale, the simulated halos
exhibit positive biasing for relatively small $\delta_{\rm m}$ in
agreement with the predictions. On the other hand, they tend to be
underpopulated for large $\delta_{\rm m}$, or {\it anti-biased}.  This
is mainly due to the exclusion effect of dark halos due to their finite
volume size which is not taken into account in the theoretical model.
Since our simulated {\it galaxies} have smaller spatial extent than the
halos, the exclusion effect is not so serious. This is clearly
illustrated in lower panels in Figure~\ref{fig:sm412}, and indeed they
show much better agreement with the theoretical model.

Turn next to a more conventional biasing parameter defined through the
two-point statistics:
\begin{equation}
 b_{\xi,i}(r)\equiv \sqrt{\frac{\xi_{ii}(r)}{\xi_{\rm mm}(r)}} ,
\end{equation}
where $\xi_{ii}(r)$ and $\xi_{\rm mm}(r)$ are two-point correlation
functions of objects $i$ and of dark matter, respectively.  While the
above biasing parameter is ill-defined where either $\xi_{ii}(r)$ or
$\xi_{\rm mm}(r)$ becomes negative, it is not the case at clustering
scales of interest ($< 10 h^{-1}$Mpc).  

\begin{figure}[hbt]
\begin{center}
\leavevmode\epsfxsize=10cm \epsfbox{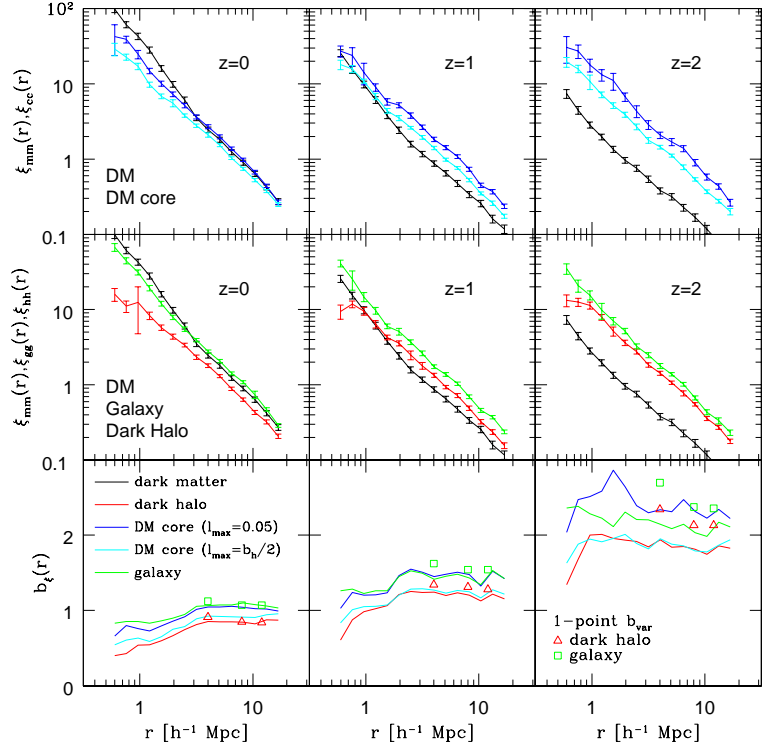}
\caption{Two-point correlation functions of dark matter, galaxies, and
 dark halos from cosmological hydrodynamical simulations.
(ref.~\cite{YTJS-01})
\label{fig:xi_bias}}
 \end{center}
\end{figure}
Figure~\ref{fig:xi_bias} shows two-point correlation functions of dark
matter, galaxies, dark halos and DM cores ({\it upper and middle
panels}), and the profiles of biasing parameters $b_{\xi}(r)$ for those
objects ({\it lower panels}) at $z=0$, 1 and 2.  In lower panels, we
also plot the parameter $b_{{\rm var},i} \equiv \sigma_i/\sigma_{\rm
m}$, which are defined in terms of the one-point statistics (variance),
for comparison on the smoothing scale $R_s=4h^{-1}$Mpc, $8h^{-1}$Mpc and
$12h^{-1}$Mpc at $r=R_s$ for each kind of objects by different symbols.
In the upper panels, we show the correlation functions of DM cores
identified with two different maximum linking length; $l_{\rm max}=0.05$
and $l_{\rm max}=b_{\rm h}/2$. Correlation functions of DM cores
identified with $l_{\rm max}=0.05$ are similar to those of galaxies. On
the other hand, those identified with $l_{\rm max}=b_{\rm h}/2$ exhibit
much weaker correlation, and are rather similar to those of dark
halos. This is due to the fact that the present algorithm of group
identification with larger $l_{\rm max}$ tends to pick up lower mass
halos which are poorly resolved in our numerical resolution.

The correlation functions of galaxies are almost unchanged with
redshift, and correlation functions of dark halos only slightly evolve
between $z=0$ and 2. By contrast, the amplitude of the dark matter
correlation functions evolves rapidly by factor of $\sim 10$ from $z=2$
to $z=0$. The biasing parameter $b_{\xi,{\rm g}}$ is larger at a higher
redshift, for example, $b_{\xi,{\rm g}}\simeq 2$--2.5 at $z=2$.  The
biasing parameter $b_{\xi,{\rm h}}$ for dark halos is systematically
lower than that of galaxies and DM cores again due to the volume
exclusion effect. At $z=0$, galaxies and DM cores are slightly
anti-biased relative to dark matter at $r\simeq 1h^{-1}$Mpc.  In lower
panels, we also plot the one-point biasing parameter $b_{{\rm var},i}
\equiv \sigma_i/\sigma_{\rm m}$ at $r=R_s$ for comparison. In general we
find that $b_{\xi,i}$ is very close to $b_{{\rm var},i}$ at $z \sim 0$,
but systematically lower than $b_{{\rm var},i}$ at higher redshifts.
\begin{figure}[htb]
\begin{center}
\leavevmode\epsfxsize=8cm \epsfbox{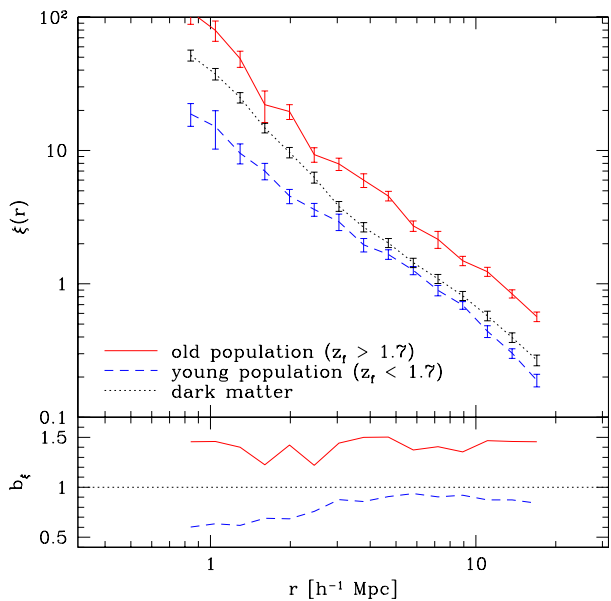} \caption{Two-point
correlation functions for the old and young populations of galaxies at
$z=0$ as well as that of dark matter distribution. The profiles of bias
parameters $b_{\xi}(r)$ for both of the two populations are also shown
in the lower panel.  (ref.~\cite{YTJS-01})
\label{fig:early_late_xi_bias}}
 \end{center}
\end{figure}

For each galaxy identified at $z=0$, we define its formation redshift
$z_{\rm f}$ by the epoch when half of its {\it cooled gas} particles
satisfy our criteria of galaxy formation.  Roughly speaking, $z_{\rm f}$
corresponds to the median formation redshift of {\it stars} in the
present-day galaxies. We divide all simulated galaxies at $z=0$ into two
populations (the young population with $z_{\rm f}<1.7$ and the old
population with $z_{\rm f}>1.7$) so as to approximate the observed
number ratio of $3/1$ for late-type and early-type galaxies.

The difference of the clustering amplitude can be also quantified by
their two-point correlation functions at $z=0$ as plotted in
Figure~\ref{fig:early_late_xi_bias}.  The old population indeed clusters
more strongly than the mass, and the young population is
anti-biased. The relative bias between the two populations $b^{\rm
rel}_{\xi,{\rm g}} \equiv \sqrt{\xi_{\rm old}/\xi_{\rm young}}$ ranges
$1.5$ and 2 for $1h^{-1}\mbox{Mpc}<r<20h^{-1}\mbox{Mpc}$, where
$\xi_{\rm young}$ and $\xi_{\rm old}$ are the two-point correlation
functions of the young and old populations.

\subsection{Halo occupation function approach for galaxy biasing} 

Since the clustering of dark matter halos is well understood now, one
can describe the galaxy biasing if the halo model is combined with the
relation between the halos and luminous objects.  This is another
approach to galaxy biasing, {\it halo occupation function} (HOF), which
becomes very popular recently.  Indeed the basic idea behind HOF has a
long history, but the model predictions have been significantly improved
with the recent accurate models for the mass function, the biasing and
the density profile of dark matter halos. We refer the readers for an
extensive review on the HOF by Cooray \& Sheth (2002)~\cite{CS-02}.
Here we briefly outline this approach.

We adopt a simple parametric form for the average number of a given
galaxy population as a function of the hosting halo mass:
\begin{equation}
\label{eq:Ng}
  N_{\rm g}(M)=
\begin{cases}
{(M/M_1)^\alpha & ($M>M_{\rm min}$)  \cr 
0 & ($M<M_{\rm min}$) } 
\end{cases} .
\end{equation}
The above statistical and empirical relation is the essential ingredient
in the current modeling characterized by the minimum mass of halos which
host the population of galaxies ($M_{\rm min}$), a normalization
parameter which can be interpreted as the critical mass above which
halos typically host more than one galaxy ($M_1$; note that $M_1$ may
exceed $M_{\rm min}$ since the above relation represents the statistical
expected value of number of galaxies), and the power-law index of the
mass dependence of the efficiency of galaxy formation ($\alpha$).  We
will put constraints on the three parameters from the observed number
density and clustering amplitude for each galaxy population.  In short,
the number density of galaxies is most sensitive to $M_1$ which changes
the average number of galaxies per halo.  The clustering amplitude on
large scales is determined by the hosting halos and thus very sensitive
to the mass of those halos, $M_{\rm min}$.  The clustering on smaller
scales, on the other hand, depends on those three parameters in a fairly
complicated fashion; roughly speaking, $M_{\rm min}$ changes the
amplitude, while $\alpha$, and to a lesser extent $M_1$ as well, changes
the slope.

With the above relation, the number density of the corresponding galaxy
population at redshift $z$ is given by
\begin{equation}
\label{eq:ng}
n_{g,z}(z)=\int_{M_{\rm min}}^{\infty} dM~n_{\rm halo}(M,z)~N_{\rm g}(M),
\end{equation}
where $n_{\rm halo}(M)$ denotes the halo mass function.

The galaxy two-point correlation function on small scales is dominated
by contributions of galaxy pairs located in the same halo. For instance,
Bullock et al. (2002)~\cite{bullock-02} adopted the mean number of
galaxy pairs $\langle N_{\rm g}(N_{\rm g}-1)\rangle(M)$ within a halo of
mass $M$ of the form:
\begin{equation}
\label{eq:ngpair}
\langle N_{\rm g}(N_{\rm g}-1)\rangle(M) =
\begin{cases}
{N_{\rm g}^2(M) & ($N_{\rm g}(M)>1$) \cr 
N_{\rm g}^2(M)\log(4 N_{\rm g}(M))/\log(4) & ($1>N_{\rm g}(M)>0.25$) \cr
0 &  ($N_{\rm g}(M)<0.25$)} 
\end{cases} .
\end{equation}

In the framework of the halo model, the galaxy power spectrum consists
of two contributions, one from galaxy pairs located in the same halo
(1-halo term) and the other from galaxy pairs located in two different
halos (2-halo term):
\begin{eqnarray}
\label{eq:P1h2h}
P_{\rm g}^{tot}(k) &=& P_{\rm g}^{1h}(k) + P_{\rm g}^{2h}(k) .
\end{eqnarray}
The 1-halo term is written as
\begin{eqnarray}
\label{eq:P1h}
\hspace*{-0.5cm}
P_{\rm g}^{1h}(k)&=&{1 \over {(2 \pi)^3 n_{g,z}^2}} 
\int dM~ n_{\rm halo}(M)~ \langle N_{\rm g}(N_{\rm g}-1) \rangle
 b(M) |y(k,M)|^p .
\end{eqnarray}
Seljak (2000)~\cite{seljak-00} chose $p=2$ for $\langle N_{\rm g}(N_{\rm
g}-1) \rangle >1$ and $p=1$ for $\langle N_{\rm g}(N_{\rm g}-1) \rangle
<1$.  The 2-halo term on the assumption of the linear halo bias
model~\cite{mw-96} reduces to
\begin{eqnarray}
\label{eq:P2h}
P_{\rm g}^{2h}(k) &=& \frac{P_{\rm lin}(k)}{n_{g,z}^2} 
 \left[\int dM~n_{\rm halo}(M) N_{\rm g}(M) b(M) y(k,M) \right]^2,
\end{eqnarray}
where $P_{\rm lin}(k)$ is the linear dark matter power spectrum, $b(M)$
is the halo bias factor, and $y(k,M)$ is the Fourier transform of the
halo dark matter profile normalized by its mass,
$y(k,M)=\tilde{\rho}(k,M)/M$~\cite{seljak-00}. 

The halo occupation formalism, although simple, provides a useful
framework in deriving constraints on galaxy formation models from large
data sets of the upcoming galaxy redshift surveys.  For example, Zehavi
et al. (2003)~\cite{zehavi-03} used the halo occupation formalism to
model departures from a power law in the SDSS galaxy correlation
function.  They demonstrated that this is due to the transition from a
large-scale regime dominated by galaxy pairs in different halos to a
small-scale regime dominated by those in the same halo.  Magliocchetti
\& Porciani (2003)~\cite{MP-03} applied the halo occupation formalism to
the 2dFGRS clustering results per spectral type of Madgwick et
al. (2003)~\cite{Madgwick-03}.  This provides constraints on the
distribution of late-type and early-type galaxies within the dark matter
halos of different mass.

\clearpage
\section{Relativistic effects observable in clustering at high redshifts}
\label{section:lightcone}

Redshift surveys of galaxies definitely serve as the central database
for observational cosmology. In addition to the existing {\it shallower}
surveys ($z<0.2$), clustering in the universe in the range $z= 1 - 3$
has been partially revealed by, for instance, the Lyman-break galaxies
and X-ray selected AGNs. In particular, the 2dF and SDSS QSO redshift
surveys promise to extend the observable scale of the universe by an
order of magnitude, up to a few Gpc. A proper interpretation of such
redshift surveys in terms of the clustering evolution, however, requires
an understanding of many cosmological effects which can be neglected for
$z\ll 1$ and thus have not been considered seriously so far.  These
cosmological {\it contaminations} include linear redshift-space
(velocity) distortion, nonlinear redshift-space (velocity)
distortion, cosmological redshift-space (geometrical) distortion, 
and cosmological light-cone effect.

We describe a theoretical formalism to incorporate those effects, in
particular the cosmological redshift-distortion and light-cone effects,
and present several specific predictions in CDM models. The details of
the material presented in this section may be found in
refs.~\cite{suto-99, YS-99, YNS-99, MJS-00, HCS-01, HYSE-01}.

\subsection{Cosmological lightcone effect on the two-point 
correlation functions}

Observing a distant patch of the universe is equivalent to observing
the past.  Due to the finite light velocity, a line-of-sight direction
of a redshift survey is along the time, as well as spatial, coordinate
axis. Therefore the entire sample does not consist of objects on a
constant-time hypersurface, but rather on a light-cone, i.e., a null
hypersurface defined by observers at $z=0$. This implies that many
properties of the objects change across the depth of the survey
volume, including the mean density, the amplitude of spatial
clustering of dark matter, the bias of luminous objects with respect
to mass, and the intrinsic evolution of the absolute magnitude and
spectral energy distribution. These aspects should be properly taken
into account in order to extract cosmological information from
observed samples of redshift surveys. 

In order to predict quantitatively the two-point statistics of objects
on the light cone, one must take account of (i) nonlinear
gravitational evolution, (ii) linear redshift-space distortion, (iii)
nonlinear redshift-space distortion, (iv) weighted averaging over the
light-cone, (v) cosmological redshift-space distortion due to the
geometry of the universe, and (vi) object-dependent clustering bias.
The effect (v) comes from our ignorance of the correct cosmological
parameters, and (vi) is rather sensitive to the objects which one has
in mind. Thus the latter two effects will be discussed in the next
subsection.

Nonlinear gravitational evolution of mass density fluctuations is now
well understood, at least for two-point statistics. In practice, we
adopt an accurate fitting formula~\cite{pd-96} for the nonlinear power
spectrum $P^{\rm R}_{\rm nl}(k,z)$ in terms of its linear counterpart.
If one assumes a scale-independent deterministic linear bias,
furthermore, the power spectrum distorted by the peculiar velocity field
is known to be well approximated by the following expression:
\begin{equation}
\label{eq:power_in_redshiftspace}
  P^{({\rm S})}(k_\perp,k_\parallel;z) 
  = b^2(z)P^{({\rm R})}_{{\rm\scriptscriptstyle mass}}(k;z)
\left[1+\beta(z) \left(\frac{k_\parallel}{k}\right)^2 \right]^2
D_{\rm vel}
\left[k_\parallel{\hbox {$\sigma_{\scriptscriptstyle {\rm P}}$}}(z)\right],
\end{equation}
where $k_\perp$ and $k_\parallel$ are the comoving wavenumber
perpendicular and parallel to the line-of-sight of an observer, and
$P^{({\rm R})}_{{\rm\scriptscriptstyle mass}}(k;z)$ is the mass power
spectrum in real space.  The second factor in the right-hand-side comes
from the linear redshift-space distortion~\cite{kaiser-87}, and the last
factor is a phenomenological correction for non-linear velocity
effect~\cite{pd-96}.  In the above, we introduce
\begin{eqnarray} 
\label{eq:betaz}
   \beta(z) \equiv {1\over b(z)} \frac{d\ln D(z)}{d\ln a} \simeq
   {1\over b(z)} \left\{\Omega^{0.6}(z) + {\lambda(z) \over 70}
\left[1+ {\Omega(z) \over 2}\right] \right\} .
\end{eqnarray}

We assume that the pair-wise velocity distribution in real space is
approximated by
\begin{equation}
\label{eq:expveldist}
  f_v(v_{12}) = {1 \over \sqrt{2}{\hbox {$\sigma_{\scriptscriptstyle
         {\rm P}}$}}} \exp\left(-{\sqrt{2}|v_{12}| \over {\hbox
         {$\sigma_{\scriptscriptstyle {\rm P}}$}}} \right) ,
\end{equation}
with ${\hbox {$\sigma_{\scriptscriptstyle {\rm P}}$}}$ being the
1-dimensional pair-wise peculiar velocity dispersion.  Then the
finger-of-god effect is modeled by the damping function,
$D_{\rm vel}\left[k_\parallel{\hbox {$\sigma_{\scriptscriptstyle {\rm
P}}$}}(z)\right]$:
\begin{equation}
\label{eq:damping}
  D_{\rm vel}[k\mu\sigma_{\scriptscriptstyle {\rm P}}]
=\frac{1}{1+\kappa^2\mu^2} ,
\end{equation}
where $\mu$ is the direction cosine in $k$-space, and the dimensionless
wavenumber $\kappa$ is related to the peculiar velocity dispersion
$\sigma_{\scriptscriptstyle {\rm P}}$ in the physical velocity units:
\begin{equation}
\kappa(z)= \frac{k(1+z)\sigma_{\scriptscriptstyle {\rm P}}(z)}
{\sqrt{2}H(z)} .
\end{equation}

Since we are mainly interested in the scales around $1h^{-1}$Mpc, we
adopt the following fitting formula throughout the analysis below which
better approximates the small-scale dispersions in physical units:
\begin{eqnarray}
\label{eq:sigma_pfit}
  \sigma_{\rm\scriptscriptstyle P}(z) \sim \left\{ 
      \begin{array}{ll}
         740 (1+z)^{-1} {\rm km/s} & \mbox{for SCDM model} \\
         650 (1+z)^{-0.8} {\rm km/s} & \mbox{for $\Lambda$CDM model}.
      \end{array}
   \right.
\end{eqnarray}

Integrating equation (\ref{eq:power_in_redshiftspace}) over $\mu$, one
obtains the direction-averaged power spectrum in redshift space:
\begin{equation}
 \frac{ P^{\rm S}_{\rm nl}(k,z)}{P^{\rm R}_{\rm nl}(k,z)} =
A(\kappa)+{2\over3}\beta(z) B(\kappa)+{1\over 5}\beta^2(z) 
C(\kappa)
\label{eq:PSnl}
\end{equation}
where
\begin{eqnarray}
 A(\kappa) &=& {{\arctan}(\kappa) \over \kappa}, \\
  B(\kappa) &=& {3\over\kappa^2}
    \left[1-{{\arctan}(\kappa) \over \kappa}\right] ,\\
  C(\kappa) &=& {5\over3\kappa^2}\left[1-{3\over\kappa^2} +
{3 \, {\arctan}(\kappa) \over \kappa^3} \right] .
\end{eqnarray}

Adopting those approximations, the direction-averaged correlation
functions on the light-cone are finally computed as
\begin{eqnarray}
\label{eq:lcrdximom}
    \xi^{\rm\scriptscriptstyle {LC}}(x_{\rm s}) 
&=& {
   {\displaystyle 
     \int_{z_{\rm min}}^{z_{\rm max}} dz 
     {dV_{\rm c} \over dz} ~[\phi(z)n_0(z)]^2
    \xi(x_{\rm s};z)
    }
\over
    {\displaystyle
     \int_{z_{\rm min}}^{z_{\rm max}} dz 
     {dV_{\rm c} \over dz}  ~[\phi(z)n_0(z)]^2
     }
} ,
\end{eqnarray}
where $z_{\rm min}$ and $z_{\rm max}$ denote the redshift range of the
survey, and
\begin{eqnarray}
\xi(x_{\rm s};z) \equiv 
{1 \over 2\pi^2}\int_0^\infty
 P^{\rm S}_{\rm nl}(k,z) {\sin kx_{\rm s} \over kx_{\rm s}} k^2\,dk .
\end{eqnarray}

Throughout the present analysis, we assume a standard Robertson --
Walker metric of the form:
\begin{equation}
ds^2 = -dt^2 + a(t)^2 
\{ d\chi^2 + S_K(\chi)^2 [d\theta^2 + \sin^2\theta d\phi^2 ] \} ,
\end{equation}
where $S_K(\chi)$  is determined by the sign of the curvature $K$ as
\begin{equation}
 S_K(\chi) = 
  \left\{ 
      \begin{array}{ll}
         \sin{(\sqrt{K}\chi)}/\sqrt{K} & \mbox{$(K>0)$} \\
         \chi & \mbox{$(K=0)$} \\
         \sinh{(\sqrt{-K}\chi)}/\sqrt{-K} & \mbox{$(K<0)$} 
      \end{array}
   \right. ,
 \\ \nonumber
\end{equation}
where the present scale factor $a_\0$ is normalized as unity, and the
spatial curvature $K$ is given as (eq.[\ref{eq:komega}])
\begin{equation}
K = H_0^2 (\Omegam + \Omegal -1) .
\end{equation}
The radial comoving distance $\chi(z)$ is computed by
\begin{eqnarray} 
\chi(z) &=& \int_t^{t_0} {dt \over a(t)}
=  \int_0^z {d z \over H(z)} .
\end{eqnarray}

The comoving angular diameter distance $D_c(z)$ at redshift $z$ is
equivalent to $S^{-1}(\chi(z))$, and, in the case of $\Omegal=0$, is
explicitly given by Mattig's formula:
\begin{equation}
D_c(z) = {1 \over H_0} {z \over 1+z}
{1+z+\sqrt{1+\Omegam z} \over 1 +\Omegam z/2 +\sqrt{1+\Omegam z}} .
\end{equation}
Then $dV_{\rm c}/dz$, the comoving volume element per unit solid
angle, is explicitly given as
\begin{eqnarray}
{dV_{\rm c} \over dz} &=& S_K^2(\chi) {d\chi \over dz} \cr
&=& { S_K^2(\chi)\over
H_0 \sqrt{\Omegam (1 + z)^3 + (1-\Omegam-\Omegal) (1 + z)^2 +
  \Omegal} } .
\end{eqnarray}

\subsection{Evaluating two-point correlation functions from N-body 
simulation data}

The theoretical modeling described above was tested against simulation
results by Hamana, Colombi \& Suto (2001)~\cite{HCS-01}.  Using
cosmological $N$-body simulations in SCDM and $\Lambda$CDM models, they
generated light-cone samples as follows; first, they adopt a distance
observer approximation and assume that the line-of-sight direction is
parallel to $Z$-axis regardless with its $(X,Y)$ position. Second, they
periodically duplicate the simulation box along the $Z$-direction so
that at a redshift $z$, the position and velocity of those particles
locating within an interval $\chi(z) \pm \Delta\chi(z)$ are dumped,
where $\Delta\chi(z)$ is determined by the output time-interval of the
original $N$-body simulation.  Finally they extract five independent
(non-overlapping) cone-shape samples with the angular radius of 1 degree
(the field-of-view of $\pi$ degree$^2$).  In this manner, they have
generated mock data samples on the light-cone continuously extending up
to $z=0.4$ (relevant for galaxy samples) and $z=2.0$ (relevant for QSO
samples), respectively from the small and large boxes.

Two-point correlation function is estimated by the conventional 
pair-count adopting the following estimator~\cite{LS-93}:
\begin{equation}
\xi(x)= \frac{DD(x)-2DR(x)+RR(x)}{RR(x)} .
\end{equation}

The comoving separation $x_{12}$ of two objects located at $z_1$ and
$z_2$ with an angular separation $\theta_{12}$ is given by
\begin{eqnarray} 
x_{12}^2 &=& x_1^2 + x_2^2 
-K x_1^2 x_2^2 (1+\cos^2\theta_{12})\nonumber\\
&&-2 x_1 x_2 \sqrt{1-Kx_1^2}\sqrt{1-Kx_2^2}\cos\theta_{12} ,
\end{eqnarray}
where $x_1\equiv D_c(z_1)$ and $x_2\equiv D_c(z_2)$. 

In redshift space, the observed redshift $z_{\rm obs}$ for each object
differs from the ``real'' one $z_{\rm real}$ due to the velocity
distortion effect:
\begin{eqnarray} 
\label{eq:zobs}
z_{\rm obs} = z_{\rm real} + (1+z_{\rm real}) v_{\rm pec} ,
\end{eqnarray}
where $v_{\rm pec}$ is the line of sight relative peculiar velocity
between the object and the observer in {\it physical} units. Then the
comoving separation $s_{12}$ of two objects in redshift space is
computed as
\begin{eqnarray} 
s_{12}^2 &=&s_1^2 + s_2^2 
-K s_1^2 s_2^2 (1+\cos^2\theta_{12})\nonumber\\
&&-2 s_1 s_2 \sqrt{1-Ks_1^2}\sqrt{1-Ks_2^2}\cos\theta_{12} ,
\end{eqnarray}
where $s_1\equiv D_c(z_{\rm obs, 1})$ and $s_2\equiv D_c(z_{\rm obs,
  2})$.

\begin{figure}[tbh]
\begin{center}
   \leavevmode \epsfxsize=9.5cm \epsfbox{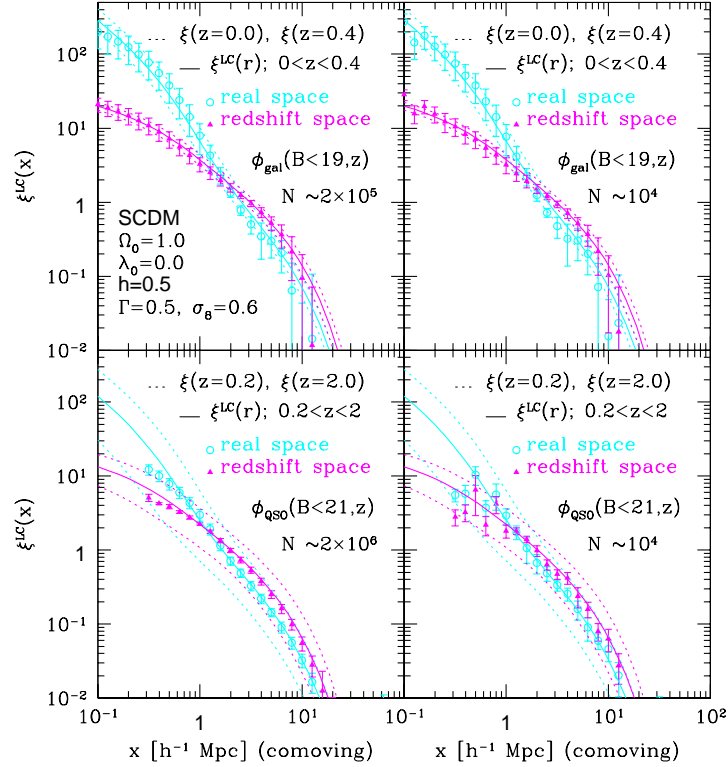}
\caption{ Mass two-point correlation functions on the light cone for
 particles with redshift-dependent selection functions in SCDM model.
{\it Upper:} $z<0.4$, {\it Lower:} $0.2<z<2.0$.
{\it Left:} with selection function whose shape
 is the same as that of the B-band magnitude limit of 
19 for galaxies (upper) and 21 for QSOs (lower). 
{\it Right:} randomly selected $N\sim 10^4$ particles from the
 particles in the left results. (ref.~\cite{HCS-01})}
\label{fig:xilc_phi}
\end{center}
\end{figure}

In properly predicting the power spectra on the light cone, the
selection function should be specified.  For galaxies, they adopt a B-band
luminosity function of the APM galaxies fitted to the Schechter
function~\cite{Loveday-92}.  For quasars, they adopt the B-band luminosity
function from the 2dF QSO survey data~\cite{boyle-00}.
To compute the B-band apparent magnitude from a quasar of absolute
magnitude $M_{\rm B}$ at $z$ (with the luminosity distance
$d_{\rm L}(z)$), they applied the K-correction:
\begin{equation}
  B = M_{\rm B} + 5 \log(d_{\rm L}(z)/ 10 {\rm pc}) 
- 2.5(1-p)\log (1+z) 
\end{equation}
for the quasar energy spectrum $L_\nu \propto \nu^{-p}$ (they used
$p=0.5$).  In practice, they adopt the galaxy selection function
$\phi_{\rm gal}(<B_{\rm lim},z)$ with $B_{\rm lim}=19$ and $z_{\rm
min}=0.01$ for the small box realizations, and the QSO selection
function $\phi_{\rm QSO}(<B_{\rm lim},z)$ with $B_{\rm lim}=21$ and
$z_{\rm min}=0.2$ for the large box realizations.  They do not introduce
the spatial biasing between selected particles and the underlying dark
matter.

\begin{figure}[tbh]
\begin{center}
   \leavevmode \epsfxsize=9.5cm \epsfbox{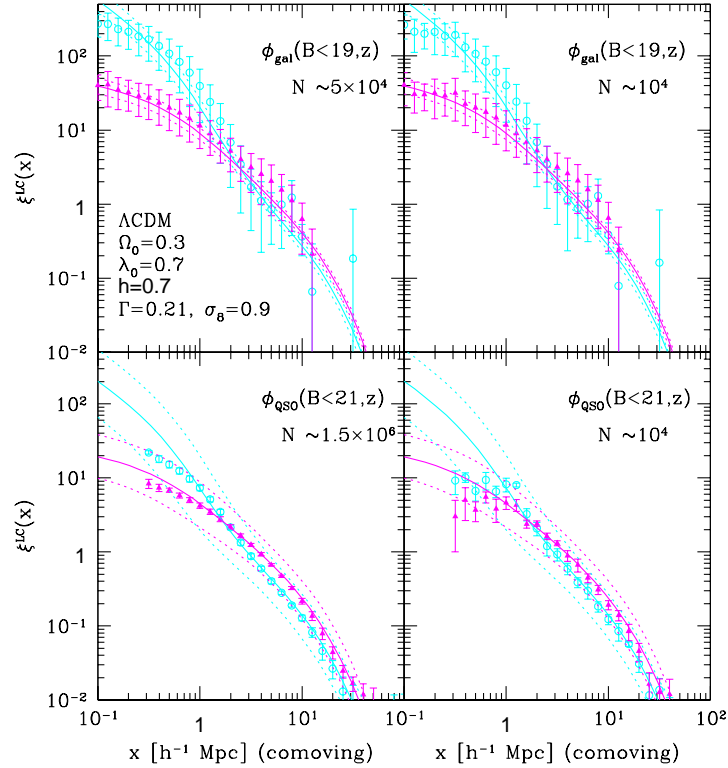}
\caption{ Same as Figure \ref{fig:xilc_phi} but for $\Lambda$CDM
model. (ref.~\cite{HCS-01})}
\label{fig:xilc_phi_lcdm}
\end{center}
\end{figure}

Figures \ref{fig:xilc_phi} and \ref{fig:xilc_phi_lcdm} plot the
two-point correlation functions in SCDM and $\Lambda$CDM, respectively,
taking account of the selection functions.  It is clear that the
simulation results and the predictions are in good agreement. 

\subsection{ Cosmological redshift-space distortion}

Consider a spherical object at high redshift.  If the wrong cosmology is
assumed in interpreting the distance-redshift relation along the line of
sight and in the transverse direction, the sphere will appear distorted.
Alcock \& Paczynski (1979)~\cite{AP-79} pointed out that this curvature
effect could be used to estimate the cosmological constant.  Matsubara
\& Suto (1996)~\cite{ms-96} and Ballinger, Peacock \& Heavens
(1996)~\cite{BPH-96} developed a theoretical framework to describe the
geometrical distortion effect (cosmological redshift distortion) in the
two-point correlation function and the power spectrum of distant
objects, respectively. Certain studies were less optimistic than others
about the possibility of measuring this AP effect.  For example,
Ballinger, Peacock and Heavens (1996) argued that the geometrical
distortion could be confused with the dynamical redshift distortions
caused by peculiar velocities and characterized by the linear theory
parameter $\beta \equiv \frac{\Omega_m^{0.6}}{b}$.  Matsubara \& Szalay
(2002, 2003)~\cite{MS-02, MS-03} showed that the typical SDSS and 2dF
samples of normal galaxies at low redshift ($\sim 0.1$) have
sufficiently low signal-to-noise, but they are too shallow to detect the
Alcock \& Paczynski effect.  On the other hand, the quasar SDSS and
2dFGRS surveys are at a useful redshift, but they are too sparse.  A
more promising sample is the SDSS Luminous Red Galaxies survey (out to
redshift $z\sim 0.5$) which turns out to be optimal in terms of both
depth and density.

While this analysis is promising, it remains to be tested if non-linear
clustering and complicated biasing (which is quite plausible for red
galaxies) would not `contaminate' the measurement of the Equation of
State.  Even if the Alcock \& Paczynski test turns out to be less
accurate than other cosmological tests (e.g., CMB and SN Ia) the effect
itself is an interesting and important ingredient in analyzing the
clustering pattern of galaxies at high redshifts.  We shall now present
the formalism for this effect.

Due to a general-relativistic effect through the geometry of the
universe, the {\it observable} separations perpendicular and parallel
to the line-of-sight direction, $x_{{\rm s}{\scriptscriptstyle\perp}} =
(c/H_0)z\delta\theta$ and $x_{{\rm s}{\scriptscriptstyle\parallel}}=
(c/H_0)\delta z$, are mapped differently to the corresponding comoving
separations in real space $x_{{\scriptscriptstyle\perp}}$ and
$x_{{\scriptscriptstyle\parallel}}$:
\begin{eqnarray}
\label{eq:x2xs}
x_{{\rm s}{\scriptscriptstyle\perp}} (z) &=& x_{\perp} cz/[H_0
  (1+z)d_{\rm\scriptscriptstyle {A}}(z)] \equiv x_{\perp}/c_\perp(z),
\\
 x_{{\rm s}{\scriptscriptstyle\parallel}} (z) &=& x_{\parallel}
  H(z)/H_0 \equiv x_{\parallel}/c_\parallel(z) ,
\end{eqnarray}
with $d_{\rm\scriptscriptstyle {A}}(z)$ being the angular diameter
distance. The difference between $c_\perp(z)$ and $c_\parallel(z)$
generates an apparent anisotropy in the clustering statistics, which
should be isotropic in the comoving space. Then the power spectrum in
cosmological redshift space, $P^{({\rm\scriptscriptstyle {CRD}} )} $,
is related to $P^{({\rm S})}$ defined in the {\it comoving} redshift
space as
\begin{equation}
\label{eq:crdrel}
  P^{({\rm\scriptscriptstyle {CRD}} )}
(k_{{\rm s}\perp},k_{{\rm s}\parallel};z) 
  =\frac{1}{c_\perp(z)^2c_\parallel(z)}
P^{({\rm S})} \left(\frac{k_{{\rm s}{\scriptscriptstyle\perp}}}{c_\perp(z)},
\frac{k_{{\rm s}{\scriptscriptstyle\parallel}}}{c_\parallel(z)};z \right) ,
\end{equation}
where the first factor comes from the Jacobian of the volume element
$dk_{{\rm s}{\scriptscriptstyle\perp}}^2 dk_{{\rm
s}{\scriptscriptstyle\parallel}}$, and $k_{{\rm s}\perp}= c_\perp(z)
k_{\perp}$ and $k_{{\rm s}\parallel}= c_\parallel(z) k_{\parallel}$ are
the wavenumber perpendicular and parallel to the line-of-sight
direction. 

\begin{figure}[tbh]
\begin{center}
    \leavevmode\epsfxsize=12cm \epsfbox{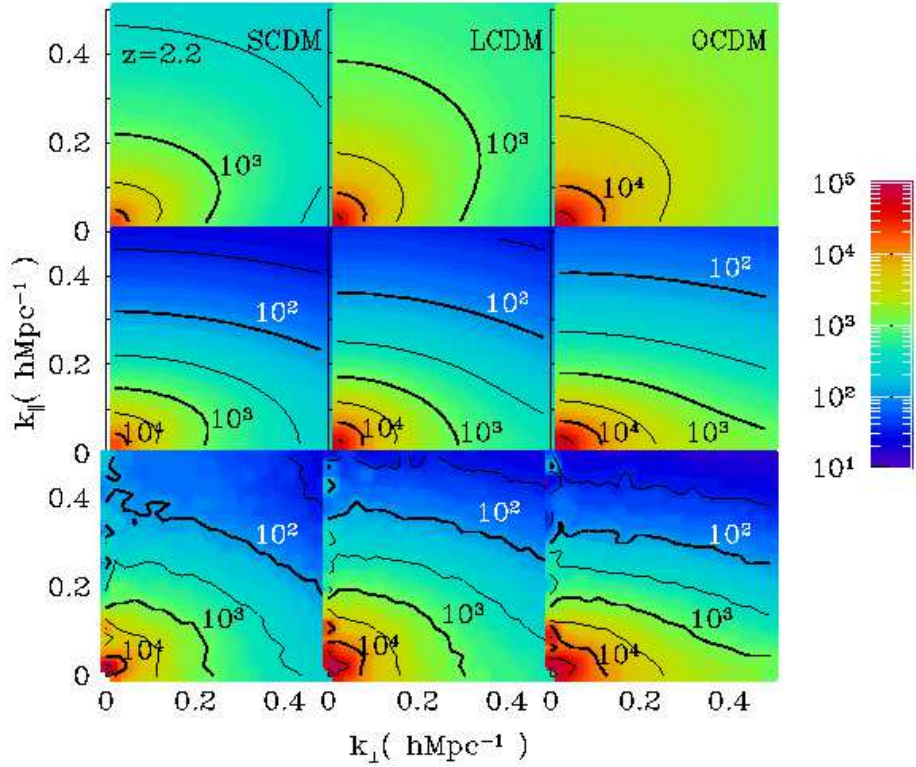}
\caption{Two-dimensional power spectra in cosmological redshift space at
 $z=2.2$.   (ref.~\cite{MJS-00}) \label{fig:power2d}}
\end{center}
\end{figure}
Using equation (\ref{eq:power_in_redshiftspace}), equation
(\ref{eq:crdrel}) reduces to
\begin{eqnarray}
\label{eq:powercrd}
P^{({\rm\scriptscriptstyle {CRD}} )}(k_{\rm s},\mu_k;z)
  &=&\frac{b^2(z)}{c_\perp(z)^2c_\parallel(z)}
P^{({\rm R})}_{\rm\scriptscriptstyle mass} \left(\frac{k_{\rm s}}{c_\perp(z)}
  \sqrt{1+\left[{1\over \eta(z)^2}-1\right]\mu_k^2} ; z \right) \cr &&
  \hspace*{-4cm} \times \left\{1+ \left[{1\over
  \eta(z)^2}-1\right]\mu_k^2 \right\}^{-2} \left\{1+ \left[{1+\beta(z)
  \over \eta(z)^2}-1\right]\mu_k^2\right\}^2 ~ \left[1+
  \frac{k_{\rm s}^2\mu_k^2{\hbox {$\sigma_{\scriptscriptstyle {\rm
  P}}$}}^2}{2c^2_\parallel(z)} \right]^{-1},
\end{eqnarray}
where 
\begin{eqnarray}
\label{eq:k2ks}
k_{\rm s} \equiv \sqrt{ k_{{\rm s}\perp}^2 + k_{{\rm s}\parallel}^2}, \quad
  \mu_k \equiv k_{{\rm s}\parallel}/k_{\rm s}, \quad
  \eta \equiv c_\parallel/c_\perp .
\end{eqnarray}

Figure \ref{fig:power2d} shows anisotropic power spectra
$P^{({\rm\scriptscriptstyle {CRD}} )}(k_{\rm s},\mu_k;z=2.2)$.  As
specific examples, we consider SCDM, LCDM and OCDM models, which have
$(\Omegam, \Omegal, h, \sigma_8)$ $= (1.0, 0.0, 0.5, 0.6)$, $(0.3, 0.7,
0.7, 1.0)$, and $(0.3, 0.0, 0.7, 1.0)$, respectively.  Clearly the
linear theory predictions ($\sigma_{\scriptscriptstyle {\rm P}}=0$; top
panels) are quite different from the results of N-body simulations
(bottom panels), indicating the importance of the nonlinear velocity
effects ($\sigma_{\scriptscriptstyle {\rm P}}$ computed according to
ref.~\cite{MJB-97}; middle panels).

\begin{figure}[thb]
\begin{center}
\leavevmode\epsfxsize=12cm \epsfbox{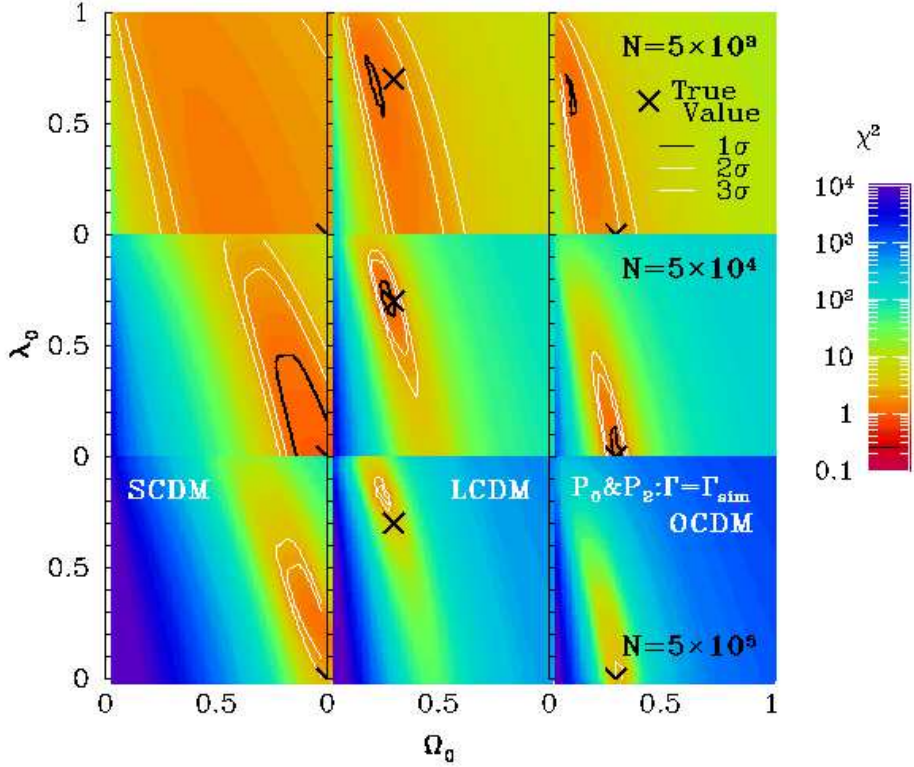}
\caption{The confidence contours on  $\Omegam$-$\Omegal$  plane
  from the $\chi^2$-analysis of the monopole and quadrupole moments of
  the power spectrum in the cosmological redshift space at $z=2.2$.
  We randomly selected $N=5\times10^3$ (upper panels), $N=5\times10^4$
  (middle panels), and $N=5\times10^5$ (lower panels) particles from
  N-body simulation. The value of $\sigma_8$ is adopted from the
  cluster abundance.  (ref.~\cite{MJS-00})
\label{fig:chi2p02ol}}
\end{center}
\end{figure}

Next we decompose the power spectrum into harmonics:
\begin{eqnarray}
\label{eq:pkmoment}
P(k,\mu_k;z) = \sum_{l: {\rm even}} P_l(k) L_l(\mu_k), \quad 
P_l(k;z) \equiv 
\frac{2l+1}{2}\int^1_{-1}d\mu_k P(k,\mu_k;z) L_l(\mu_k) ,
\end{eqnarray}
where $L_l(\mu_k)$ are the $l$-th order Legendre polynomials.
Similarly, the two-point correlation function is decomposed as
\begin{eqnarray}
\label{eq:ximoment}
\xi(x,\mu_x;z) = \sum_{l: {\rm even}} \xi_l(x) L_l(\mu_x), \quad 
\xi_l(x;z) \equiv 
\frac{2l+1}{2}\int^1_{-1}d\mu_x \xi(x,\mu_x;z) L_l(\mu_x) ,
\end{eqnarray}
using the direction cosine, $\mu_x$, between the separation vector and
the line-of-sight.  The above multipole moments satisfy the following
relations:
\begin{eqnarray}
\label{eq:pk2xi}
\xi_l(x;z) &=& {1 \over 2\pi^2 i^l} \int_0^\infty P_l(k;z) j_l(kx) k^2dk, \\
\label{eq:xi2pk}
P_l(k;z) &=& 4\pi i^l \int_0^\infty \xi_l(x;z) j_l(kx) x^2dx, 
\end{eqnarray}
with $j_l(kx)$ being spherical Bessel functions. Substituting
$P^{({\rm\scriptscriptstyle {CRD}} )}(k_{\rm s},\mu_k;z)$ in equation
(\ref{eq:pkmoment}) yields $P^{({\rm\scriptscriptstyle {CRD}} )}_
l(k_{\rm s};z)$, and then $\xi^{({\rm\scriptscriptstyle {CRD}} )}({\bf
  x_s};z)$ can be computed from equation (\ref{eq:pk2xi}).

Comparison of the monopoles and quadrupoles from simulations and model
predictions exhibits how the results are sensitive to the cosmological
parameters, which in turn may put potentially useful constraints on
$(\Omegam, \Omegal)$.  Figure \ref{fig:chi2p02ol} indicates the
feasibility, which interestingly results in a constraint fairly
orthogonal to that from the Supernovae Ia Hubble diagram.

\subsection{Two-Point Clustering Statistics 
on a Light-Cone in Cosmological Redshift Space}

In order to explore the relation between the two-point
statistics on a constant-time hypersurface in real space and that
on a light-cone hypersurface in cosmological redshift space,
we simply consider the case of the deterministic, linear, and
scale-independent bias:
\begin{eqnarray}
\label{eq:biasdef}
 \delta({\bf x},z)  = 
     b(z)~\delta_{\rm m}({\bf x},z) .
\end{eqnarray}
In what follows, we explicitly use the subscript, mass, to indicate the
quantities related to the mass density field, and those without the
subscript correspond to objects satisfying equation (\ref{eq:biasdef}).

Using equation (\ref{eq:powercrd}), the two-point correlation function
in the cosmological redshift space, $\xi^{({\rm\scriptscriptstyle {CRD}}
)}(x_{{\rm s}\perp},x_{{\rm s}\parallel};z)$, is computed as
\begin{eqnarray}
\xi^{({\rm\scriptscriptstyle {CRD}} )}({\bf x_s};z)
&=& {1 \over (2\pi)^3} 
\int P^{({\rm\scriptscriptstyle {CRD}} )}({\bf k_{\rm s}};z) 
\exp(-i{\bf k_{\rm s}}\cdot{\bf x_s}) d^3k_{\rm s} \cr
&=& {1 \over (2\pi)^3} \int 
  P^{({\rm S})}({\bf k};z) \exp(-i{\bf k}\cdot{\bf x}) d^3k \cr
&=& \xi^{({\rm S})}(c_\perp x_{{\rm s}\perp},c_\parallel x_{{\rm s}\parallel};z) ,
\end{eqnarray}
where $\xi^{({\rm S})}(x_{\perp}, x_{\parallel};z)$ is the redshift-space
correlation function defined through equation
(\ref{eq:power_in_redshiftspace}).

Since $P_l^{({\rm\scriptscriptstyle {CRD}} )}(k_{\rm s};z)$ and
$\xi_l^{({\rm\scriptscriptstyle {CRD}} )} (x_{\rm s};z)$ are defined in
redshift space, the proper weight should be
\begin{eqnarray}
\label{eq:weight}
d^3s^{({\rm\scriptscriptstyle {CRD}} )}
~[\phi(z)n_0^{{\rm\scriptscriptstyle {CRD}} }(z)]^2
=d^3x{[\phi(z)n_0^{{\rm\scriptscriptstyle {com}} }(z)]^2
~c_\perp(z)^2c_\parallel(z)},
\end{eqnarray}
where $n_0^{{\rm\scriptscriptstyle {CRD}} }(z)$ and
$n_0^{{\rm\scriptscriptstyle {com}} }(z)$ denote the number densities of
the objects in cosmological redshift space and comoving space,
respectively, and $\phi(z)$ is the selection function determined by
the observational target selection and the luminosity function of the
objects. Then, the final expressions~\cite{SMY-00} reduce to
\begin{eqnarray}
\label{eq:lccrdpkmom}
    P^{({\rm\scriptscriptstyle {LC}} ,{\rm\scriptscriptstyle {CRD}} )}_l(k_{\rm s}) 
&=& {
   {\displaystyle 
     \int_{z_{\rm min}}^{z_{\rm max}} dz 
     {dV_{\rm c} \over dz} ~[\phi(z)n_0^{{\rm\scriptscriptstyle {com}} }(z)]^2
    {c_\perp(z)^2c_\parallel(z)} P_l^{({\rm\scriptscriptstyle {CRD}} )}(k_{\rm s};z)
    }
\over
    {\displaystyle
     \int_{z_{\rm min}}^{z_{\rm max}} dz 
     {dV_{\rm c} \over dz}  ~[\phi(z)n_0^{{\rm\scriptscriptstyle {com}} }(z)]^2
          {c_\perp(z)^2c_\parallel(z)}
     }
} , \\
\label{eq:lccrdximom}
    \xi^{({\rm\scriptscriptstyle {LC}} ,{\rm\scriptscriptstyle {CRD}} )}_l(x_{\rm s}) 
&=& {
   {\displaystyle 
     \int_{z_{\rm min}}^{z_{\rm max}} dz 
     {dV_{\rm c} \over dz} ~[\phi(z)n_0^{{\rm\scriptscriptstyle {com}} }(z)]^2
          {c_\perp(z)^2c_\parallel(z)}
    \xi^{{\rm\scriptscriptstyle {CRD}} }_l(x_{\rm s};z)
    }
\over
    {\displaystyle
     \int_{z_{\rm min}}^{z_{\rm max}} dz 
     {dV_{\rm c} \over dz}  ~[\phi(z)n_0^{{\rm\scriptscriptstyle {com}} }(z)]^2
            {c_\perp(z)^2c_\parallel(z)}
     }
} ,
\end{eqnarray}
where $z_{\rm min}$ and $z_{\rm max}$ denote the redshift range of the
survey, $dV_{\rm c}/dz = d_{\rm\scriptscriptstyle {C}} ^2(z)/H(z)$ is
the comoving volume element per unit solid angle.

Note that $k_{\rm s}$ and $x_{\rm s}$, defined in
$P_l^{({\rm\scriptscriptstyle {CRD}} )}(k_{\rm s};z)$ and
$\xi^{{\rm\scriptscriptstyle {CRD}} }_ l(x_{\rm s};z)$, are related to
their comoving counterparts at $z$ through equations (\ref{eq:k2ks}) and
(\ref{eq:x2xs}), while those in $P^{({\rm\scriptscriptstyle {LC}}
,{\rm\scriptscriptstyle {CRD}} )}_ l(k_{\rm s})$ and
$\xi^{({\rm\scriptscriptstyle {LC}} , {\rm\scriptscriptstyle {CRD}}
)}_l(x_{\rm s})$ are not specifically related to any comoving wavenumber
and separation. Rather, they correspond to the quantities averaged over
the range of $z$ satisfying the observable conditions $x_{\rm
s}=(c/H_0)\sqrt{\delta z^2 + z^2\delta\theta^2}$ and $k_{\rm
s}=2\pi/x_{\rm s}$.

Let us show specific examples of the two-point clustering statistics on
a light-cone in cosmological redshift space. We consider SCDM and LCDM
models, and take into account the selection functions relevant to the
upcoming SDSS spectroscopic samples of galaxies and quasars by adopting
the {\it B}-band limiting magnitudes of 19 and 20, respectively.

\begin{figure}[htb]
\begin{center}
  \leavevmode\epsfxsize=11cm \epsfbox{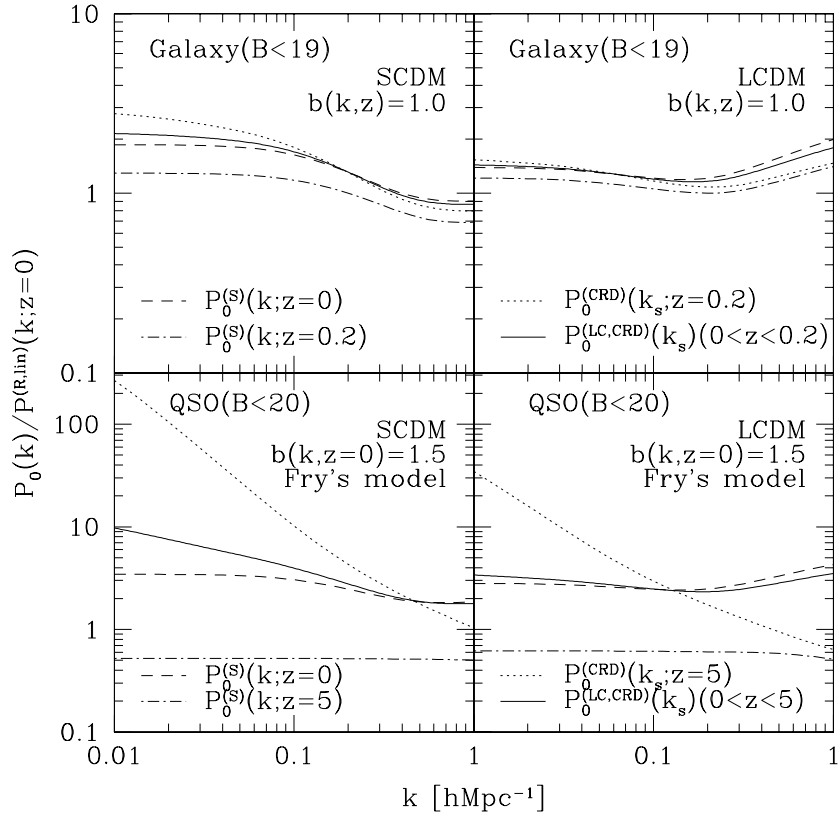}
\caption{Light-cone and cosmological redshift-space distortion
  effects on angle-averaged power spectra. (ref.~\cite{SMY-00})
\label{fig:pk_lccrd}
}
\end{center}
\end{figure}

Figure \ref{fig:pk_lccrd} compares the predictions for the
angle-averaged (monopole) power spectra under various
approximations. The upper and lower panels adopt the selection functions
appropriate for galaxies in $0<z<{z_{\rm max}}=0.2$ and QSOs in
$0<z<{z_{\rm max}}=5$, respectively. The left and right panels present
the results in SCDM and LCDM models. For simplicity we adopt a
scale-independent linear bias model~\cite{fry-96}:
\begin{equation} 
  b(z)= 1 +{1\over D(z)} [b(k,z=0)-1],
\label{FryM}
\end{equation}
with $b(k,z=0)=1$ and $1.5$ for galaxies and quasars, respectively.

The upper and lower panels correspond to magnitude-limited samples of
galaxies ($B<19$ in $0<z<{z_{\rm max}}=0.2$; no bias model) and QSOs
($B<20$ in $0<z<{z_{\rm max}}=5$; Fry's linear bias model),
respectively.  We present the results normalized by the real-space power
spectrum in linear theory $P^{\rm (R,lin)}(k;z)$ ~\cite{bbks-86}, and
$P^{\rm (S)}_0(k;z=0)$, $P^{\rm (S)}_0(k;z={z_{\rm max}})$,
$P^{({\rm\scriptscriptstyle {CRD}} )}_ 0(k_{\rm s};z={z_{\rm max}})$,
and $P^{({\rm\scriptscriptstyle {LC}} , {\rm\scriptscriptstyle {CRD}}
)}_0(k_{\rm s})$ are computed using the nonlinear power
spectrum~\cite{pd-96}.

Consider first the results for the galaxy sample (upper panels). On
linear scales ($k < 0.1h~$Mpc$^{-1}$), $P^{\rm (S)}_0(k;z=0)$ plotted in
dashed lines is enhanced relative to that in real space, mainly due to a
linear redshift-space distortion [the Kaiser factor in equation
(\ref{eq:power_in_redshiftspace})]. For nonlinear scales, the nonlinear
gravitational evolution increases the power spectrum in real space,
while the finger-of-god effect suppresses that in redshift space. Thus,
the net result is sensitive to the shape and the amplitude of the
fluctuation spectrum, itself; in the LCDM model that we adopted, the
nonlinear gravitational growth in real space is stronger than the
suppression due to the finger-of-god effect. Thus, $P^{\rm
(S)}_0(k;z=0)$ becomes larger than its real-space counterpart in linear
theory. In the SCDM model, however, this is opposite and $P^{\rm
(S)}_0(k;z=0)$ becomes smaller.

The power spectra at $z=0.2$ (dash-dotted lines) are smaller than those
at $z=0$ by the corresponding growth factor of the fluctuations, and one
might expect that the amplitude of the power spectra on the light-cone
(solid lines) would be in-between the two.  While this is correct, if we
use the comoving wavenumber, the actual observation on the light-cone in
the cosmological redshift space should be expressed in terms of $k_{\rm
s}$ [equation (\ref{eq:k2ks})].  If we plot the power spectra at $z=0.2$
taking into account the geometrical distortion,
$P^{({\rm\scriptscriptstyle {CRD}} )}_0(k_{\rm s};z=0.2)$ in the dotted
lines becomes significantly larger than $P^{\rm
(S)}_0(k;z=0.2)$. Therefore, $P^{({\rm\scriptscriptstyle {LC}} ,
{\rm\scriptscriptstyle {CRD}} )}_ 0(k_{\rm s})$ should take a value
between those of $P^{({\rm\scriptscriptstyle {CRD}} )}_0(k_{\rm s};z=0)
= P^{\rm (S)}_0(k;z=0)$ and $P^{({\rm\scriptscriptstyle {CRD}} )}_
0(k_{\rm s};z=0.2)$. This explains the qualitative features shown in the
upper panels of Figure \ref{fig:pk_lccrd}.  As a result, both the
cosmological redshift-space distortion and the light-cone effect
substantially change the predicted shape and amplitude of the power
spectra, even for the galaxy sample~\cite{NMS-98}.  The results for the
QSO sample can be basically understood in a similar manner, except that
the evolution of the bias makes a significant difference, since the
sample extends to much higher redshifts.

\begin{figure}[htb]
\begin{center}
   \leavevmode\epsfxsize=11cm \epsfbox{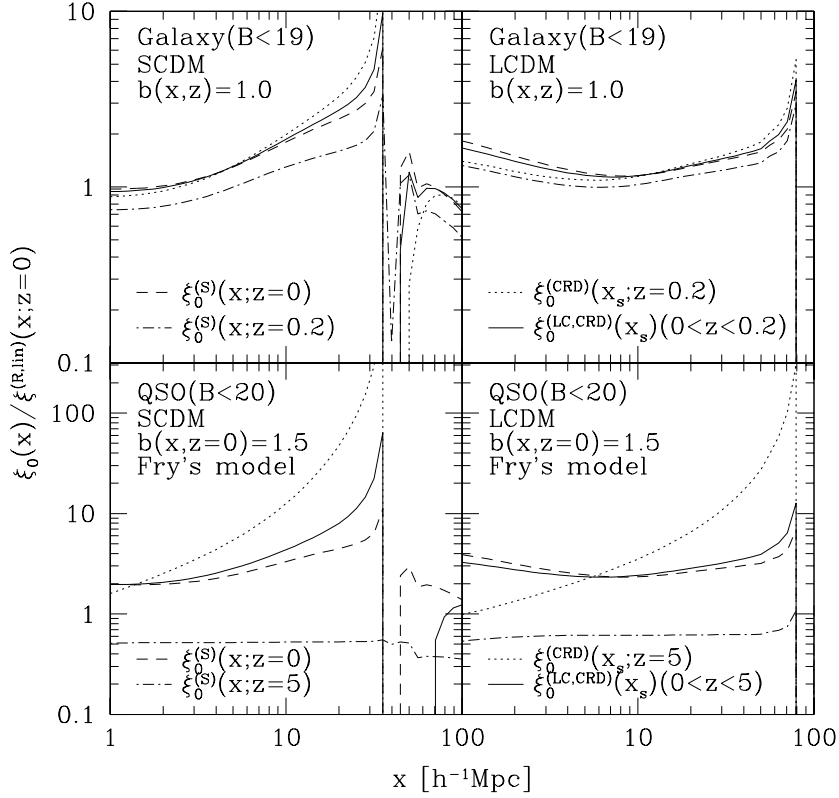}
\caption{Same as Figure \ref{fig:pk_lccrd} on angle-averaged two-point
correlation functions.  (ref.~\cite{SMY-00})
\label{fig:xi_lccrd}
}
\end{center}
\end{figure}

Figure \ref{fig:xi_lccrd} shows the results for the angle-averaged
(monopole) two-point correlation functions, exactly corresponding to
those in Figure \ref{fig:pk_lccrd}. The results in this figure can also
be understood by an analogy of those presented in Figure
\ref{fig:pk_lccrd} at $k \sim 2\pi/x$. Unlike the power spectra,
however, two-point correlation functions are not positive definite.  The
funny features in Figure \ref{fig:xi_lccrd} on scales larger than $30
h^{-1}$Mpc ($100 h^{-1}$Mpc) in SCDM (LCDM) originate from the fact that
$\xi^{\rm (R,lin)}(x,z=0)$ becomes negative there.

In fact, since the resulting predictions are sensitive to the bias,
which is unlikely to quantitatively be specified by theory, the present
methodology will find two completely different applications.  For
relatively shallower catalogues, like galaxy samples, the evolution of
bias is not supposed to be so strong. Thus, one may estimate the
cosmological parameters from the observed degree of the redshift
distortion, as has been conducted conventionally. Most importantly, we
can correct for the systematics due to the light-cone and geometrical
distortion effects, which affect the estimate of the parameters by $\sim
10$\%.  Alternatively, for deeper catalogues like high-redshift quasar
samples, one can extract information on the object-dependent bias only by
correcting the observed data on the basis of our formulae. 

In a sense, the former approach uses the light-cone and geometrical
distortion effects as real cosmological signals, while the latter
regards them as inevitable, but physically removable, noise. In both
cases, the present methodology is important in properly interpreting the
observations of the universe at high redshifts.

\clearpage
\section{Recent results from 2dF and SDSS}
\label{section:2dFSDSS}

\subsection{The latest galaxy redshift surveys}

Redshifts surveys in the 1980s and the 1990s (e.g the CfA, IRAS and Las
campanas surveys) measured thousands to tens of thousands galaxy
redshifts.  Multifibre technology now allows us to measure redshifts of
millions of galaxies. Below we summarize briefly the properties of the
main new surveys, 2dFGRS, SDSS, 6dF, VIRMOS, DEEP2 and we discuss key
results from 2dFGRS and SDSS.  Further analysis of these surveys is
currently underway.

\subsubsection{The 2dF galaxy redshift survey \label{subsubsec:2dF}}

\begin{figure}[hbt]
\begin{center}
\leavevmode\epsfxsize=12cm\epsfbox{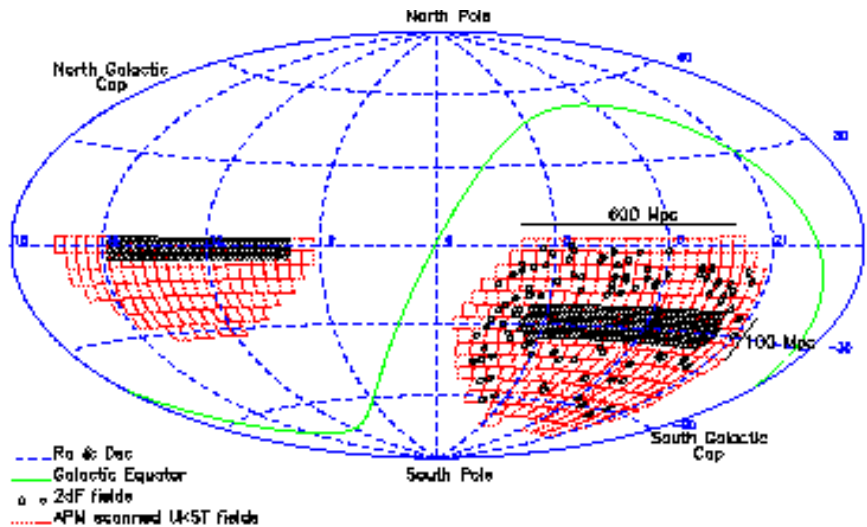} 
\caption{The 2dFGRS fields (small circles) superimposed on the APM
catalogue area (dotted outlines of Sky Survey plates). }
\label{fig:2dfmap}
\end{center}
\end{figure}

\begin{figure}[tbh]
\begin{center}
\leavevmode\epsfxsize=12cm \epsfbox{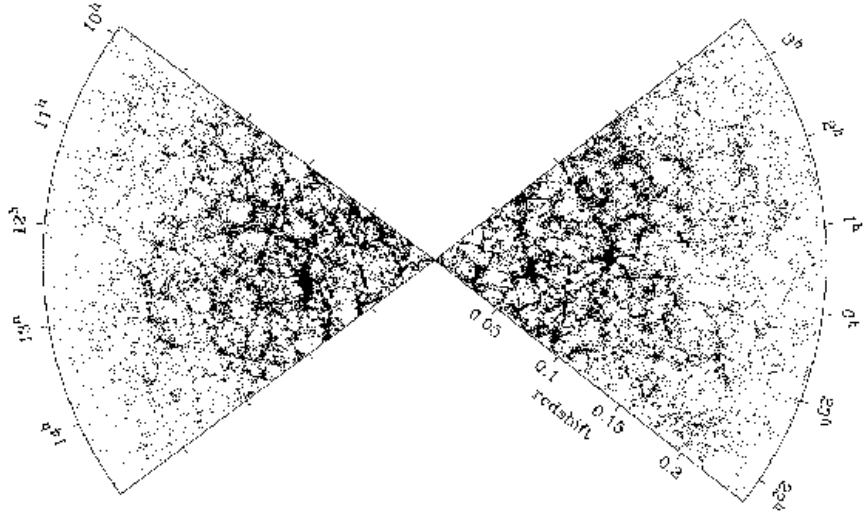} 
\caption{The distribution of 63,000 2dFGRS galaxies in the NGP ({\it Left}) 
and SGP ({\it Right}) strips.
}  \label{fig:2dfslice}
\end{center}
\end{figure}

The Anglo-Australian 2 degree Field Galaxy Redshift (2dFGRS)\footnote{
http://www.mso.anu.edu.au/2dFGRS/} has recently been completed with
redshifts for 230,000 galaxies selected from the APM catalogue December
2002) down to an extinction corrected magnitude limit of $b_J<19.45$.
The main survey regions are two declination strips, in the northern and
southern Galactic hemispheres, and also 100 random fields, covering in
total about 1800 deg$^2$ (Figs. \ref{fig:2dfmap} and
\ref{fig:2dfslice}).  
The median redshift of the
2dFGRS is ${\bar z} \sim 0.1$ (see refs.~\cite{colless-03, peacock-03}
for reviews).

\subsubsection{The SDSS galaxy redshift survey \label{subsubsec:SDSS}}

The SDSS (Sloan Digital Sky Survey) is a US-Japan-Germany joint project
to image a quarter of the Celestial Sphere at high Galactic latitude as
well as to obtain spectra of galaxies and quasars from the imaging
data\footnote{http://www.sdss.org/}. The dedicated 2.5 meter telescope
at Apache Point Observatory is equipped with a multi-CCD camera with
five broad bands centered at $3561$, $4676$, $6176$, $7494$, and
$8873${\,\AA}.  For further details of SDSS, see refs.~\cite{york-00,%
strauss-02}

\begin{figure}[hbt]
\begin{center}
 \leavevmode\epsfxsize=12cm\epsfbox{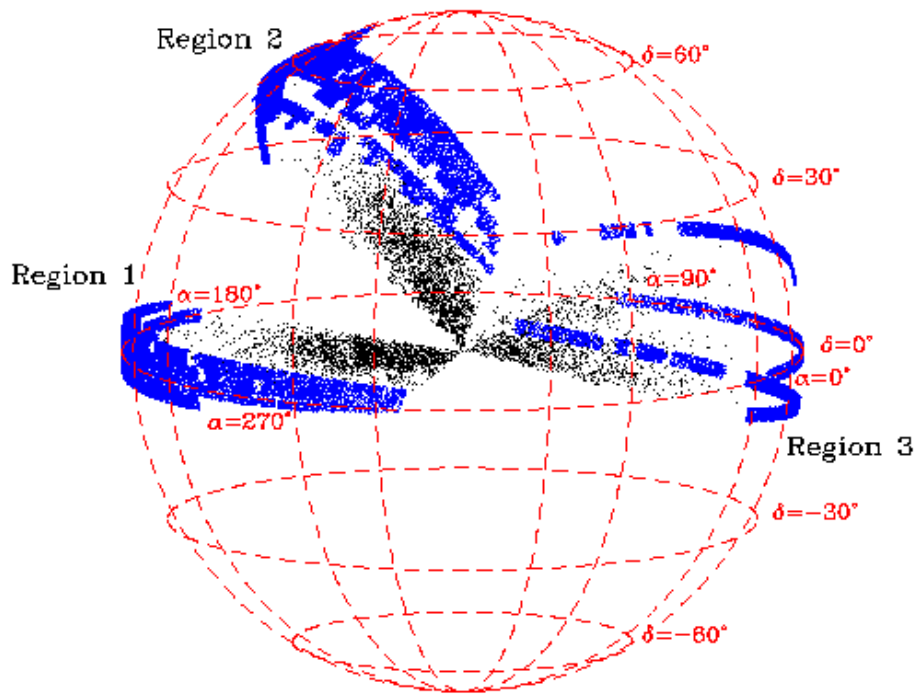}
\caption{3D redshift-space map centered on us, and its
projection on the celestial sphere of SDSS galaxy subset, including
the three main regions. (ref.~\cite{hikage-03})  
\label{fig:sdssmap_equator}}
\end{center}
\end{figure}

\begin{figure}[hbt]
\begin{center}
 \leavevmode\epsfxsize=6cm \epsfbox{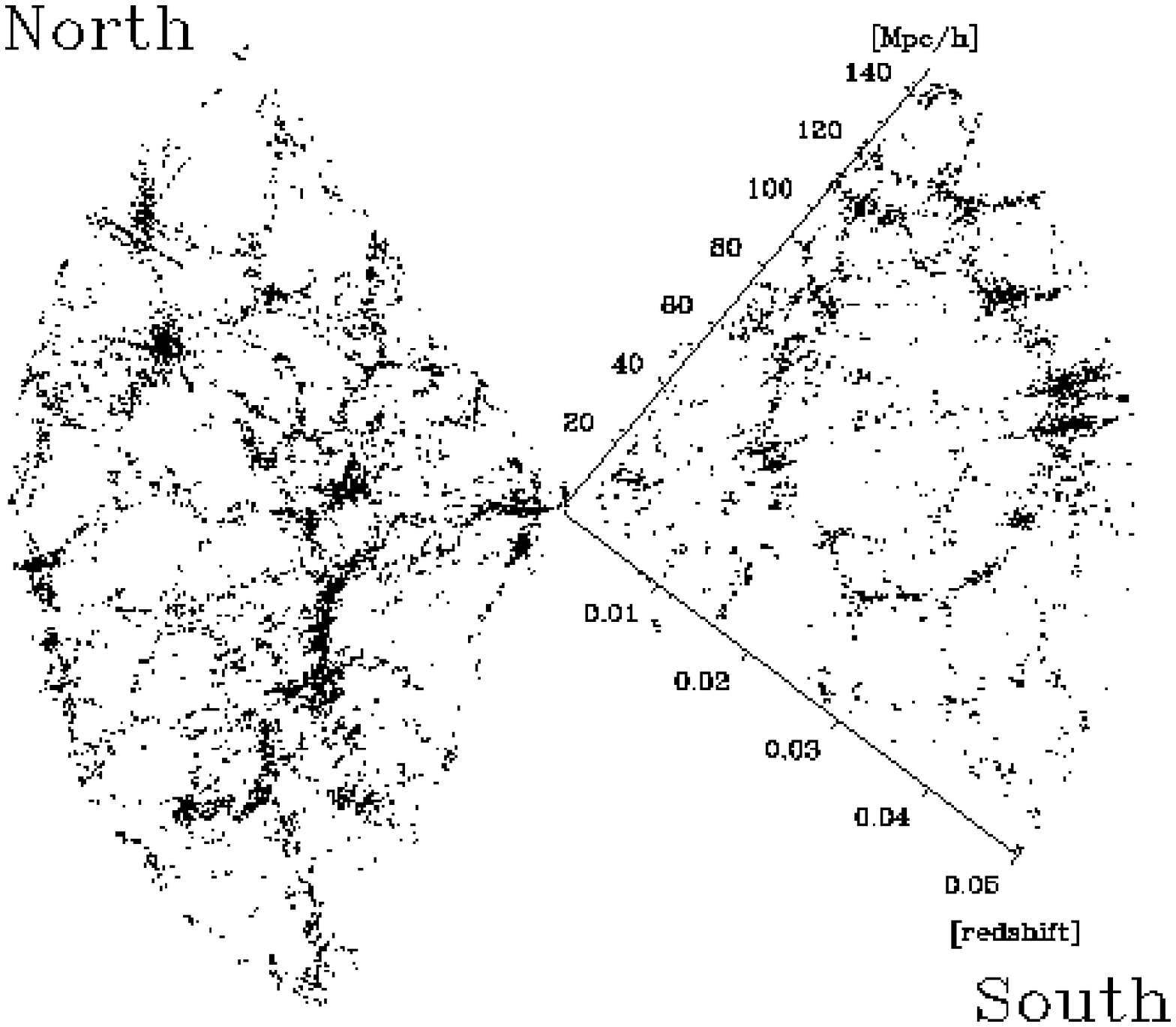}
 \leavevmode\epsfxsize=6cm \epsfbox{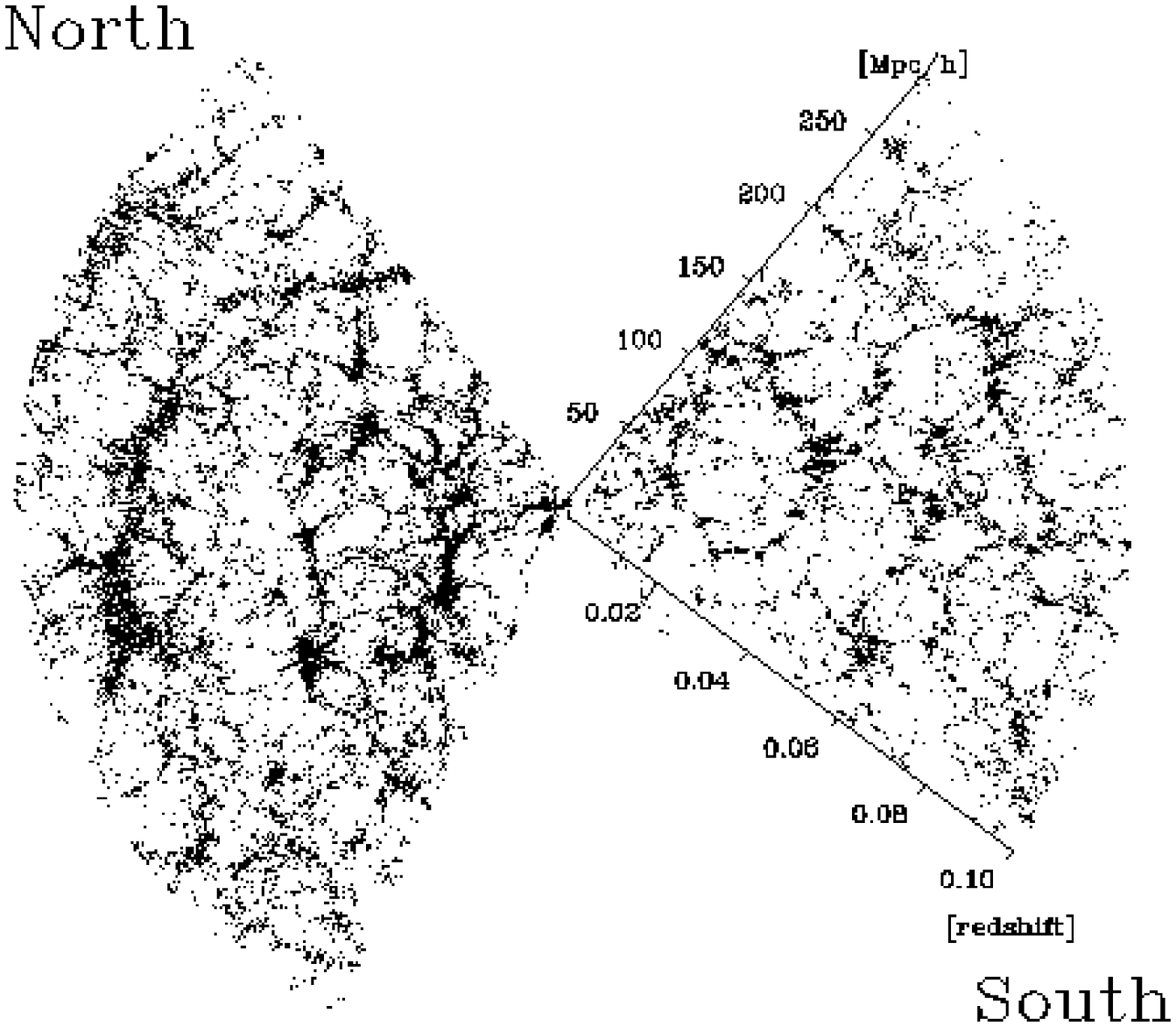}
\vspace*{0.5cm}
 \leavevmode\epsfxsize=11.5cm \epsfbox{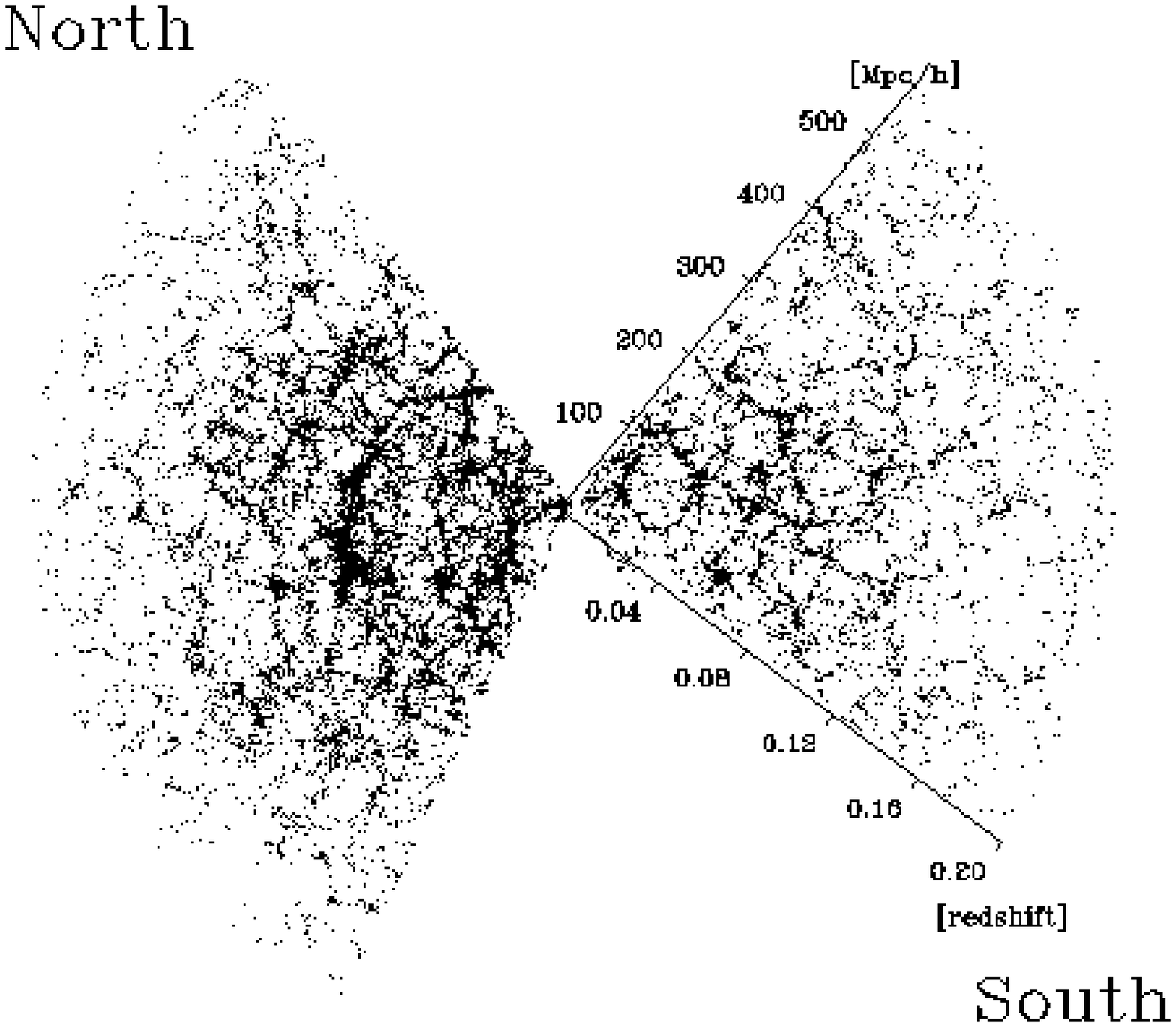}
\caption{Redshift slices of SDSS galaxy data around the
  equatorial plane. The redshift limits and the thickness of the
  planes are: {\it Upper-left} $z<0.05$, $10h^{-1}$Mpc; {\it
  Upper-right} $z<0.1$, $15h^{-1}$Mpc; {\it Lower} $z<0.2$,
  $20h^{-1}$Mpc.  The size of points has been adjusted. Note that the
  data for the Southern part are sparser than those for the Northern
  part, especially for thick slices. (ref.~\cite{hikage-03})}
\label{fig:sdssslice}
\end{center}
\end{figure}

The latest map of the SDSS galaxy distribution, together with a typical
slice, are shown in Figures \ref{fig:sdssmap_equator}, and
\ref{fig:sdssslice} ~\cite{hikage-03}.  The three--dimensional map
centered on us in equatorial coordinate system is shown Figure
\ref{fig:sdssmap_equator}.  Redshift slices of galaxies centered around
the equatorial plane with various redshift limits and thicknesses of
planes are shown in Figure \ref{fig:sdssslice}: $z<0.05$ with thickness
of $10h^{-1}$Mpc centered around the equatorial plane in the upper-left
panel; $z<0.1$ with thickness of $15h^{-1}$Mpc in the upper-right panel;
$z<0.2$ with thickness of $20h^{-1}$Mpc in the lower panel.

\clearpage

\subsubsection{The 6dF  galaxy redshift survey}

The 6dF (6-degree Field)\footnote{http://www.mso.anu.edu.au/6dFGS/} is a
survey of redshifts and peculiar velocities of galaxies selected
primarily in the Near Infrared from new 2MASS (Two Micron All Sky
Survey) catalogue\footnote{http://www.ipac.caltech.edu/2mass/}. One goal
is to measure redshifts of more than 170,000 galaxies over nearly the
whole Southern sky.  Another exciting aim of the survey is to measure
peculiar velocities (using 2MASS photometry and 6dF velocity
dispersions) of about 15,000 galaxies out to $150h^{-1}$Mpc.  The high
quality data of this survey could revive peculiar velocities as a
cosmological probe (which was very popular about 10-15 years ago).
Observations have so far obtained nearly 40,000 redshifts and completion
is expected in 2005.

\subsubsection{The DEEP  galaxy redshift survey}

The DEEP survey is a two-phased project using the Keck telescopes to
study the properties and distribution of high redshift galaxies
\footnote{http://deep.berkeley.edu/~marc/deep/}.  Phase 1 used the LRIS
spectrograph to study a sample of $\sim 1000$ galaxies to a limit of
I=24.5.  Phase 2 of the DEEP project will use the new DEIMOS
spectrograph to obtain spectra of $\sim 65,000$ faint galaxies with
redshifts $z \sim 1$. The scientific goals are to study the evolution of
properties of galaxies and the evolution of the clustering of galaxies
compared to samples at low redshift.  The survey is designed to have the
fidelity of local redshift surveys such as the LCRS survey, and to be
complementary to ongoing large redshift surveys such as the SDSS project
and the 2dF survey.  The DEIMOS/DEEP or DEEP2 survey will be executed
with resolution R~4000, and we therefore expect to measure linewidths
and rotation curves for a substantial fraction of the target galaxies.
DEEP2 will thus also be complementary to the VLT/VIRMOS project, which
will survey more galaxies in a larger region of the sky but with much
lower spectral resolution and with fewer objects at high redshift.

\subsubsection{The VIRMOS  galaxy redshift survey}

The on-going Franco-Italian VIRMOS project\footnote{
http://www.astrsp-mrs.fr/virmos/} has delivered the VIMOS spectrograph
for the European Southern Observatory Very Large Telescope
(ESO-VLT). VIMOS is a VIsible imaging Multi-Object Spectrograph with
outstanding multiplex capabilities: with 10 arcsec slits, spectra can be
taken of 600 objects simultaneously. In integral field mode, a
6400-fibre Integral Field Unit (IFU) provides spectroscopy for all
objects covering a 54x54 arcsec$^2$ area. VIMOS therefore provides
unsurpassed efficiency for large surveys.  The VIRMOS project consists
of: Construction of VIMOS, and a Mask Manufacturing Unit for the
ESO-VLT.  The VIRMOS-VLT Deep Survey (VVDS), a comprehensive imaging and
redshift survey of the deep universe based on more than 150,000
redshifts in four 4 sq.-degree fields.

\subsection{Cosmological parameters from 2dFGRS}

\subsubsection{The Power spectrum of 2dF Galaxies on large scales}

An initial estimate of the convolved, redshift-space power spectrum of
the 2dFGRS was determined by Percival et al. (2002)~\cite{percival-02}
for a sample of 160,000 redshifts.  On scales $0.02h/{\rm Mpc} <k<0.15
h/{\rm Mpc}$, the data are fairly robust and the shape of the power
spectrum is not significantly affected by redshift-space distortion or
non-linear effects, while its overall amplitude is increased due to the
linear redshift-space distortion effect (see \S
\ref{section:lightcone}).  

If one fits the $\Lambda$-CDM model predictions to the 2dFGRS power
spectrum (Fig. \ref{fig:neutrinomass}) over the above range in $k$, one
can constrain the cosmological parameters. For instance, assuming a
Gaussian prior on the Hubble constant $h=0.7\pm0.07$ (from
ref.~\cite{freedman-01}), Percival et al. (2002)~\cite{percival-02}
obtained the 68 percent confidence limits on the shape parameter
$\omegam h=0.20 \pm 0.03$, and the baryon fraction $\omegab/\omegam=0.15
\pm 0.07$.  For a fixed set of cosmological parameters, i.e., $n=1,
\omegam = 1 - \omegal = 0.3$, $\Omega_{\rm b} h^{2} = 0.02$ and
$h=0.70$, the rms mass fluctuation amplitude of 2dFGRS galaxies smoothed
over a top-hat radius of $8h^{-1}$Mpc in redshift space turned out to be
$\sigma_{8{\rm g}}^S (L_s,z_s) \approx 0.94$.

\subsubsection{An upper limit on neutrino masses}

\begin{figure}[tbh]
\begin{center}
\leavevmode\epsfysize=8cm \epsfbox{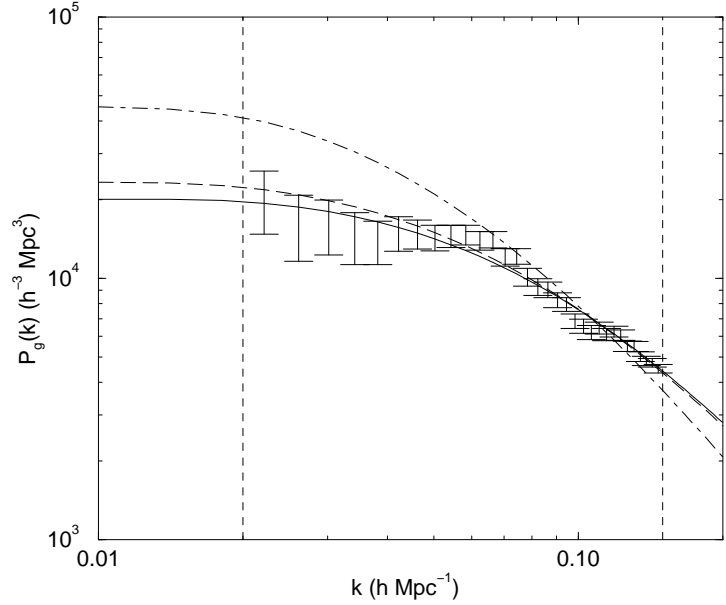} 
\caption {The power spectrum of the 2dFGRS.  The points with error bars
show the measured 2dFGRS power spectrum measurements in redshift space,
convolved with the window function.  Also plotted are linear CDM models
with neutrino contribution of $\Omega_{\nu} = 0, 0.01$ and 0.05 (bottom
to top lines) The other parameters are fixed to the concordance model.
The good fit of the linear theory power spectrum at $k > 0.15 \hompc$ is
due to a conspiracy between the non-linear gravitational growth and the
Finger-of-God smearing.  (refs.~\cite{percival-02, Elgaroy-02})}
\label{fig:neutrinomass}
\end{center}
\end{figure}

The recent results of atmospheric and solar neutrino
oscillations~\cite{fukuda-00, ahmad-01} imply non-zero mass-squared
differences of the three neutrino flavours.  While these oscillation
experiments do not directly determine the absolute neutrino masses, a
simple assumption of the neutrino mass hierarchy suggests a lower limit
on the neutrino mass density parameter, $\Omega_{\nu} = m_{{\nu},{\rm
tot}} h^{-2} /(94 eV) \approx 0.001$.  Large scale structure data can
put an upper limit on the ratio $\Omega_{\nu}/\omegam$ due to the
neutrino 'free streaming' effect~\cite{hu-98}.  By comparing the 2dF
galaxy power-spectrum of fluctuations with a four-component model
(baryons, cold dark matter, a cosmological constant and massive
neutrinos) it was estimated that $\Omega_{\nu}/\omegam < 0.13$ (95\%
CL), or with concordance prior of $\omegam =0.3$, $\Omega_{\nu} <
0.04$, or an upper limit of $\sim 2$ eV on the total neutrino mass,
assuming a prior of $h \approx 0.7$ ~\cite{Elgaroy-02, EL-03} (see
Fig.~\ref{fig:neutrinomass}).  In order to minimize systematic effects
due to biasing and non-linear growth, the analysis was restricted to the
range $ 0.02 < k < 0.15 \hompc$.  Additional cosmological data sets
bring down this upper limit by a factor of two~\cite{wmap-spergel}.

\subsubsection {Combining 2dFGRS \& CMB}

While the CMB probes the fluctuations in matter, the galaxy redshift
surveys measure the perturbations in the light distribution of
particular tracer (e.g. galaxies of certain type).  Therefore, for a
fixed set of cosmological parameters, a combination of the two can
constrain better cosmological parameters, and it can also provide
important information on the way galaxies are `biased' relative to the
mass fluctuations,
 
The CMB fluctuations are commonly represented by the spherical harmonics
$C_{\ell}$.  The connection between the harmonic $\ell$ and $k$ is
roughly
\begin{equation}
\ell \simeq k\, {\frac{2c}{H_0 \,\omegam^{0.4}}}\;  
\end{equation} 
for a spatially-flat universe.  For $\omegam =0.3$ the 2dFGRS range
$0.02 < k < 0.15 \hompc$ corresponds approximately to $ 200 < \ell <
1500$, which is well covered by the recent CMB experiments.

Recent CMB measurements have been used in combination with the 2dF power
spectrum.  Efstathiou et al. (2002)~\cite{efstathiou-02} showed that
2dFGRS+CMB provide evidence for a positive Cosmological Constant
$\omegal \sim 0.7$ (assuming $w=-1$), independently of the
studies of Supernovae Ia.  As explained in ref.~\cite{percival-02}, the
shapes of the CMB and the 2dFGRS power spectra are insensitive to Dark
Energy. The main important effect of the Dark Energy is to alter the
angular diameter distance to the last scattering, and thus the position
of the first acoustic peak. Indeed, the latest result from a combination
of WMAP with 2dFGRS and other probes gives $h = 0.71^{+0.04}_{-0.03}$,
$\omegab h^2 = 0.0224\pm0.0009$, $\omegam h^2 =
0.135^{+0.008}_{-0.009}$, $\sigma_8 = 0.84 \pm 0.04$, $\Omega_{\rm tot} =
1.02 \pm 0.02$, and $w <-0.78 $ (95\% CL, assuming $w \ge
-1$)~\cite{wmap-spergel}.

\subsubsection{Redshift-space distortion}

An independent measurement of cosmological parameters on the basis of
2dFGRS comes from redshift-space distortions on scales $ \simlt 10
\Mpc$; a correlation function $\xi(\pi,\sigma)$ in parallel and
transverse pair separations $\pi$ and $\sigma$.  As described in \S
\ref{section:lightcone}, the distortion pattern is a combination of the
coherent infall, parameterized by $\beta = \omegam^{0.6}/b$ and random
motions modelled by an exponential velocity distribution function
(eq.[\ref{eq:expveldist}]).  This methodology has been applied by many
authors. For instance, Peacock et al. (2001)~\cite{peacock-01} derived
$\beta(L_s=0.17,z_s=1.9L_*) = 0.43 \pm 0.07$ and Hawkins et
al. (2003)~\cite{hawkins-03} obtained $\beta(L_s=0.15,z_s=1.4L_*) = 0.49
\pm 0.09$ and velocity dispersion $\sigma_{\scriptscriptstyle {\rm P}} =
506 \pm 52$ km/sec.  Using the full 2dF+CMB likelihood function on the
$(b, \omegam$) plane, Lahav et al. (2002)~\cite{lahav-02} derived a
slightly larger (but consistent within the quoted error-bars) value,
$\beta(L_s=0.17,z_s=1.9L_*) \simeq 0.48 \pm 0.06$.

\subsubsection{The bi-spectrum and higher moments}

It is well established that important  information 
on the non-linear growth of structure is encoded at the high order
moments, e.g. the skewness or its Fourier version, the bi-spectrum.
Verde et al. ~\cite{verde-02}) 
computed the bi-spectrum of 2dFGRS and used it to measure 
the bias parameter of the galaxies.
They assumed  a specific quadratic biasing model:
$$
\delta_g = b_1 \delta_m +  {\frac{1}{2}}  b_2 \delta_m^2.
$$
By analysing 80 million triangle configurations 
in the wavenumber range $0.1 < k < 0.5 h$/Mpc they found 
$b_1 = 1.04 \pm 0.11$ and $b_2 = -0.054 \pm 0.08$, in support of
no biasing on large scale.
This is a non-trivial result, as the analysis covers non-linear scales.
Baugh et al.~\cite{baugh-04} and Croton et al.~\cite{croton-04})  
measured the moments of the galaxy count probability distribution function 
in 2dFGRG up to order $p=6$ (order $p=2$ is the variance, $p=3$ is 
the skewness, etc.). They demonstrated 
the hierarchical scaling of the averaged 
$p$-point galaxy correlation functions.
However, they found that the higher moments are  strongly affected 
by the presence of two massive superclusters in the 2dFGRS volume.
This poses the question of whether 2dFGRS is a 'fair sample' for high order 
moments.

\subsection {Luminosity and spectral-type dependence of galaxy clustering} 

Although biasing was commonly neglected until the early 1980s, it has
become evident observationally that on scales $ \simlt 10 \Mpc$
different galaxy populations exhibit different clustering amplitudes,
the so-called morphology-density relation~\cite{Dressler-80}. As
discussed in \S \ref{section:bias}, galaxy biasing is naturally
predicted from a variety of theoretical considerations as well as direct
numerical simulations~\cite{kaiser-84, mw-96, DL-99, TS-00, TMJS-01,%
YTJS-01}. Thus in this subsection, we summarize the extent to which the
galaxy clustering is dependent on the luminosity, spectral-type and
color of the galaxy sample from the 2dFGRS and SDSS.

\subsubsection {2dFGRS: Clustering per luminosity and spectral type} 

\begin{figure}[htb]
\begin{center}
\leavevmode\epsfxsize=8cm\epsfbox{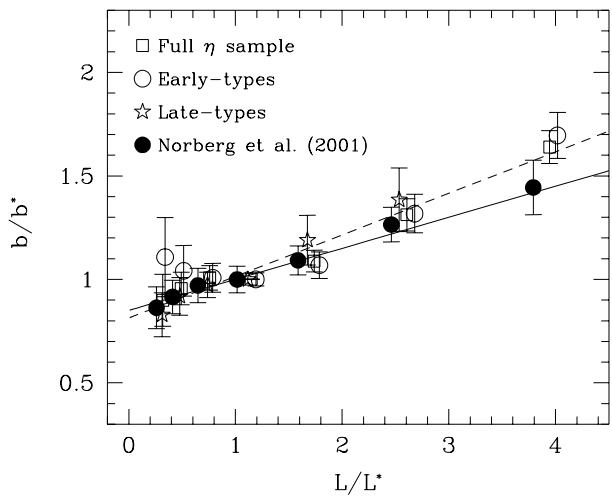} 
\caption{The variation of the galaxy biasing parameter with luminosity,
relative to an $L_*$ galaxy for the full sample and for subsamples of
early and late spectra types (ref.~\cite{Norberg-02}).  }
\label{fig:2df_b_var}
\end{center}
\end{figure}

Madgwick et al. (2003)~\cite{Madgwick-03} applied the Principal
Component Analysis to compress each galaxy spectrum into one quantity,
$\eta \approx 0.5\;pc_1 + pc_2$.  Qualitatively, $\eta$ is an indicator
of the ratio of the present to the past star formation activity of each
galaxy.  This allows one to divide the 2dFGRS into $\eta$-types, and to
study e.g. luminosity functions and clustering per type.  Norberg et
al. (2002)~\cite{Norberg-02} showed that, at all luminosities,
early-type galaxies have a higher bias than late-type galaxies, and that
the biasing parameter, defined here as the ratio of the galaxy to matter
correlation function, $b \equiv \sqrt{\xi_g/\xi_m}$ varies as $b/b_* =
0.85 + 0.15 L/L_*$.  Figure \ref{fig:2df_b_var} indicates that for $L_*$
galaxies, the real space correlation function amplitude of $\eta$
early-type galaxies is $\sim 50 \% $ higher than that of late-type
galaxies.

\begin{figure}[hbt]
\begin{center}
\leavevmode\epsfxsize=12cm\epsfbox{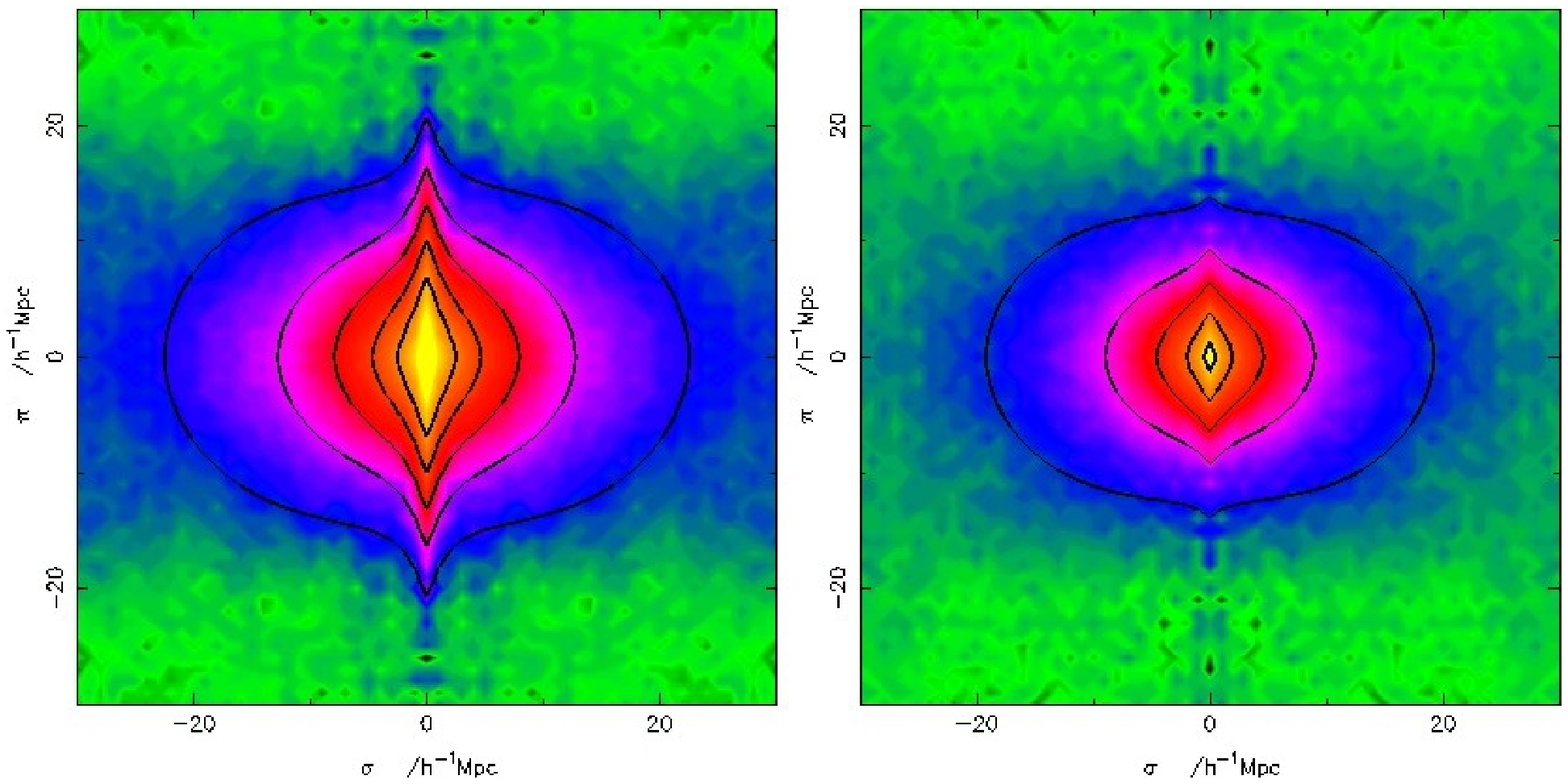} 
\caption{The two point correlation function $\xi(\sigma,\pi)$ plotted
for passively (left) and actively (right) star-forming galaxies.  The line
contours levels show the best-fitting model
(ref.~\cite{Madgwick-03}).}  
\label{fig:2df_xisigmapi_type}
\end{center}
\begin{center}
\leavevmode\epsfysize=8cm \epsfbox{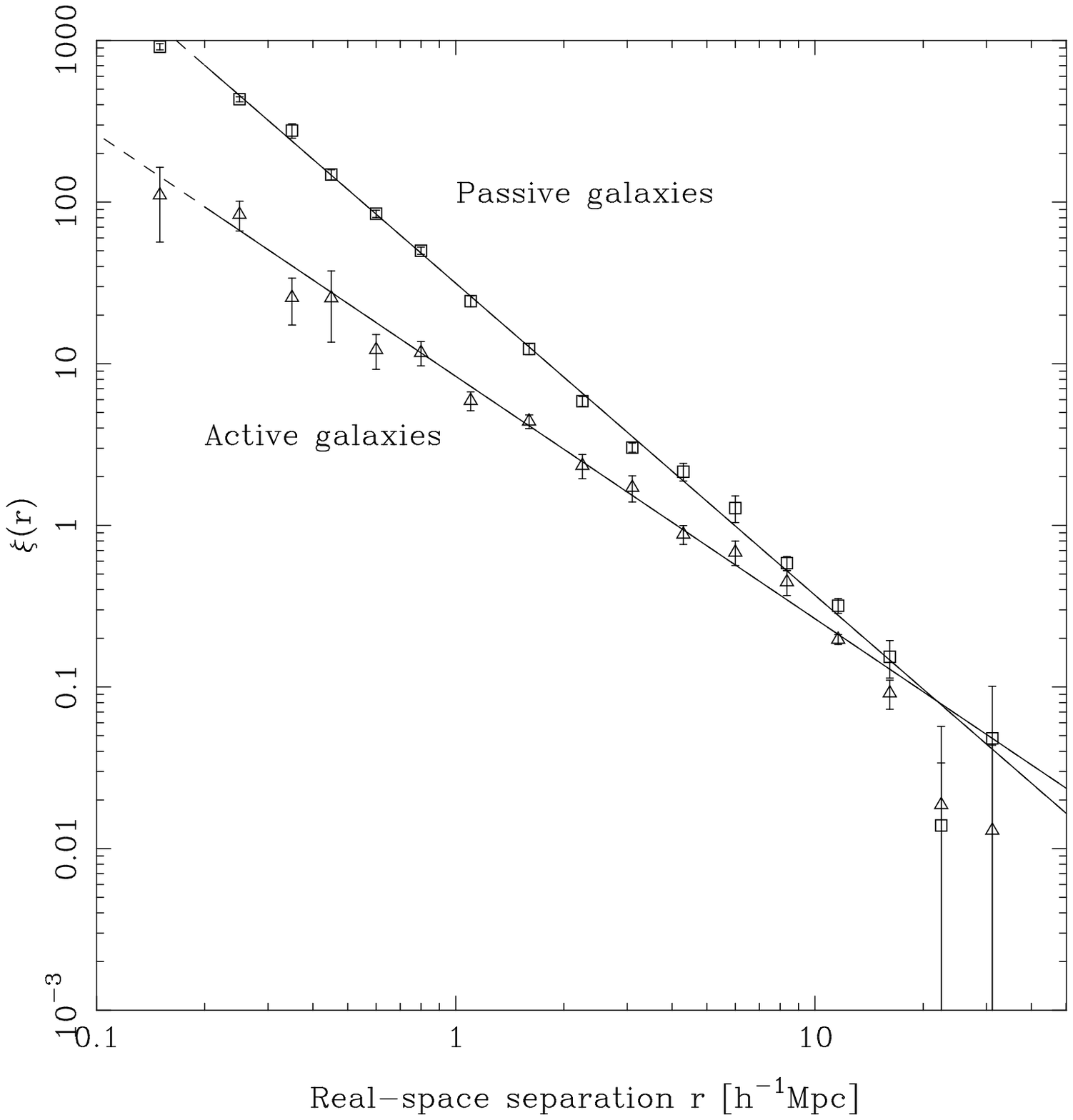} 
\caption{The correlation function for early and late spectral types.
The solid lines show best-fitting models, whereas the dashes lines are
extrapolations of these lines (ref.~\cite{Madgwick-03}).}  
\label{fig:2df_xitype}
\end{center}
\end{figure}

Figure \ref{fig:2df_xisigmapi_type} shows the redshift-space correlation
function in terms of the line-of-sight and perpendicular to the
line-of-sight separation $\xi(\sigma,\pi)$.  The correlation function
calculated from the most passively (`red', for which the present rate of
star formation is less than 10 \% of its past averaged value) and
actively (`blue') star-forming galaxies.  The clustering properties of
the two samples are clearly distinct on scales $\simlt 10 \Mpc$.  The
`red' galaxies display a prominent `finger-of-god' effect and also have
a higher overall normalization than the `blue' galaxies.  This is a
manifestation of the well-known morphology-density relation.  By fitting
$\xi(\pi, \sigma)$ over the separation range $8-20 h^{-1}$Mpc for each
class, it was found that $\beta_{active} = 0.49 \pm 0.13$,
$\beta_{passive} = 0.48 \pm 0.14$ and corresponding pairwise velocity
dispersions $\sigma_{\scriptscriptstyle {\rm P}}$ of $416 \pm 76$ km/sec
and $612 \pm 92$ km/sec ~\cite{Madgwick-03}.  At small separations, the
real space clustering of passive galaxies is stronger than that of
active galaxies, the slopes $\gamma$ are respectively 1.93 and 1.50
(Fig. \ref{fig:2df_xitype}) and the relative bas between the two classes
is a declining function of separation.  On scales larger than 10
$h^{-1}$Mpc the biasing ratio is approaching unity.

Another statistic was applied recently by Wild et al.~\cite{wild-04} and 
Conway et al.\cite{conway-04}, of a joint counts-in-cells
on 2dFGRS galaxies, classified by both color and spectral type. Exact linear
bias is ruled out on all scales.
The counts are better fitted to a bivariate lognormal distribution.
On small scales there is evidence
for stochasticity. Further investigation of galaxy formation models 
is required to understand the origin of the stochasticity. 

\clearpage

\subsubsection {SDSS: Two-point correlation functions 
 per luminosity and color \label{subsubsec:sdss-2pt}}

Zehavi et al. (2002)~\cite{zehavi-02} analyzed the Early Data Release
(EDR) sample of the SDSS 30,000 galaxies to explore the clustering of
per luminosity and color.  The inferred real-space correlation function
is well described by a single power-law; $\xi(r)=(r/6.1\pm0.2 h^{-1}{\rm
Mpc})^{-1.75\pm0.03}$, for $0.1 h^{-1} {\rm Mpc}\leq r\leq 16 h^{-1}
{\rm Mpc}$.  The galaxy pairwise velocity dispersion is
$\sigma_{12}\approx 600\pm100 km/sec$ for projected separations $0.15
h^{-1} {\rm Mpc} \leq r_p \leq 5 h^{-1} {\rm Mpc}$.  When divided by
color, the red galaxies exhibit a stronger and steeper real-space
correlation function and a higher pairwise velocity dispersion than do
the blue galaxies.  In agreement with 2dFGRS there is clear evidence for
scale-independent luminosity bias at $r \sim 10 h^{-1} {\rm
Mpc}$. Subsamples with absolute magnitude ranges centered on $M_*-1.5$,
$M_*$, and $M_*+1.5$ have real-space correlation functions that are
parallel power laws of slope $\approx -1.8$ with correlation lengths of
approximately $7.4 h^{-1} {\rm Mpc}$, $6.3 h^{-1} {\rm Mpc}$, and $4.7
h^{-1} {\rm Mpc}$, respectively.

\begin{figure}[htb]
\begin{center}
\leavevmode\epsfxsize=5.5cm \epsfbox{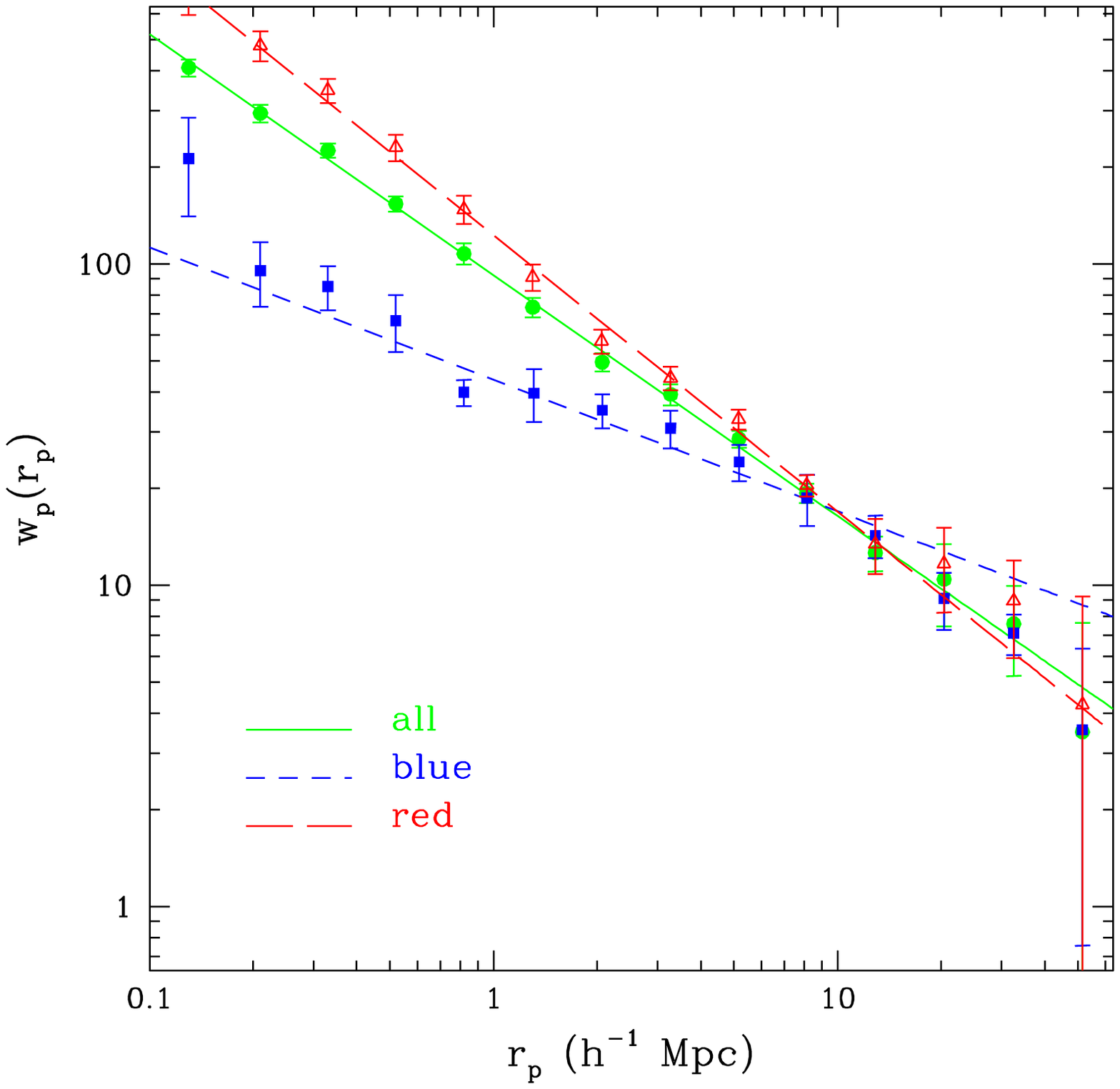} 
\vspace*{0.5cm}
\leavevmode\epsfxsize=5.5cm \epsfbox{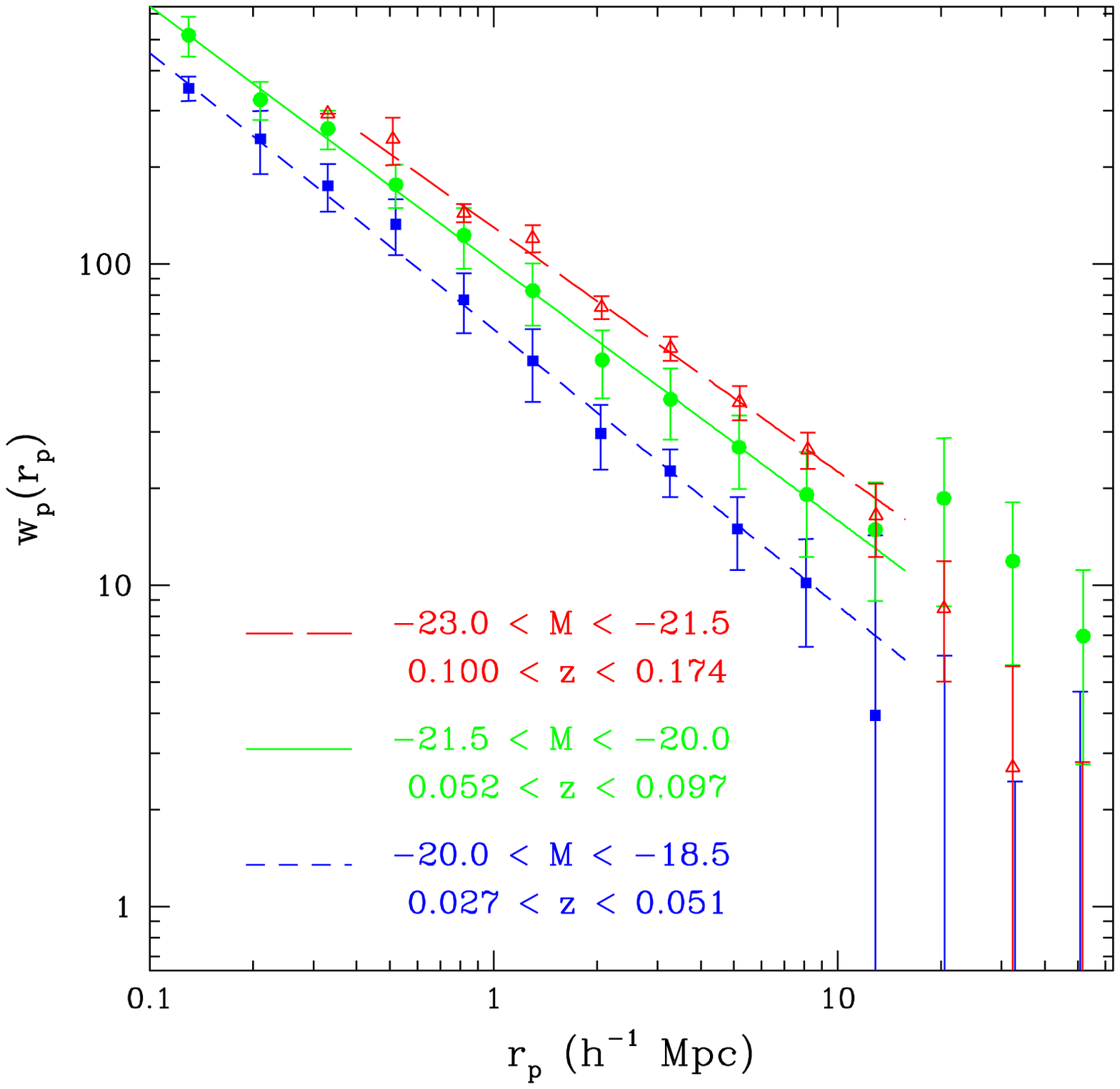} 
\caption{{\it Upper:} The
SDSS (EDR) projected correlation function for blue (squares), red
(triangles) and the full sample, with best-fitting models over the range
$0.1 < r_p < 16 h^{-1}$Mpc.  {\it Lower:} The SDSS (EDR) projected
correlation function for three volume-limited samples, with absolute
magnitude and redshift ranges as indicated and best-fitting power-law
models. (ref.~\cite{zehavi-02})} \label{fig:sdss_zehavi}
\end{center}
\end{figure}

Figures \ref{fig:2df_xitype} and \ref{fig:sdss_zehavi} pose an
interesting challenge to the theory of galaxy formation, to explain why
the correlation functions per luminosity bins have similar slope, while
the slope for early type galaxies is steeper than for late type.

\subsubsection {SDSS: Three-point correlation functions and the
   nonlinear biasing of galaxies per luminosity and color}

Let us move next to the the three-point correlation functions (3PCF) of
galaxies, which are the lowest-order unambiguous statistic to
characterize non-Gaussianities due to nonlinear gravitational evolution
of dark matter density fields, formation of luminous galaxies and their
subsequent evolution.  The determination of the 3PCF of galaxies was
pioneered by Peebles \& Groth (1975) and Groth \& Peebles
(1977)~\cite{PG1975,GP1977} using the Lick and Zwicky angular catalogs
of galaxies. They found that the 3PCF $\zeta (r_{12}, r_{23}, r_{31})$
obeys the {\it hierarchical relation}:
\begin{eqnarray}
  \zeta (r_{12}, r_{23}, r_{31}) 
 = Q_{\rm r} \, [\xi(r_{12})\xi(r_{23}) + 
           \xi(r_{23})\xi(r_{31}) + \xi(r_{31})\xi(r_{12})] 
\end{eqnarray}
with $Q_{\rm r}$ being a constant. The value of $Q_{\rm r}$ in real
space de--projected from these angular catalogues is $1.29 \pm 0.21$ for
$r < 3 h^{-1}{\rm Mpc}$.  Subsequent analyses of redshift catalogs
confirmed the hierarchical relation, at least approximately, but the
value of $Q_z$ (in redshift space) appears to be smaller, $0.5\sim1$.

As we have seen in \S \ref{subsubsec:sdss-2pt}, galaxy clustering is
sensitive to the intrinsic properties of the galaxy samples under
consideration, including their morphological types, colors, and
luminosities. Nevertheless the previous analyses were not able to
examine those dependences of 3PCFs because of the limited number of
galaxies. Indeed Kayo et al. (2004)~\cite{kayo-04} are the first to
perform the detailed analysis of 3PCFs explicitly taking account of the
morphology, color and luminosity dependence.  They constructed
volume-limited samples from a subset of the SDSS galaxy redshift data,
`Large--scale Structure Sample 12'. Specifically they divided each
volume limited sample into color subsamples of red (blue) galaxies,
which consist of 7949 (8329), 8930 (8155), and 3706 (3829) galaxies for
$-22<M_r-5\log h<-21$, $-21<M_r-5\log h<-20$, and $-20<M_r-5\log h<-19$,
respectively.

\begin{figure}[hbt]
\begin{center}
 \leavevmode\epsfxsize=7.5cm\epsfbox{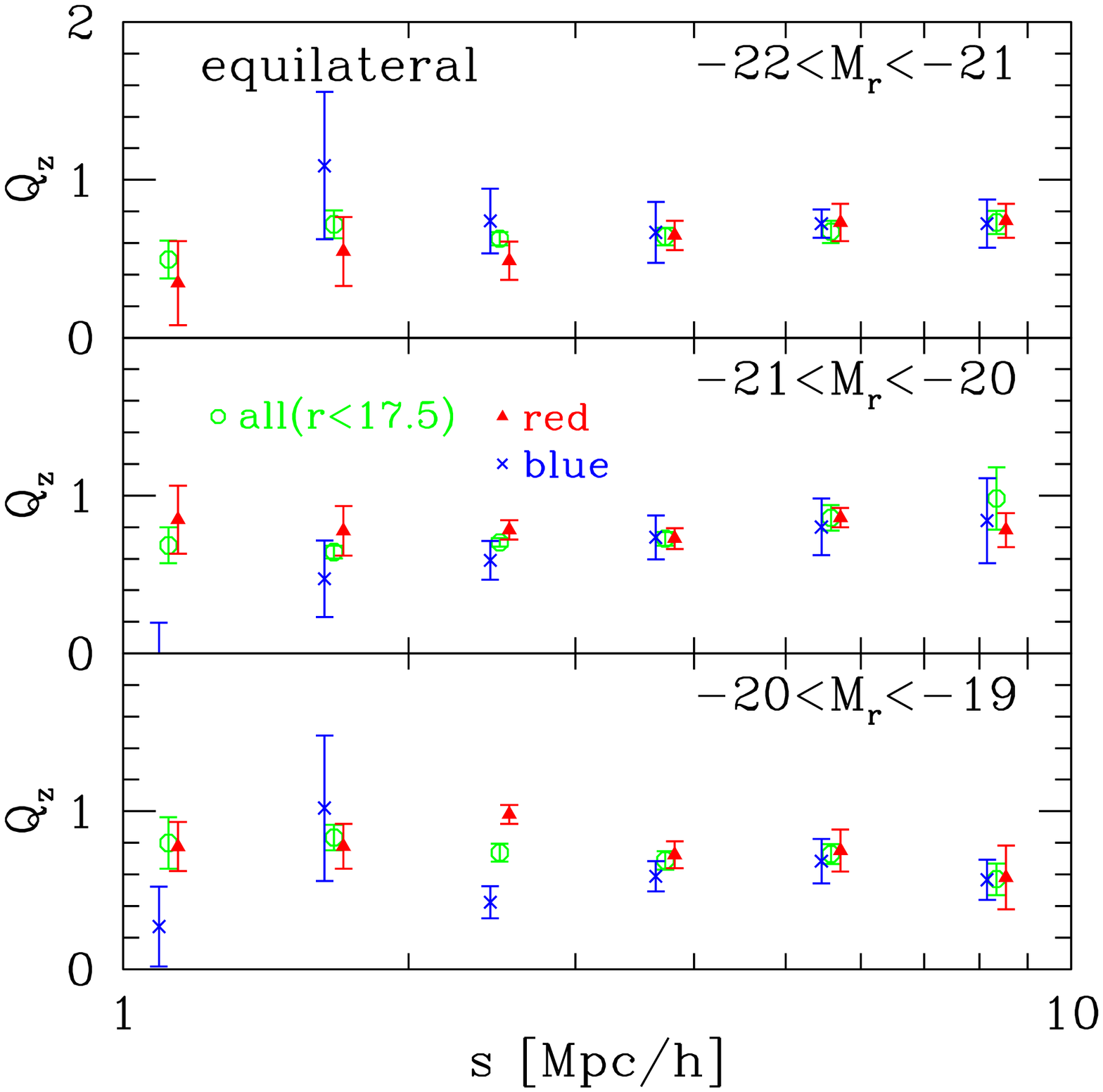}
\caption{Dimensionless amplitude of the three-point correlation
 functions of SDSS galaxies in redshift space. 
The galaxies are classified according to their colors; all
 galaxies in open circles, red galaxies in solid triangles, and blue
 galaxies in crosses.  (ref.~\cite{kayo-04}). \label{fig:Qcolor}}
\end{center}
\begin{center}
 \leavevmode\epsfxsize=7.5cm\epsfbox{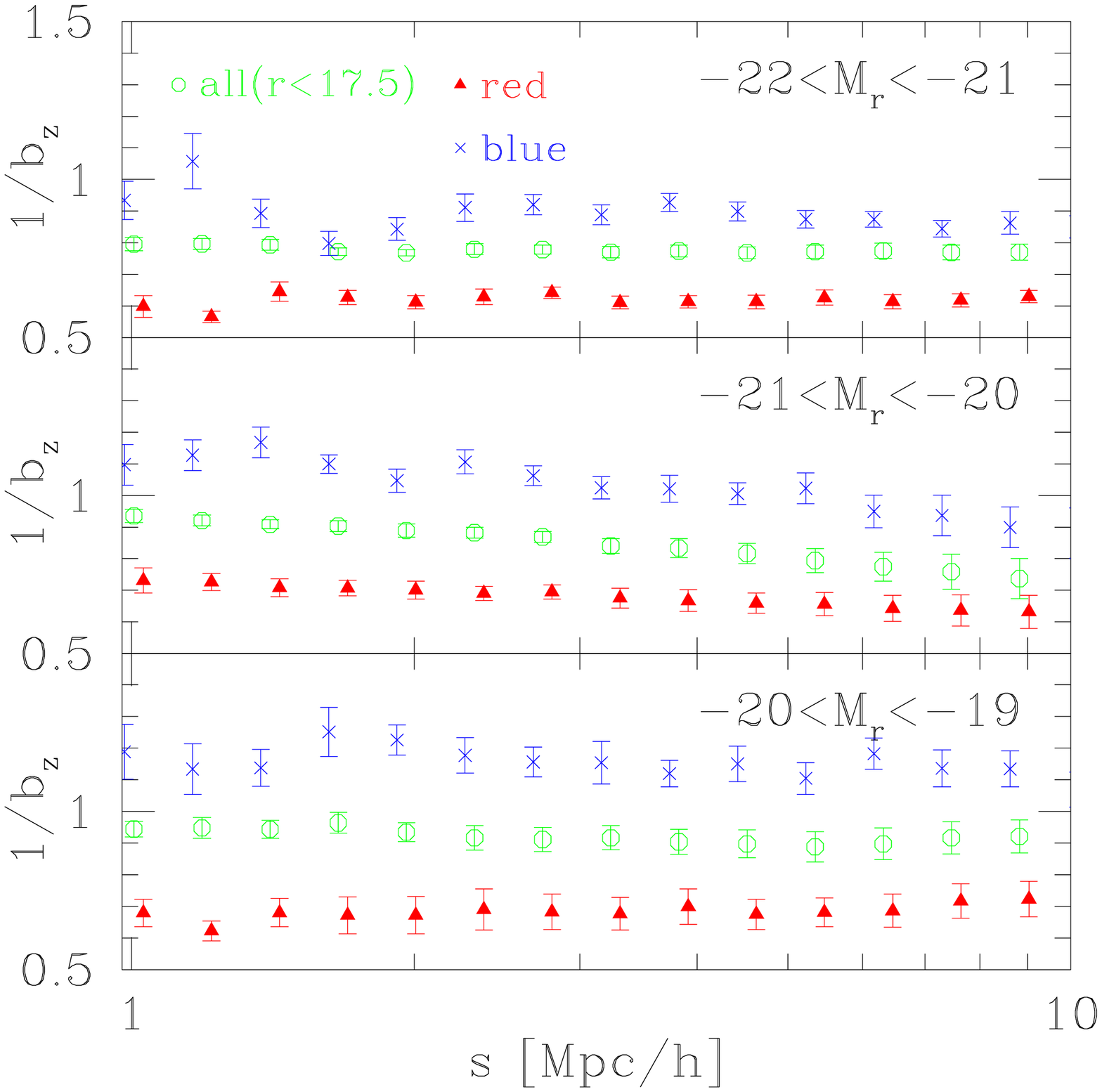}
\caption{Same as Fig.\ref{fig:Qcolor} but for
 the inverse of the biasing
 parameter defined through the two-point correlation functions.
(ref.~\cite{kayo-04}). \label{fig:binvcolor}}
\end{center}
\end{figure}

Figure \ref{fig:Qcolor} indicates the dimensionless amplitude of the
3PCFs of SDSS galaxies in redshift space:
\begin{eqnarray}
 Q_z(s_{12}, s_{23}, s_{31}) 
 \equiv  \frac{\zeta(s_{12}, s_{23}, s_{31})}
  {\xi(s_{12})\xi(s_{23})+\xi(s_{23})\xi(s_{31})+\xi(s_{31})\xi(s_{12})}
\label{eq:Qdef}
\end{eqnarray}
for the equilateral triplets of galaxies.  The overall conclusion is
that $Q_z$ is almost scale-independent and ranges between 0.5 and 1.0,
and that no systematic dependence is noticeable on luminosity and color.
This implies that the 3PCF itself does depend on the galaxy properties
since two-point correlation functions (2PCFs) exhibit clear dependence on
luminosity and color.  Previous simulations and theoretical models~
\cite{suto-93,MS1994,M1994b,TJ2003} indicate that $Q$ decreases with
scale in both real and redshift spaces.  This trend is not seen in the
observational results.

\clearpage

In order to demonstrate the expected dependence in
the current samples, they compute the biasing parameters estimated from
the 2PCFs:
\begin{eqnarray}
 b_{z, i}(s) \equiv 
\sqrt{\frac{\xi_{z,i}(s)}{\xi_{z,{\rm \Lambda CDM}}(s)}} ,
\end{eqnarray} 
where the index $i$ runs over each sample of galaxies with different
colors and luminosities.  The predictions of the mass 2PCFs in
redshift space, $\xi_{z,{\rm \Lambda CDM}}(s)$, in the $\Lambda$ cold dark
matter model are computed following ref.~\cite{HCS-01}.

As an illustrative example, consider a simple bias model in which the
galaxy density field $\delta_{{\rm g}, i}$ for the $i$-th population of
galaxies is given by
\begin{eqnarray}
\label{eq:quadratic_bias}
\delta_{{\rm g}, i} = b_{{\rm g}, i (1)}\delta_{\rm mass}
+ b_{{\rm g}, i (2)}\delta_{\rm mass}^2.
\end{eqnarray} 
If both $b_{{\rm g}, i (1)}$ and $b_{{\rm g}, i (1)}$ are constant and
mass density field $\delta_{\rm mass} \ll 1$, equation (\ref{eq:Qdef})
implies that
\begin{eqnarray}
\label{eq:Q_quadratic_bias}
Q_{{\rm g}, i} = \frac{1}{b_{{\rm g}, i(1)}}
Q_{\rm mass} + \frac{b_{{\rm g}, i(2)}}{b_{{\rm g}, i(1)}^2} .
\end{eqnarray} 
Thus the linear bias model ($b_{{\rm g}, i(2)}=0$) simply implies that
$Q_{{\rm g}, i}$ is inversely proportional to $b_{{\rm g}, i(1)}$, which
is plotted in Figure \ref{fig:binvcolor}.  Comparison of Figures
\ref{fig:Qcolor} and \ref{fig:binvcolor} indicates that the biasing in
the 3PCFs seems to compensate the difference of $Q_{\rm g}$ purely due
to that in the 2PCFs.

Such behavior is unlikely to be explained by any simple model inspired
by the perturbative expansion like equation
(\ref{eq:quadratic_bias}). Rather it indeed points to a kind of
regularity or universality of the clustering hierarchy behind galaxy
formation and evolution processes.  Thus the galaxy biasing seems much
more complex than the simple deterministic and linear model. More
precise measurements of 3PCFs and even higher-order statistics with
future SDSS datasets would be indeed valuable to gain more specific
insights into the empirical biasing model.

\subsection {Topology of the universe: analysis of
  SDSS galaxies in terms of Minkowski functionals} 

All the observational results presented in the preceding subsections
were restricted to the two-point statistics. As emphasized in \S
\ref{section:statistics}, the clustering pattern of galaxies has much
richer content than the two-point statistics can probe.  Historically
the primary goal of the topological analysis of galaxy catalogues: was
to test Gaussianity of the primordial density fluctuations.  Although
the major role for that goal has been superseded by the CMB map
analysis~\cite{wmap-komatsu}, the proper characterization of the
morphology of large-scale structure beyond the two-point statistics is
of fundamental importance in cosmology.  In order to illustrate a
possibility to explore the topology of the universe by utilizing the new
large surveys, we summarize the results of the Minkowski Functionals
(MF) analysis of SDSS galaxy data ~\cite{hikage-03}.

In an apparent-magnitude limited catalogue of galaxies, the average
number density of galaxies decreases with distance because only
increasingly bright galaxies are included in the sample at larger
distance.  With the large redshift surveys it is possible to avoid this
systematic change in both density and galaxy luminosity by construct
volume--limited samples of galaxies, with cuts on both
absolute--magnitude and redshift.  This is in particular useful for
analyses such as MF and was carried out in the analysis shown here.

\begin{figure}[hbt]
\begin{center}
 \leavevmode\epsfxsize=12cm\epsfbox{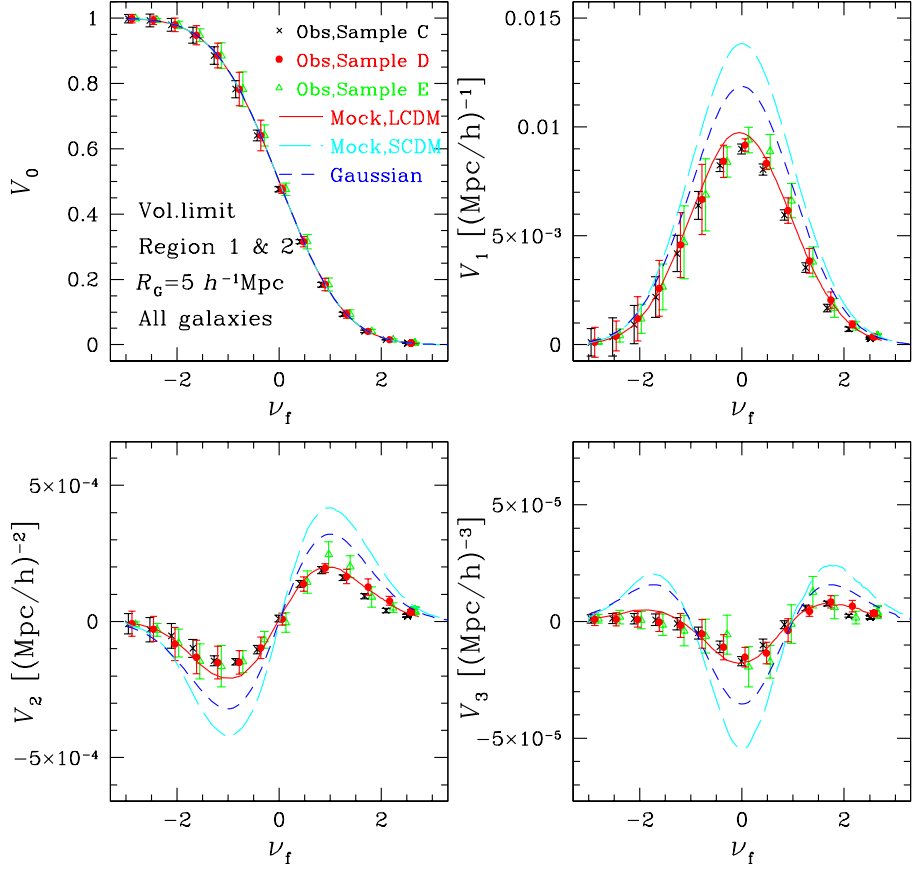}
\caption{MFs as a function of $\nu_{\rm f}$ for $R_{\rm G}=5h^{-1}$Mpc
for SDSS data.  Averaged MFs of the mock samples are plotted for LCDM
(solid lines) and SCDM (long dashed lines).  Gaussian model predictions
(eqs.[\ref{eq:minkowski_v0}] to [\ref{eq:minkowski_v3}]) are also
plotted with short dashed lines.  The results favor the LCDM model with
random--Gaussian initial conditions. (ref.~\cite{hikage-03}).}
\label{fig:volmagvfr}
\end{center}
\end{figure}

Figure \ref{fig:volmagvfr} shows the MFs as a function of $\nu_{\rm f}$
defined from the volume fraction~\cite{GMD-86}:
\begin{equation}
\label{eq:nuf}
f=\frac{1}{\sqrt{2\pi}}\int^\infty_{\nu_{\rm f}} e^{-x^2/2}dx .
\end{equation}
This is intended to map the threshold so that the volume fraction on the
high-density side of the isodensity surface is identical to the volume
in regions with density contrast $\delta =\nu_{\rm f} \sigma$, for a Gaussian
random field with r.m.s. density fluctuations $\sigma$.  If the evolved
density field may approximately have a good one-to-one correspondence
with the initial random-Gaussian field, then this transformation removes
the effect of evolution of the PDF of the density field.  Under this
assumption, the MFs as a function of volume fraction would be sensitive
only to the topology of the isodensity contours rather than evolution
with time of the density threshold assigned to a contour. While the
limitations of the approximation of monotonicity in the relation between
initial and evolved density fields are well recognized~\cite{kayo-01},
we plot the result in this way for simplicity.

The good match between the observed MFs and the mock predictions based
on the LCDM model with the initial random--Gaussianity, as illustrated
in Figure~\ref{fig:volmagvfr}, might be interpreted to imply that the
primordial Gaussianity is confirmed.  A more conservative interpretation
is that, given the size of the estimated uncertainties, these data do
not provide evidence for initial non--Gaussianity, i.e., the data are
{\it consistent} with primordial Gaussianity.  Unfortunately due to the
statistical limitation of the current SDSS data, it is not easy to put
more quantitative statement concerning the initial Gaussianity.
Moreover, in order to go further and place more quantitative constraints
on primordial Gaussianity with upcoming data, one needs a more precise
and reliable theoretical model for the MFs which properly describes the
nonlinear gravitational effect possibly as well as galaxy biasing beyond
the simple mapping on the basis of the volume fraction.  In fact, galaxy
biasing is a major source of uncertainty for relating the observed MFs
to those obtained from the mock samples for dark matter distributions.
If LCDM is the correct cosmological model, the good match of the MFs for
mock samples from the LCDM simulations to the observed SDSS MFs may
indicate that nonlinearity in the galaxy biasing is relatively small, at
least small enough that it does not significantly affect the MFs (the
MFs as a function of $\nu_{\sigma}$ remain unchanged for the linear
biasing).
\clearpage

\subsection{ Other statistical measures}

In this section, we have presented the results on the basis of
particular statistical measures including the two-point correlation
functions, power spectrum, redshift distortion, and Minkowski
functionals.  Of course there are other useful approaches in analysing
redshift surveys; the void probability function, counts-in-cells,
Voronoi cells, percolation, and minimal spanning trees.  Another area
not covered here is optimal reconstruction of density field (e.g, using
the Wiener filtering).  The reader is referred to a good summary of
those and other methodology in the book by Martinez \& Saar
(2002)~\cite{martinez-saar-02}.

Admittedly the results that we presented here are rather observational
and phenomenological, and far from being well-understood
theoretically. It is quite likely that when other on-going and future
surveys are be analysed in great detail, the nature of galaxy clustering
will be revealed in a much more quantitative manner. They are supposed
to act as a bridge between cosmological framework and galaxy formation
operating in the universe. While the proper understanding of physics of
galaxy formation is still far away, the future redshift survey data will
present interesting challenges for constructing models of galaxy
formation.

\clearpage
\section{Discussion}
\label{section:discussion}

As a classical probe, galaxy redshift surveys still remain an important
tool for studying cosmology and galaxy formation.  On the large scales
($ > 10 h^{-1}$ Mpc or so) they nicely complement the Cosmic Microwave
Background, Supernovae Ia and Gravitational Lensing in quantifying in
detail the cosmological model.  On the small scales ($< 10 h^{-1}$ Mpc)
the clustering patterns of different galaxy types (defined by structural
or spectral properties) provide important constraints on models of
biased galaxy formation.

The redshift surveys mainly constrain $\Omega_{\rm m}$ via both redshift
distortion (which also depends on biasing) and the shape of the
$\Lambda$-CDM power spectrum, which depends on the primordial spectrum,
the product $\Omega_{\rm m} h$ and also the baryon density $\Omega_{\rm
b}$.  Redshift surveys at a given epoch are not sensitive to the Dark
Energy (or the Cosmological Constant, in a specific case), but combined
with the CMB they can constrain the cosmic Equation of State.

A good example of the importance of the redshift surveys in Cosmology is
given by the recent WMAP analysis of cosmological parameters, where the
estimation of certain parameters was much improved by adding the 2dF
power spectrum of fluctuations (\cite{wmap-spergel}) or the SDSS power
spectrum (\cite{sdss-tegmark}).  This is illustrated in
Table~\ref{best_param} by contrasting the WMAP alone derived parameters
from those derived from WMAP+SDSS (\cite{sdss-tegmark}).  Such results
are sensitive to the assumed parameter space and priors, but for
simplicity we quote here the results for the simple six-parameter
model. In this analysis it was assumed that the universe is flat, the
fluctuations are adiabatic, no gravity waves, no running tilt of the
spectral index, negligible neutrino masses and that the Dark Energy is
in the form of Einstein's Cosmological constant ($w=-1$).  Within the
$\Lambda$-CDM model this scenario can be characterised by the six
parameters given in Table~\ref{best_param} based on \cite{sdss-tegmark}.
The WMAP data used are both the temperature and polarization
fluctuations. It can bee seen that by adding the SDSS information more
than halves the WMAP-only error bars on some of the parameters.  These
results are in good agreement with the joint analysis WMAP+2dF
(\cite{wmap-spergel}).

\begin{table}[htb]
\small \caption{Derived 6 cosmological parameters from 
WMAP alone (temperature \& polarization) vs. WMAP+SDSS, 
from Tables 2 and 3 of Tegmark et al. (2003).\cite{sdss-tegmark}
The quoted error bars correspond to 1-$\sigma$. \label{best_param}}
\begin{tabular}{lllccccc}
Symbol & WMAP alone & WMAP+SDSS & description \\ \hline
$\Omega_\Lambda$ & $0.75^{+0.10}_{-0.10}$ & $0.699^{+0.042}_{-0.045}$ 
& dimensionless cosmological constant\\
$\Omega_b h^2 $ & $0.0245^{+0.0050}_{-0.0019}$ &
 $0.0232^{+0.0013}_{-0.0010}$ 
& baryon density parameter\\
$\Omega_{\rm m} h^2 $ & $0.140^{+0.020}_{-0.018}$ & $0.1454^{+0.0091}_{-0.0082}$ & total matter  density parameter\\ 
$\sigma_8$ & $0.99^{+0.19}_{-0.14}$  & $0.917^{+0.090}_{-0.072}$ &  
mass fluctuation amplitude \\
&&&at $8h^{-1}$Mpc sphere (linear theory) \\
$n_s$ & $1.02^{+0.16}_{-0.06}$ & $0.977^{+0.039}_{-0.025}$ 
&  primordial scalar spectral index \\
&&& at $k = 0.05$/Mpc\\
$\tau$ & $0.21^{+0.24}_{-0.11}$ & $0.124^{+0.083}_{-0.057}$ 
& re-ionization optical depth
\\ \hline
\end{tabular}
\end{table}

We emphasize that 
these parameters
were fitted assuming the $\Lambda$-CDM model.
While the
degree of such phenomenological successes of the $\Lambda$-CDM model is
truly amazing, there are many fundamental open questions:

\begin{itemize}

\item[(i)] Both components of the model, $\Lambda$ and CDM, have not been
directly measured.  Are they `real' entities or just `epicycles'?
Historically epicycles were actually quite useful in forcing observers
to improve their measurements and theoreticians to think about better
models!

\item[(ii)] `The Old Cosmological Constant problem': Why is $\Omega_\Lambda$
at present so small relative to what is expected from Early Universe
physics?

\item[(iii)] `The New Cosmological Constant problem': Why is $\Omega_{\rm m} \sim
      \Omega_\Lambda$ at the present-epoch ? Why is $w \sim -1$ ?
Do we need to introduce new physics 
or to invoke the Anthropic Principle to explain it ?

\item[(iv)] There are still open problems in $\Lambda$-CDM on the small scales
e.g. galaxy profiles and satellites.

\item[(v)]  
Could other (yet unknown) models fit the data equally well ?

\item[(vi)]
Where does the field go from here ?
Should the activity focus on refinement of 
the cosmological parameters within $\Lambda$-CDM, 
or on introducing entirely new paradigms ?
\end{itemize}

Even if $\Lambda$CDM turns out to be the correct model, it is not yet
the ``end of cosmology''.  Beyond the `zero-th order' task of finding
the cosmological parameters of the FRW model, we would like to
understand the non-linear growth of mass density fluctuations and then
the formation and evolution of luminous objects.  The wealth of data
of galaxy images and spectra in the new surveys calls for the development
of more detailed models of galaxy formation.  This is important so the
comparison of the measurements (e.g. correlation function per spectral
type or colour) and the models could be done on equal footing, with
the goal of constraining scenarios of galaxy formation.
There is also room for new statistical methods to quantify the 
`cosmic web' of filaments, clusters of voids, for effective comparison
with the simulations.
It may well be that in the future the cosmological parameters will be fixed
by the CMB, SN Ia and other probes. Then, for fixed cosmological parameters,
one may use redshift surveys primarily to study galaxy biasing and 
evolution with cosmic epoch.


\clearpage
\section{Acknowledgements}
\label{section:acknowledgements}

We thank Joachim Wambsganss and Bernard Schutz for inviting us to write
the present review article. O.L. thanks members of the 2dFGRS team and
the Leverhulme Quantitative Cosmology group for helpful discussions.
Y.S. thanks all his students and collaborators for over many years, in
particular, Thomas Buchert, Takashi Hamana, Chiaki Hikage, Yipeng Jing,
Issha Kayo, Hiromitsu Magira, Takahiko Matsubara, Hiroaki Nishioka, Jens
Schmalzing, Atsushi Taruya, Kazuhiro Yamamoto, and Kohji Yoshikawa among
others, for enjoyable and fruitful collaborations whose results form
indeed the important elements in this review. Y.S. is also grateful for
the hospitality at Institute of Astronomy, University of Cambridge,
where the most of present review was put together and written up for
completion. O.L. acknowledges a PPARC Senior Research Fellowship.
We also thank Idit Zehavi for permitting us to use Figure
\ref{fig:sdss_zehavi}.

Numerical simulations were carried out at ADAC (the Astronomical Data
Analysis Center) of the National Astronomical Observatory, Japan
(project ID: mys02, yys08a).  This research was also supported in part
by the Grants--in--Aid from Monbu--Kagakusho and Japan Society of
Promotion of Science.

\bibliography{}

\end{document}